\documentclass[12pt,  a4paper]{article}
\usepackage[top=1in, bottom=1.5in,left=1in,right=1in]{geometry}
\usepackage{amsmath,amssymb,amsfonts, amsthm}
\usepackage{upgreek}
\usepackage{physics}
\usepackage{bm}
\usepackage{dsfont}
\usepackage{tabularx}
\usepackage[final, hidelinks]{hyperref}
\usepackage{diagbox}
\hypersetup{allcolors=black}
\usepackage{enumitem} 

\usepackage{booktabs,siunitx,threeparttable,float} 

\usepackage[ruled,vlined]{algorithm2e}

\usepackage{graphicx}
\usepackage{placeins} 
\usepackage{flafter} 
\usepackage{needspace}

\usepackage{caption}
\usepackage{xcolor}

\newtheorem{theorem}{Theorem}[section]

\newtheorem{lemma}{Lemma}[section]

\newtheorem{assumption}{Assumption}[section]

\renewcommand{\epsilon}{\varepsilon}

\usepackage[round]{natbib}
\usepackage[utf8]{inputenc}
\usepackage[toc]{appendix}
\author{Ruolei Zhu\thanks{Department of Economics, University College London. E-mail: \href{mailto:ruolei.zhu.19@ucl.ac.uk}{ruolei.zhu.19@ucl.ac.uk}.}}
\title{Demand Estimation for Rationally Inattentive Consumers}
\date{October 21, 2026\\[0.4em]
{\small\emph{\href{https://drive.google.com/file/d/1uPjpgyVCV_8fJIIBoQmqBC1PCxkzb7Mi/view?usp=sharing}{Click here for the latest version}}}}

\begin{document}
\hypersetup{pageanchor=false}
\maketitle
\thispagestyle{empty}

\begin{abstract}
We develop a rational-inattention BLP (RI--BLP) model for demand estimation from aggregate market shares, allowing for endogenous prices and unobserved preference heterogeneity. We establish conditions for uniqueness of market-share inversion and construct a simulated RI--BLP GMM estimator with instrumental-variable moments and market-share constraints. Under regularity and simulation-accuracy conditions, the estimator is consistent and asymptotically normal as the number of markets grows, with the number of products fixed. Monte Carlo experiments show accurate recovery of mean utility coefficients, with less precise estimates of dispersion parameters. In a canned-tuna application, RI--BLP and logit yield similar average own-price elasticities. Unlike logit, the fitted RI--BLP model implies that a price increase can reduce demand even for similar products and shift demand toward the outside option.

\end{abstract}

\clearpage
\pagenumbering{arabic}
\hypersetup{pageanchor=true}

\section{Introduction}
\label{sec:introduction}

One of the most widely used frameworks in structural demand estimation is the BLP model developed by \citet{berry1995}. It is a random-coefficients discrete-choice model that is typically estimated using aggregate market-level data, although it is derived from an underlying individual-level random utility framework. In each market $t=1,\dots,T$, consumer $i$ chooses an alternative from $\mathcal J=\{0,1,\dots,J\}$ to maximize utility. Here, $j=0$ denotes the outside option, and $j=1,\dots,J$ denote the inside products.\footnote{Our model allows choice sets to vary across markets, with $\mathcal J_t=\{0,1,\dots,J_t\}$. This changes the dimensions of the market-specific vectors but leaves the basic model and estimation formulation unchanged. For simplicity, our analysis uses a common choice set $\mathcal J=\{0,1,\dots,J\}$ across markets.} In the standard BLP model, utility from choosing an inside product $j$ is given by
\[
u_{ijt}=\beta_i^{\prime}x_{jt}-\alpha_i p_{jt}+\xi_{jt}+\varepsilon_{ijt}.
\]

Here, $x_{jt}$ is a vector of product characteristics for product $j$ in market $t$, and $p_{jt}$ is its price. The consumer-specific coefficients $\beta_i$ and $\alpha_i$ capture preferences for product characteristics and price sensitivity, respectively. The term $\xi_{jt}$ is a product-market demand shifter that is common across consumers and observed by consumers and firms, but unobserved by the Econometrician. The term $\varepsilon_{ijt}$ is an individual-specific utility shock observed by consumer $i$, but unobserved by firms and the Econometrician. Firms may respond to the demand shifter $\xi_{jt}$ when setting prices, so equilibrium prices $p_{jt}$ can be correlated with $\xi_{jt}$, making price endogenous in the demand equation.

The BLP framework builds on traditional random utility models by incorporating random coefficients that capture heterogeneous preferences, together with product-market demand shifters. However, the standard BLP model assumes that consumers know all product characteristics relevant to utility when making their choices. This assumption may be restrictive in differentiated-product markets, where consumers face many options and may not be fully informed before choosing. Rational inattention provides a natural framework for relaxing this assumption. In this framework, consumers may directly observe prices and some product characteristics while choosing how much costly attention to devote to learning about other utility-relevant characteristics.

Recent applications of rational-inattention discrete-choice models include insurance choice \citep{brown2024endogenous}, consumer-welfare analysis \citep{joo2023rational}, and online food-delivery demand \citep{natan2025choice}. All three use individual-level choice data. We instead develop a rational-inattention BLP (RI--BLP) model for demand estimation from aggregate market shares. The model retains unobserved preference heterogeneity, product-market demand shifters, and endogenous prices while allowing consumers to acquire costly information about uncertain utility.

Our utility structure is related to the market-level rational-inattention model of \citet[Section~4]{liao2024}. That model includes individual-level additive taste shocks and allows utility coefficients to differ across observed demographic groups, but keeps these coefficients common within each group. Our model additionally allows consumer-specific random coefficients on product characteristics and price.

We begin by decomposing consumer $i$'s utility from choosing alternative $j\in\mathcal J$ in market $t$ as
\[
u_{ijt}=u_{0,ijt}+u_{1,ijt}.
\]
The component $u_{0,ijt}$ is observed by the consumer before information acquisition, while $u_{1,ijt}$ is the latent utility component about which the consumer may acquire costly information. Following the rational-inattention discrete-choice framework of \citet{matejka2015rational}, consumers choose an information-acquisition strategy and a product-choice rule to maximize expected utility net of information costs. For inside products $j=1,\dots,J$, a general RI--BLP specification is
\begin{equation}
\begin{aligned}
u_{0,ijt}&=\beta_{0,i}^{\prime}x_{0,jt}-\alpha_i p_{jt}+\xi_{0,jt}+\varepsilon_{0,ijt},\\
u_{1,ijt}&=\beta_1^{\prime}x_{1,jt}+\xi_{1,jt}+\varepsilon_{1,ijt}.
\end{aligned}
\label{eq:Udef_general}
\end{equation}

Here, $x_{0,jt}$ and $x_{1,jt}$ are vectors of product characteristics, and $p_{jt}$ is price. The product-market demand shifters $\xi_{0,jt}$ and $\xi_{1,jt}$ enter $u_{0,ijt}$ and $u_{1,ijt}$, respectively. The terms $\varepsilon_{0,ijt}$ and $\varepsilon_{1,ijt}$ are individual-level additive utility shifters/shocks entering the corresponding components. BLP-style heterogeneity is captured by the consumer-specific coefficients $\beta_{0,i}$ and $\alpha_i$, which are known to the consumer before information acquisition. The coefficient $\beta_1$ is common across consumers. In particular, setting $\beta_{0,i}=\beta_0$ and $\alpha_i=\alpha$ for every consumer gives the common-coefficient specification. The general specification also includes standard BLP as a full-information special case.\footnote{Set $x_{0,jt}=x_{jt}$, $\beta_{0,i}=\beta_i$, $\xi_{0,jt}=\xi_{jt}$, and $u_{1,ijt}=0$, with the additive-shock distribution corresponding to standard BLP under the same outside-option normalization. All utility is then observed by the consumer before choice. This nesting statement concerns the general specification in Equation~\eqref{eq:Udef_general}, before omitting the baseline demand shifter as described below.}

The timing of the model is as follows. At stage $\tau=0$, firms set prices. At stage $\tau=1$, consumers observe $u_{0,ijt}$ but face uncertainty about $u_{1,ijt}$ and choose an information-acquisition strategy and a product-choice rule. At stage $\tau=2$, consumers receive the information generated by their chosen strategy and choose among alternatives using their product-choice rule. These choices aggregate into market shares. At stage $\tau=3$, the Econometrician observes aggregate market-level data. We next describe the information available to firms, consumers, and the Econometrician.

On the supply side, firms observe the realizations of $x_{0,jt}$, $x_{1,jt}$, $\xi_{0,jt}$, and $\xi_{1,jt}$ when setting prices. Because both demand shifters affect demand and are observed by firms when prices are set, equilibrium prices may be correlated with them, making price endogenous in the demand equation.

From the consumer's perspective, $u_{0,ijt}$ is observed before information acquisition, while $u_{1,ijt}$ contains utility-relevant uncertainty. The consumer holds a subjective prior $\pi_{u_1}$ and chooses an information-acquisition strategy to maximize expected utility net of information costs. The consumer then chooses using information that need not fully reveal $u_{1,ijt}$.

From the Econometrician's perspective, $x_{0,jt}$, $x_{1,jt}$, $p_{jt}$, and aggregate market shares $s_{jt}$ are observed. The consumer-specific coefficients $\beta_{0,i}$ and $\alpha_i$, demand shifters $\xi_{0,jt}$ and $\xi_{1,jt}$, and utility shocks $\varepsilon_{0,ijt}$ and $\varepsilon_{1,ijt}$ are unobserved. We treat the joint data-generating distribution $F_{\varepsilon}$ of the additive shocks as known. We specify a parametric family for the random-coefficient distribution. We jointly estimate the population means, the heterogeneity parameters, and the common coefficient $\beta_1$. Appendix~\ref{appendix:information_structure} summarizes the information structure for firms, consumers, and the Econometrician.

To simplify the analysis, we set $\xi_{0,jt}=\varnothing$ and retain $\xi_{1,jt}$ in the latent utility component for the remainder of the paper. The remaining product-market demand shifter $\xi_{1,jt}$ is unobserved by the Econometrician, as is $\xi_{jt}$ in standard BLP. However, standard BLP assumes that consumers know $\xi_{jt}$ when they choose. In our model, $\xi_{1,jt}$ enters the latent utility component $u_{1,ijt}$, about which consumers are initially uncertain and may acquire information. Allowing such uncertainty is more realistic when the demand shifter captures aspects of product quality, such as reliability, durability, and future maintenance costs, that are difficult to assess before purchase.

This simplification preserves the model's main information channel. The specification still allows for asymmetric information among consumers, firms, and the Econometrician. This asymmetry enters through $x_{1,jt}$ and may arise because firms and the Econometrician have better access to product information, data, or expert knowledge, while consumers may acquire information about these characteristics through costly attention. The specification also includes $\varepsilon_{1,ijt}$ to capture consumers' uncertainty about how well a product suits their individual tastes or needs. By contrast, the model in \citet{liao2024} includes neither the information asymmetry represented by $x_{1,jt}$ nor a latent individual-level shock $\varepsilon_{1,ijt}$.

For the RI--BLP specification, our identification analysis establishes conditions under which the unobserved demand shifters $\xi_{1,jt}$ can be uniquely recovered from observed market shares $s_{jt}$ for given structural parameters and a given subjective prior $\pi_{u_1}$. The argument builds on the analysis of conditional choice probabilities (CCPs) in \citet{kristensenzhu2026RI}.

For estimation, we address price endogeneity using instruments for price. Prices or insurance premiums enter the empirical specifications of \citet{brown2024endogenous}, \citet{joo2023rational}, and \citet{natan2025choice}, but their demand estimation does not use instrumental-variable (IV) or control-function corrections for price endogeneity. We use generalized method of moments (GMM) to estimate the structural parameters and recover the unobserved demand shifters $\xi_{1,jt}$ from aggregate market shares. We present a share-inversion formulation and, under the stated conditions, an equivalent share-constrained formulation written as a mathematical program with equilibrium constraints (MPEC).\footnote{GMM defines the statistical estimator, whereas MPEC refers to its constrained programming formulation. For the general MPEC approach to structural estimation, see \citet{su2012constrained}; for its application to BLP-type random-coefficients demand models, see \citet{dube2012improving}.} We obtain a simulated GMM estimator by approximating the integrals in the RI problem and market-share equations.

Following the many-market framework of \citet{freyberger2015}, our asymptotic analysis lets $T\to\infty$ with $J$ fixed. Under the stated identification and regularity conditions, we establish consistency and asymptotic normality of the exact GMM estimator. These results extend to the simulated estimator under additional simulation-accuracy conditions.

Computing the simulated estimator involves two main tasks. First, evaluating structural market shares requires solving the consumer's inner rational-inattention problem. We solve the simulated inner problem using Newton--BA, a hybrid of Newton--Raphson and the Blahut--Arimoto (BA) algorithm guided by the Karush--Kuhn--Tucker (KKT) conditions \citep{zhu2026RIestimation}. Second, we use trust-region methods for GMM searches with share inversion and IPOPT \citep{wachter2006implementation} to solve the MPEC formulation when further constrained optimization is required.

We assess finite-sample performance using two Monte Carlo designs. In the common-coefficient design, we estimate the common utility coefficients. In the random-coefficient design, we add random coefficients on an observed characteristic and price and estimate their dispersion parameters jointly with the mean utility coefficients. The mean utility coefficients $\beta_0$, $\alpha$, and $\beta_1$ are recovered accurately in both designs, although the dispersion parameters are estimated less precisely. Confidence-interval coverage is generally close to 95\% in larger samples, with some overcoverage for the dispersion parameter associated with $\beta_{0,i}$.

We then apply the common-coefficient specification to Dominick's canned-tuna data and compare RI--BLP with an aggregate-logit IV benchmark using the same sample, controls, and price instrument. The models yield similar average own-price elasticities but different substitution patterns across products.

The remainder of the paper proceeds as follows. Section~\ref{sec:model} presents the RI--BLP model and characterizes the consumer's information-acquisition problem. Section~\ref{sec:share_inversion_identification} studies market-share inversion and identification of the structural parameters and unobserved demand shifters. Section~\ref{sec:estimation} develops the GMM estimator and establishes its asymptotic properties. Section~\ref{sec:computation} describes the computational procedures. Section~\ref{sec:monte_carlo} presents Monte Carlo experiments, and Section~\ref{sec:tuna_application} applies the model to tuna demand. Section~\ref{sec:conclusion} concludes.

\section{The Model}
\label{sec:model}

This section formally introduces the RI--BLP model used for estimation. The rational-inattention component follows the discrete-choice framework of \citet{matejka2015rational}. We specify the utility decomposition, characterize the consumer's rational-inattention problem, introduce the maintained parametric specification, and derive the aggregate market-share equation linking the structural model to the data. Throughout, all finite-dimensional vectors are column vectors.

Consumer $i$'s utility from choosing alternative $j\in\mathcal J$ in market $t$ is decomposed as $u_{ijt}=u_{0,ijt}+u_{1,ijt}$. The first component $u_{0,ijt}$ is observed by the consumer before information acquisition, while the second component $u_{1,ijt}$ is uncertain before information acquisition. For each consumer $i$ in market $t$, let $u_{0,it}:=(u_{0,ijt})_{j\in\mathcal J}\in\mathbb R^{J+1}$ and $u_{1,it}:=(u_{1,ijt})_{j\in\mathcal J}\in\mathcal U_1\subseteq\mathbb R^{J+1}$ denote the corresponding utility vectors across alternatives.

We assume that consumers within each market share the same public information and hold a common subjective prior $\pi_{u_1}$ over $u_{1,it}$.\footnote{Market-specific public information and any parameter dependence of the subjective prior are suppressed in the notation.} The prior may depend on structural parameters and reflect observed prices and other public information. We impose no particular distributional family and allow the prior to be misspecified. Conditional on public information and holding structural parameters fixed, we adopt the additive specification of \citet[Section~2.1]{kristensenzhu2026RI}, under which the prior over $u_{1,it}$ does not depend on $u_{0,it}$. Thus, differences in $u_{0,it}$ affect consumers' information acquisition and product choices, but not their prior beliefs within the market.

For a given $u_{0,it}$, the consumer chooses a conditional choice rule with probabilities $P_j(u_{1,it}\mid u_{0,it})$ for all alternatives $j\in\mathcal J$. We represent this rule by the $J$-dimensional vector of inside-product CCPs $P(\cdot\mid u_{0,it}):=(P_j(\cdot\mid u_{0,it}))_{j=1}^{J}$. The outside-option probability is recovered residually as $P_0(u_{1,it}\mid u_{0,it}):=1-\sum_{j=1}^{J}P_j(u_{1,it}\mid u_{0,it})$. For every realization of $u_{1,it}$, feasibility requires $P(u_{1,it}\mid u_{0,it})\in\Delta(\mathcal J)$, where $\Delta(\mathcal J):=\{p\in\mathbb R^J:p_j\geq0,\ j=1,\ldots,J,\ \text{and}\ \sum_{j=1}^{J}p_j\leq1\}$.

Averaging these CCPs over the subjective prior, while holding $u_{0,it}$ fixed, gives the integrated (ex ante) inside-product CCP vector $P^0(u_{0,it}):=\int_{\mathcal U_1}P(u_{1,it}\mid u_{0,it})\,d\pi_{u_1}(u_{1,it})\in\Delta(\mathcal J)$. This averaging relation is the Bayes-consistency condition. The integrated outside-option probability is $P_0^0(u_{0,it}):=1-\sum_{j=1}^{J}P_j^0(u_{0,it})$.

The consumer solves
\begin{equation}
\max_{P(\cdot\mid u_{0,it})}\int_{\mathcal U_1}\sum_{j\in\mathcal J}P_j(u_{1,it}\mid u_{0,it})\bigl(u_{0,ijt}+u_{1,ijt}\bigr)\,d\pi_{u_1}(u_{1,it})-\lambda\mathcal I(P(\cdot\mid u_{0,it})),
\label{eq:RI_problem}
\end{equation}
where $\lambda>0$ is the unit cost of information, treated as a known constant. With probability arguments suppressed and $0\ln0:=0$, Shannon mutual information is
\begingroup
\small
\[
\mathcal I(P(\cdot\mid u_{0,it}))=-\sum_{j\in\mathcal J}P_j^0\ln P_j^0+\int_{\mathcal U_1}\sum_{j\in\mathcal J}P_j\ln P_j\,d\pi_{u_1}(u_{1,it}).
\]
\endgroup

By Theorem~1 of \citet{matejka2015rational}, the solution to Equation~\eqref{eq:RI_problem} implies the following generalized multinomial-logit form for the optimal CCPs:
\begin{equation}
P_j^*(u_{1,it}\mid u_{0,it};\theta)=\frac{P_j^{0*}(u_{0,it};\theta)\exp\bigl((u_{0,ijt}+u_{1,ijt})/\lambda\bigr)}{\sum_{\ell\in\mathcal J}P_\ell^{0*}(u_{0,it};\theta)\exp\bigl((u_{0,i\ell t}+u_{1,i\ell t})/\lambda\bigr)},\quad j=1,\dots,J.
\label{eq:logitccp}
\end{equation}
Here, $\theta$ denotes the vector of unknown structural parameters; dependence on the fixed information cost $\lambda$ is suppressed in the CCP notation. The optimal integrated CCPs satisfy the Bayes-consistency condition
\begin{equation}
P_j^{0*}(u_{0,it};\theta)=\int_{\mathcal U_1}P_j^*(u_{1,it}\mid u_{0,it};\theta)\,d\pi_{u_1}(u_{1,it}),\qquad j=1,\dots,J.
\label{eq:bayes_consistency}
\end{equation}

Following \citet[Lemma~2]{matejka2015rational}, substituting the optimal CCP form in Equation~\eqref{eq:logitccp} into the RI objective and using Bayes consistency in Equation~\eqref{eq:bayes_consistency} yields an equivalent maximization over the integrated CCP vector $P^0\in\Delta(\mathcal J)$, with objective
\begin{equation}
Q^0(P^0;u_{0,it};\theta):=\lambda\int_{\mathcal U_1}\ln\left[\sum_{j\in\mathcal J}P_j^0\exp\bigl((u_{0,ijt}+u_{1,ijt})/\lambda\bigr)\right]\,d\pi_{u_1}(u_{1,it}).
\label{eq:Q0_reduced}
\end{equation}
The optimal integrated CCP vector satisfies $P^{0*}(u_{0,it};\theta)\in\arg\max_{P^0\in\Delta(\mathcal J)}Q^0(P^0;u_{0,it};\theta)$. Given $P^{0*}(u_{0,it};\theta)$, the optimal CCPs are recovered from Equation~\eqref{eq:logitccp}.

To state the optimality conditions, fix $u_{0,it}$ and define the exponential payoff function $r_j(u_{1,it}):=\exp\bigl((u_{0,ijt}+u_{1,ijt})/\lambda\bigr)$ for each $j\in\mathcal J$. For any $P^0\in\Delta(\mathcal J)$, define
\begin{equation}
\kappa_j(P^0;u_{0,it},\theta):=\int_{\mathcal U_1}\frac{r_j(u_{1,it})}{\sum_{\ell\in\mathcal J}P_\ell^0r_\ell(u_{1,it})}\,d\pi_{u_1}(u_{1,it}),\qquad j\in\mathcal J.
\label{eq:inner_kkt_ratio}
\end{equation}
The Lagrangian for maximizing $Q^0(P^0;u_{0,it};\theta)$ subject to $P_j^0\geq0$ for $j\in\mathcal J$ and $\sum_{j\in\mathcal J}P_j^0=1$ yields the following KKT conditions at an optimum $P^{0*}$ \citep[Proposition~1]{caplin2019rational}:
\begin{equation}
\kappa_j\bigl(P^{0*}(u_{0,it};\theta);u_{0,it},\theta\bigr)\leq1,\qquad j\in\mathcal J,
\label{eq:inner_kkt_inequality}
\end{equation}
and
\begin{equation}
P_j^{0*}(u_{0,it};\theta)\left[\kappa_j\bigl(P^{0*}(u_{0,it};\theta);u_{0,it},\theta\bigr)-1\right]=0,\qquad j\in\mathcal J.
\label{eq:inner_kkt_complementary_slackness}
\end{equation}
Because the objective is concave and the constraints are linear, these conditions are necessary and sufficient for global optimality. We refer to Equation~\eqref{eq:inner_kkt_inequality} as the KKT inequality and to Equation~\eqref{eq:inner_kkt_complementary_slackness} as the complementary-slackness condition. Together, they require the KKT ratio $\kappa_j(\cdot)$ to equal one for every alternative assigned positive probability and to be no greater than one for an alternative assigned zero probability. The alternatives $j\in\mathcal J$ with $P_j^{0*}(u_{0,it};\theta)>0$ form the consumer's \emph{consideration set}. The KKT conditions accommodate both interior and boundary solutions.

However, $Q^0(\cdot;u_{0,it};\theta)$ is not strictly concave on $\Delta(\mathcal J)$ in general, so the KKT conditions guarantee optimality but not uniqueness.\footnote{Strict concavity can fail when two distinct integrated CCP vectors generate the same weighted combination of the exponential payoff functions. Reallocating probability between such vectors leaves the logarithmic aggregator, and hence $Q^0$, unchanged. This can occur when the exponential payoff functions are affinely dependent; it is not caused simply by allowing some CCPs to equal zero.} Following \citet{matejka2015rational} and \citet{caplin2019rational}, we impose the following sufficient condition for strict concavity.

\begin{assumption}[Affine independence]
\label{ass:affine_independence}
For each market $t$, candidate parameter vector $\theta$, and baseline utility vector $u_{0,it}$, the exponential payoff functions $\{r_j(\cdot)\}_{j\in\mathcal J}$ are affinely independent in $L^1(\mathcal U_1,\pi_{u_1})$.
\end{assumption}

Under Assumption~\ref{ass:affine_independence}, $Q^0(\cdot;u_{0,it};\theta)$ is strictly concave on $\Delta(\mathcal J)$. The optimal integrated CCP vector is therefore unique, and we denote it by $P^{0*}(u_{0,it};\theta):=\bigl(P_j^{0*}(u_{0,it};\theta)\bigr)_{j=1}^{J}\in\mathbb R^J$.

We next relate these choice probabilities to aggregate market shares. We first write the utility components for inside products $j=1,\ldots,J$ under the specification $\xi_{0,jt}=\varnothing$ adopted in Section~\ref{sec:introduction}:
\[
\begin{gathered}
u_{0,ijt}=\beta_{0,i}^{\prime}x_{0,jt}-\alpha_i p_{jt}+\varepsilon_{0,ijt},\\
u_{1,ijt}=\beta_1^{\prime}x_{1,jt}+\xi_{1,jt}+\varepsilon_{1,ijt}.
\end{gathered}
\]
Here, $x_{0,jt}\in\mathbb R^{K_0}$ and $x_{1,jt}\in\mathbb R^{K_1}$ are column vectors of product characteristics, and $p_{jt}$ is the product price. The product-market demand shifter $\xi_{1,jt}$ enters only $u_{1,ijt}$. The consumer-specific coefficients $\beta_{0,i}$ and $\alpha_i$, with respective population means $\beta_0$ and $\alpha$, remain fixed across products. To retain a common subjective prior over $u_1$, we restrict random coefficients to $u_0$ and keep $\beta_1$ common across consumers.

Adapting the standard BLP notation, we define the baseline index $\delta_{0,jt}$, the latent index $\delta_{1,jt}$, and the individual-specific deviation $\mu_{0,ijt}$ induced by random coefficients as
\[
\begin{gathered}
\delta_{0,jt}:=\beta_0^{\prime}x_{0,jt}-\alpha p_{jt},\\
\delta_{1,jt}:=\beta_1^{\prime}x_{1,jt}+\xi_{1,jt},\\
\mu_{0,ijt}:=(\beta_{0,i}-\beta_0)^{\prime}x_{0,jt}-(\alpha_i-\alpha)p_{jt},
\end{gathered}
\]
for inside products $j=1,\ldots,J$. Thus, the utility components can be written as
\[
\begin{gathered}
u_{0,ijt}=\delta_{0,jt}+\mu_{0,ijt}+\varepsilon_{0,ijt},\\
u_{1,ijt}=\delta_{1,jt}+\varepsilon_{1,ijt}.
\end{gathered}
\]

By \citet[Lemma~2.3]{kristensenzhu2026RI}, common utility shifts leave optimal CCPs unchanged. We therefore express both utility components relative to the outside option and normalize $u_{0,i0t}=u_{1,i0t}=0$, with all terms in their outside-option decompositions set to zero. Hereafter, $u_{0,it}$ and $u_{1,it}$ denote the $J$-dimensional vectors of relative inside-product utilities, and $\mathcal U_1\subseteq\mathbb R^J$ denotes the domain of $u_{1,it}$. The subjective prior $\pi_{u_1}$, payoff functions $r_j$, and Assumption~\ref{ass:affine_independence} are understood in this normalized representation.

We specify a finite-dimensional family $F_{\beta_0,\alpha}(\cdot;\theta)$ for the data-generating distribution of the random-coefficient vector $(\beta_{0,i}^{\prime},\alpha_i)^{\prime}$.\footnote{Equivalently, the distribution is known up to finite-dimensional parameters. For example, in the scalar case, specifying $\beta_{0,i}\sim\mathcal N(\beta_0,\sigma_x^2)$ fixes the normal family, while the mean $\beta_0$ and standard deviation $\sigma_x$ remain to be estimated.} The distribution is parameterized by the population means $\beta_0$ and $\alpha$ and the heterogeneity parameters $\vartheta$, which capture variation in the consumer-specific coefficients around these means. Together with the common coefficient $\beta_1$, these parameters form the unknown structural parameter vector $\theta:=(\beta_0^{\prime},\alpha,\beta_1^{\prime},\vartheta^{\prime})^{\prime}\in\Theta\subseteq\mathbb R^{d_\theta}$. Under the common-coefficient specification, $\beta_{0,i}=\beta_0$ and $\alpha_i=\alpha$ for every consumer, so $\mu_{0,ijt}=0$ and $\theta=(\beta_0^{\prime},\alpha,\beta_1^{\prime})^{\prime}$.

Let $\varepsilon_{0,it}:=(\varepsilon_{0,ijt})_{j=1}^{J}$, $\varepsilon_{1,it}:=(\varepsilon_{1,ijt})_{j=1}^{J}$, and $\varepsilon_{it}:=(\varepsilon_{0,it}^{\prime},\varepsilon_{1,it}^{\prime})^{\prime}\in\mathbb R^{2J}$. Let $F_{\varepsilon}$ denote the joint data-generating distribution of $\varepsilon_{it}$ on $\mathcal E_t\subseteq\mathbb R^{2J}$.\footnote{Throughout, we use $\pi$ for consumers' subjective distributions and $F$ for data-generating distributions.} We impose the following independence assumption.

\begin{assumption}[Independence]
\label{ass:riblp_independence}
For each consumer $i$ and market $t$, the random-coefficient vector $(\beta_{0,i}^{\prime},\alpha_i)^{\prime}$, the additive-shock vector $\varepsilon_{it}$, and the market-level vector $(x_{0,t}^{\prime},x_{1,t}^{\prime},p_t^{\prime},\xi_{1,t}^{\prime})^{\prime}$ are mutually independent.
\end{assumption}
Here, $x_{0,t}:=\bigl(x_{0,1t}^{\prime},\ldots,x_{0,Jt}^{\prime}\bigr)^{\prime}\in\mathbb R^{JK_0}$, $x_{1,t}:=\bigl(x_{1,1t}^{\prime},\ldots,x_{1,Jt}^{\prime}\bigr)^{\prime}\in\mathbb R^{JK_1}$, $p_t:=(p_{1t},\ldots,p_{Jt})^{\prime}\in\mathbb R^J$, and $\xi_{1,t}:=(\xi_{1,1t},\ldots,\xi_{1,Jt})^{\prime}\in\mathbb R^J$. The assumption restricts dependence between these three vectors, not within them. Prices and product characteristics may remain correlated with $\xi_{1,t}$.

Let $\delta_{0,t}:=(\delta_{0,jt})_{j=1}^{J}$, $\delta_{1,t}:=(\delta_{1,jt})_{j=1}^{J}$, and $\mu_{0,it}:=(\mu_{0,ijt})_{j=1}^{J}$ denote column vectors in $\mathbb R^J$. We now obtain the model-implied market share for each inside product $j=1,\ldots,J$ by integrating the optimal realized CCP in Equation~\eqref{eq:logitccp} over the data-generating distributions of the random coefficients and additive shocks.\footnote{For each candidate specification, events assigned zero probability by the prior must also have zero probability under the data-generating distribution of latent utility, conditional on the consumer's information before acquisition.} Under Assumption~\ref{ass:riblp_independence}, the market-share equation is
\begin{equation}
\begin{aligned}
s_{jt}
&=\sigma_j(\delta_{0,t},\delta_{1,t};\theta)\\
&=\int_{\mathbb R^{K_0+1}}\int_{\mathcal E_t}P_j^*\bigl(\delta_{1,t}+\varepsilon_{1,it}\mid\delta_{0,t}+\mu_{0,it}+\varepsilon_{0,it};\theta\bigr)\,dF_{\varepsilon}(\varepsilon_{it})\,dF_{\beta_0,\alpha}(\beta_{0,i},\alpha_i;\theta).
\end{aligned}
\label{eq:market_share}
\end{equation}
Here, $\sigma_j(\cdot)$ denotes the model-implied market-share function, and $s_{jt}$ denotes the corresponding market share in market $t$.\footnote{Let $\mathcal X_t:=\bigl(x_{0,t}^{\prime},x_{1,t}^{\prime},p_t^{\prime},\xi_{1,t}^{\prime}\bigr)^{\prime}$. By the conditional law of total probability, a market share is the average realized CCP over the conditional data-generating distribution of individual heterogeneity given $\mathcal X_t$. Assumption~\ref{ass:riblp_independence} implies that this distribution factors as $dF_{\beta_0,\alpha,\varepsilon\mid\mathcal X_t}(\beta_{0,i},\alpha_i,\varepsilon_{it}\mid\mathcal X_t)=dF_{\beta_0,\alpha}(\beta_{0,i},\alpha_i;\theta)\,dF_{\varepsilon}(\varepsilon_{it})$, which yields Equation~\eqref{eq:market_share}.} The notation for $\sigma_j$ suppresses dependence on the observed characteristics $x_{0,t}$ and prices $p_t$, which are held fixed during share inversion. The outside share is determined residually by $s_{0t}=1-\sum_{j=1}^{J}s_{jt}$.

At the market level, the Econometrician observes shares $s_{jt}$ for all alternatives $j\in\mathcal J$, together with prices $p_{jt}$ and product characteristics $(x_{0,jt},x_{1,jt})$ for $j=1,\dots,J$. Consumer-specific coefficients $\beta_{0,i}$ and $\alpha_i$, additive-shock realizations $\varepsilon_{it}$, and product-market demand shifters $\xi_{1,jt}$ are unobserved.

For identification and estimation, we take $F_{\varepsilon}$ as known and the market-specific prior $\pi_{u_1}$ as known at each candidate $\theta$. The structural parameters $\theta$ remain unknown. Each demand shifter enters the market-share equations through the latent index $\delta_{1,jt}=\beta_1^{\prime}x_{1,jt}+\xi_{1,jt}$. The next section formalizes the prior and shock-distribution assumptions. It then studies how to recover the candidate-implied latent indices $\delta_{1,t}$ and demand shifters $\xi_{1,t}$ by share inversion and identify $\theta$ using IV moment restrictions.

\section{Share Inversion and Identification}
\label{sec:share_inversion_identification}

Building on the aggregate market-share equation in Equation~\eqref{eq:market_share}, this section studies share inversion and identification. Let $\theta^*$ denote the true value of $\theta$. We examine how the market-share equations and instrumental-variable moment restrictions jointly identify $\theta^*$ and recover the baseline index $\delta_{0,t}^*$, latent index $\delta_{1,t}^*$, and product-market demand shifter $\xi_{1,t}$.

The identification argument evaluates each candidate $\theta$ through the following sequence:
\[
\theta
\ \longmapsto\ 
\delta_{0,t}(\beta_0,\alpha)
\ \longmapsto\ 
\delta_{1,t}(\theta)
\ \longmapsto\ 
\xi_{1,t}(\theta)
\ \xrightarrow{\mathrm{IV}}\ 
\theta^*.
\]
We now explain each step of this identification argument. First, at each candidate $\theta$, observed prices and product characteristics determine the baseline index $\delta_{0,t}(\beta_0,\alpha):=\bigl(\beta_0^{\prime}x_{0,jt}-\alpha p_{jt}\bigr)_{j=1}^{J}$. By contrast, the relation $\delta_{1,jt}=\beta_1^{\prime}x_{1,jt}+\xi_{1,jt}$ does not determine $\delta_{1,t}$ because $\xi_{1,t}$ is unobserved.

Next, we study share inversion for the latent index. Let $s_t:=(s_{1t},\ldots,s_{Jt})^{\prime}\in\mathbb R^J$ and write $\sigma_t(\cdot):=\bigl(\sigma_1(\cdot),\ldots,\sigma_J(\cdot)\bigr)^{\prime}\in\mathbb R^J$ for the model-implied inside-share vector. At each candidate $\theta$, we hold $\delta_{0,t}(\beta_0,\alpha)$ and the subjective prior fixed when varying $\delta_{1,t}$. The candidate-implied latent index $\delta_{1,t}(\theta)$ must satisfy the market-share equations
\begin{equation}
s_t=\sigma_t\bigl(\delta_{0,t}(\beta_0,\alpha),\delta_{1,t};\theta\bigr).
\label{eq:share_inverse_system}
\end{equation}
We impose the following conditions on the subjective prior and additive-shock distribution.

\begin{assumption}[Known prior]
\label{ass:known_prior}
For each market and candidate parameter value, the common subjective prior $\pi_{u_1}$ is known.
\end{assumption}

\begin{assumption}[Additive-shock distribution]
\label{ass:riblp_fullsupport}
For every market $t$, the joint additive-shock distribution $F_{\varepsilon}$ is known and has full support on $\mathcal E_t=\mathbb R^{2J}$.
\end{assumption}

To state the share-inversion result, we write $\operatorname{int}\Delta(\mathcal J)$ for the interior of $\Delta(\mathcal J)$, where $p_j>0$ for $j=1,\ldots,J$ and the residual outside share $1-\sum_{j=1}^{J}p_j$ is positive.

\begin{theorem}[Share injectivity]
\label{thm:riblp_delta1_injective}
Suppose Assumptions~\ref{ass:affine_independence}, \ref{ass:riblp_independence}, \ref{ass:known_prior}, and~\ref{ass:riblp_fullsupport} hold, and the optimal integrated CCP vector $P^{0*}(u_{0,it};\theta)$ is continuous in $u_{0,it}$. Fix a market $t$, a subjective prior $\pi_{u_1}$, a candidate parameter vector $\theta$, and the baseline index $\delta_{0,t}(\beta_0,\alpha)$. Then the mapping
\[
\delta_{1,t}\in\mathbb R^J
\ \longmapsto\ 
\sigma_t\bigl(\delta_{0,t}(\beta_0,\alpha),\delta_{1,t};\theta\bigr)\in\operatorname{int}\Delta(\mathcal J)
\]
is one-to-one. All model-implied shares, including the residual outside share, are strictly positive. The share Jacobian with respect to $\delta_{1,t}$ is symmetric and positive definite at every $\delta_{1,t}\in\mathbb R^J$.
\end{theorem}

Our share mapping $\sigma_t$ corresponds to the structural CCP mapping $\Pi(u_0,u_1,\lambda)$ in \citet{kristensenzhu2026RI}, specialized to a common information cost and averaged over the random coefficients $\beta_{0,i}$ and $\alpha_i$. Conditional on these coefficients, their arguments $u_0$ and $u_1$ correspond to $\delta_{0,t}+\mu_{0,it}$ and $\delta_{1,t}$, respectively. In the proof, the full-support condition in Assumption~\ref{ass:riblp_fullsupport} ensures that $u_{0,it}$ falls with positive probability in the regular set defined in \citet{kristensenzhu2026RI}, where all integrated CCPs, including that of the outside option, are positive. This yields strictly positive market shares and a positive-definite share Jacobian with respect to $\delta_{1,t}$, establishing injectivity. The proof is in Appendix~\ref{appendix:riblp_delta1_injective}.

Individual CCPs may nevertheless be zero outside the regular set. Without full support, the underlying RI model can generate a zero market share for a product excluded from almost every consumer's consideration set. We maintain the full-support condition in Assumption~\ref{ass:riblp_fullsupport} throughout this paper. Relaxing this condition to allow zero market shares is a possible extension of the analysis, but is not considered here.

For share inversion, Theorem~\ref{thm:riblp_delta1_injective} guarantees uniqueness, not existence, so $\delta_{1,t}(\theta)$ is defined only when Equation~\eqref{eq:share_inverse_system} has a solution. At $\theta^*$, however, the true latent-index vector, denoted by $\delta_{1,t}^*$, satisfies Equation~\eqref{eq:share_inverse_system}, so a solution exists.

Given a solution $\delta_{1,t}(\theta)$, the candidate-implied demand-shifter vector is
\begin{equation}
\xi_{1,t}(\theta):=\delta_{1,t}(\theta)-X_{1,t}\beta_1,
\label{eq:candidate_demand_shifter}
\end{equation}
where $X_{1,t}\in\mathbb R^{J\times K_1}$ has $j$th row $x_{1,jt}^{\prime}$.

Finally, we use IV moment restrictions to identify $\theta^*$. Because firms observe $\xi_{1,jt}$ when setting prices, equilibrium prices may be correlated with this unobserved demand component. Let $z_{1,jt}\in\mathbb R^q$ be the vector of $q$ instruments, comprising exogenous product characteristics and excluded instruments for price. Let $Z_{1,t}\in\mathbb R^{J\times q}$ have $j$th row $z_{1,jt}^{\prime}$. The product-level moment vector is $z_{1,jt}\xi_{1,jt}(\theta)\in\mathbb R^q$. Summing over products in market $t$ gives $Z_{1,t}^{\prime}\xi_{1,t}(\theta)=\sum_{j=1}^{J}z_{1,jt}\xi_{1,jt}(\theta)$. The expectation in the next assumption averages this within-market sum over the population of markets. We impose the following high-level global and local identification conditions.

\begin{assumption}[Global identification]
\label{ass:riblp_global_identification}
For the maintained specification of the subjective prior and the imposed value of $\lambda$,
\begin{equation}
\mathbb E\!\left[Z_{1,t}^{\prime}\xi_{1,t}(\theta)\right]=0
\quad\Longleftrightarrow\quad
\theta=\theta^*
\label{eq:riblp_global_identification}
\end{equation}
for every $\theta\in\Theta$ such that $\delta_{1,t}(\theta)$ exists almost surely.
\end{assumption}

At $\theta^*$, Equation~\eqref{eq:riblp_global_identification} imposes population instrument orthogonality. The converse requires $\theta^*$ to be the unique parameter satisfying that condition and therefore gives global identification.

\begin{assumption}[Local rank]
\label{ass:riblp_local_identification}
The population moment Jacobian at $\theta^*$ has full column rank:
\begin{equation}
\operatorname{rank}\!\left(
\left.
\frac{\partial}{\partial\theta^{\prime}}\mathbb E\!\left[Z_{1,t}^{\prime}\xi_{1,t}(\theta)\right]
\right|_{\theta=\theta^*}
\right)
=d_\theta.
\label{eq:riblp_local_rank}
\end{equation}
\end{assumption}

The local share-smoothness and moment-bound conditions in Assumption~\ref{ass:exact_gmm_asymptotics}(iv) establish differentiability of the population moments and permit differentiation under the expectation. Together with those conditions, Assumption~\ref{ass:riblp_local_identification} gives regular local identification and requires at least as many instruments as structural parameters, $q\geq d_\theta$. The rank condition rules out any nonzero parameter direction along which all population IV moments remain unchanged to first order at $\theta^*$. Appendix~\ref{appendix:joint_identification_rank} derives the moment Jacobian under joint differentiability of the share mapping.

Once $\theta^*$ is identified, the recovered objects are
\[
\delta_{0,t}^*:=\delta_{0,t}(\beta_0^*,\alpha^*),
\qquad
\delta_{1,t}^*:=\delta_{1,t}(\theta^*),
\qquad
\xi_{1,t}:=\xi_{1,t}(\theta^*)=\delta_{1,t}^*-X_{1,t}\beta_1^*.
\]

\section{Estimation}
\label{sec:estimation}

This section develops an IV--GMM estimator using aggregate market shares. We present its share-inversion and share-constrained formulations, which are equivalent under the stated conditions. We then introduce the simulation approximation, establish consistency and asymptotic normality, and construct a consistent variance estimator.

\subsection{The RI--BLP GMM Estimator}

We begin with the sample moments corresponding to the IV restrictions in Section~\ref{sec:share_inversion_identification}, using the product-characteristic matrix $X_{1,t}$ and instrument matrix $Z_{1,t}$ defined there. For any candidate index $\delta_{1,t}\in\mathbb R^J$, product $j$ contributes $z_{1,jt}(\delta_{1,jt}-x_{1,jt}^{\prime}\beta_1)$ to the IV moment. We average these contributions within each market and then across markets. For $\delta_1:=(\delta_{1,1}^{\prime},\ldots,\delta_{1,T}^{\prime})^{\prime}\in\mathbb R^{JT}$, define
\begingroup
\small
\begin{equation}
\begin{aligned}
m_t(\theta,\delta_{1,t})&:=\frac{1}{J}Z_{1,t}^{\prime}\bigl(\delta_{1,t}-X_{1,t}\beta_1\bigr)=\frac{1}{J}\sum_{j=1}^J z_{1,jt}\bigl(\delta_{1,jt}-x_{1,jt}^{\prime}\beta_1\bigr)\in\mathbb R^q,
\\
\overline m_T(\theta,\delta_1)&:=\frac{1}{T}\sum_{t=1}^T m_t(\theta,\delta_{1,t})=\frac{1}{TJ}\sum_{t=1}^T\sum_{j=1}^J z_{1,jt}\bigl(\delta_{1,jt}-x_{1,jt}^{\prime}\beta_1\bigr)\in\mathbb R^q.
\end{aligned}
\label{eq:market_moment}
\end{equation}
\endgroup
Using these moments, GMM estimation can proceed either by recovering the latent indices $\delta_1(\theta)$ at each candidate $\theta$ or by estimating $\theta$ and $\delta_1$ jointly subject to the market-share equations.

\paragraph{Share-inversion GMM.}
Assume that the share equations have a solution for every candidate $\theta$ and sample market $t$. Under the conditions of Theorem~\ref{thm:riblp_delta1_injective}, this solution is unique. Share-inversion GMM then proceeds in two stages:

\emph{Stage 1: Share inversion.} For each candidate $\theta$ and market $t=1,\ldots,T$, solve $s_t=\sigma_t\bigl(\delta_{0,t}(\beta_0,\alpha),\delta_{1,t}(\theta);\theta\bigr)$ for $\delta_{1,t}(\theta)$. The recovered index gives $\xi_{1,t}(\theta)=\delta_{1,t}(\theta)-X_{1,t}\beta_1$, which enters the GMM moments.

\emph{Stage 2: GMM minimization.} Let $\delta_1(\theta)\in\mathbb R^{JT}$ stack the recovered market-specific indices $\delta_{1,t}(\theta)$. Substitution into Equation~\eqref{eq:market_moment} gives $m_t(\theta):=m_t(\theta,\delta_{1,t}(\theta))$ and $\overline m_T(\theta):=\overline m_T(\theta,\delta_1(\theta))$. The estimator minimizes the GMM objective formed from these moments:
\begin{equation}
\widehat\theta_T\in\arg\min_{\theta\in\Theta}\overline m_T(\theta)^{\prime}\widehat W_T\overline m_T(\theta),
\label{eq:two_stage_gmm}
\end{equation}
where $\widehat W_T\in\mathbb R^{q\times q}$ is a symmetric positive-definite sample weighting matrix, held fixed during parameter optimization.

\paragraph{Share-constrained GMM.}
Share-inversion GMM requires solving the market-share equations at each candidate $\theta$. Following \citet{su2012constrained} and \citet{dube2012improving}, an alternative is to minimize the GMM objective jointly over $\theta$ and the latent indices $\delta_1$, subject to the market-share equations:
\begingroup
\small
\begin{equation}
\begin{aligned}
&(\widehat\theta_T,\widehat\delta_{1,T})\in\arg\min_{\theta\in\Theta,\;\delta_1\in\mathbb R^{JT}}\quad\overline m_T(\theta,\delta_1)^{\prime}\widehat W_T\overline m_T(\theta,\delta_1)
\\
&\text{subject to}\quad s_t=\sigma_t\bigl(\delta_{0,t}(\beta_0,\alpha),\delta_{1,t};\theta\bigr),\qquad t=1,\ldots,T.
\end{aligned}
\label{eq:gmm_mpec_exact}
\end{equation}
\endgroup
Here, $\widehat\delta_{1,T}\in\mathbb R^{JT}$ stacks the estimated latent indices across all $T$ markets. In the MPEC formulation, we optimize $\theta$ and $\delta_1$ jointly, subject to the market-share equations. This eliminates the need for a separate share inversion at each candidate $\theta$.

Share-inversion GMM and share-constrained GMM are equivalent formulations of the same estimator. Since the share equations have a unique solution at each candidate $\theta$, the constraints in Equation~\eqref{eq:gmm_mpec_exact} require $\delta_1=\delta_1(\theta)$. Substituting this solution into the constrained objective gives the criterion in Equation~\eqref{eq:two_stage_gmm}, so both formulations have the same minimizing values of $\theta$.

\subsection{The Simulated RI--BLP GMM Estimator}

The RI objective integrates under the subjective prior $\pi_{u_1}$, while market shares average the realized CCPs under the data-generating distribution $F_{\beta_0,\alpha}(\cdot;\theta)\otimes F_{\varepsilon}$. These integrals generally have no closed-form expressions, so we approximate them by averages over finitely many draws to construct a simulated GMM estimator. Let $R_T$ denote the number of joint draws of random coefficients and additive shocks used for the outer market-share integral, and $S_T$ the number of subjective-prior draws used for each outer draw. The numbers of draws are indexed by the number of markets $T$ to allow them to increase with sample size.

Each joint population draw represents one simulated consumer, so we use $r$ in place of $i$ in the simulated coefficients, shocks, and utilities. For each market $t$ and draw $r=1,\ldots,R_T$, let $(\beta_{0,t}^{(r)},\alpha_t^{(r)},\varepsilon_t^{(r)})$ be a joint draw from $F_{\beta_0,\alpha}(\cdot;\theta)\otimes F_{\varepsilon}$, with $\varepsilon_t^{(r)}=(\varepsilon_{0,t}^{(r)\prime},\varepsilon_{1,t}^{(r)\prime})^{\prime}$. The random-coefficient deviation is $\mu_{0,jt}^{(r)}:=(\beta_{0,t}^{(r)}-\beta_0)^{\prime}x_{0,jt}-(\alpha_t^{(r)}-\alpha)p_{jt}$. Collect these deviations in $\mu_{0,t}^{(r)}$. The realized utility vectors are $u_{0,t}^{(r)}:=\delta_{0,t}+\mu_{0,t}^{(r)}+\varepsilon_{0,t}^{(r)}$ and $u_{1,t}^{(r)}:=\delta_{1,t}+\varepsilon_{1,t}^{(r)}$.

For each outer draw, take $S_T$ draws $u_{1,t}^{\pi,(s,r)}$ from the subjective prior $\pi_{u_1}$, with $s=1,\ldots,S_T$. The simulated integrated CCP vector is obtained by replacing the integral in Equation~\eqref{eq:Q0_reduced} with the sample average over these draws:
\begingroup
\small
\begin{equation}
\widehat P_{rt}^{0*}(u_{0,t}^{(r)};\theta)\in\arg\max_{P^0\in\Delta(\mathcal J)}\frac{\lambda}{S_T}\sum_{s=1}^{S_T}\ln\!\left[\sum_{\ell\in\mathcal J}P_\ell^0\exp\!\left(\bigl(u_{0,\ell t}^{(r)}+u_{1,\ell t}^{\pi,(s,r)}\bigr)/\lambda\right)\right].
\label{eq:simulated_ri_problem}
\end{equation}
\endgroup
As in the model, $P_0^0=1-\sum_{j=1}^J P_j^0$, and the outside utility components are zero in both the realized and subjective-prior draws. The simulated outside integrated CCP is $\widehat P_{0rt}^{0*}(u_{0,t}^{(r)};\theta):=1-\sum_{j=1}^J\widehat P_{jrt}^{0*}(u_{0,t}^{(r)};\theta)$. For notational simplicity, we suppress dependence on $S_T$ in the simulated-CCP notation.

The simulated RI problem has a solution by continuity on the compact simplex. If the exponential payoff vectors evaluated at the prior draws are affinely independent, the objective is strictly concave and its solution $\widehat P_{rt}^{0*}$ is unique. For each fixed market $t$, population draw $r$, and candidate parameter $\theta$, sample affine independence holds with probability approaching one as $S_T\to\infty$, under Assumption~\ref{ass:affine_independence} and consistency of the prior-draw rule.\footnote{Fix $t$, $r$, and $\theta$ and suppress these indices in $r_j$, with $r_0=1$. Under Assumption~\ref{ass:affine_independence}, the $J\times J$ matrix with $(j,k)$ entry $\int_{\mathcal U_1}[r_j(u_1)-1][r_k(u_1)-1]/[1+\sum_{\ell=1}^J(r_\ell(u_1)-1)^2]\,d\pi_{u_1}(u_1)$ is positive definite. Its sample approximation using consistent averages of these bounded integrands is therefore positive definite with probability approaching one. This implies full column rank of the $S_T\times J$ matrix with $(s,j)$ entry $r_j(u_{1,t}^{\pi,(s,r)})-1$, which is equivalent to sample affine independence. For the Gaussian prior transformations in Sections~\ref{sec:monte_carlo} and~\ref{sec:tuna_application}, the bounded integrands are continuous on the interior of the uniform draw domain and hence Riemann integrable. The scrambled Sobol nets have vanishing discrepancy as the number of draws increases through powers of two, so their averages converge. This result is pointwise in $(t,r,\theta)$, not uniform.}

Substituting the simulated integrated CCPs and the realized utility draw into Equation~\eqref{eq:logitccp} gives, for $j=1,\ldots,J$,
\begingroup
\small
\[
\widehat P_{jrt}^*(u_{1,t}^{(r)}\mid u_{0,t}^{(r)};\theta):=\frac{\widehat P_{jrt}^{0*}(u_{0,t}^{(r)};\theta)\exp\bigl((u_{0,jt}^{(r)}+u_{1,jt}^{(r)})/\lambda\bigr)}{\sum_{\ell\in\mathcal J}\widehat P_{\ell rt}^{0*}(u_{0,t}^{(r)};\theta)\exp\bigl((u_{0,\ell t}^{(r)}+u_{1,\ell t}^{(r)})/\lambda\bigr)}.
\]
\endgroup
The outside-option realized CCP is $\widehat P_{0rt}^*(\cdot):=1-\sum_{j=1}^J\widehat P_{jrt}^*(\cdot)$. We then approximate the outer integral in Equation~\eqref{eq:market_share} by averaging the simulated realized CCPs over the $R_T$ outer draws:
\begingroup
\small
\begin{equation}
\sigma_{j,t,R_T,S_T}^{\mathrm{sim}}(\delta_{0,t},\delta_{1,t};\theta):=\frac{1}{R_T}\sum_{r=1}^{R_T}\widehat P_{jrt}^*(u_{1,t}^{(r)}\mid u_{0,t}^{(r)};\theta),\qquad j=1,\ldots,J.
\label{eq:simulated_share}
\end{equation}
\endgroup
Write $\sigma_{t,R_T,S_T}^{\mathrm{sim}}(\cdot):=\bigl(\sigma_{1,t,R_T,S_T}^{\mathrm{sim}}(\cdot),\ldots,\sigma_{J,t,R_T,S_T}^{\mathrm{sim}}(\cdot)\bigr)^{\prime}\in\mathbb R^J$ for the simulated inside-share vector.

Replacing exact model-implied shares with simulated shares gives simulated versions of both GMM formulations. Suppose that the simulated share equations have a unique solution $\delta_{1,t}^{\mathrm{sim}}(\theta)$ for every candidate $\theta$ and sample market $t$. Let $\delta_1^{\mathrm{sim}}(\theta)$ stack these solutions. Substituting them into Equation~\eqref{eq:market_moment} gives $m_t^{\mathrm{sim}}(\theta):=m_t\bigl(\theta,\delta_{1,t}^{\mathrm{sim}}(\theta)\bigr)$ and $\overline m_T^{\mathrm{sim}}(\theta):=\overline m_T\bigl(\theta,\delta_1^{\mathrm{sim}}(\theta)\bigr)$. We omit $R_T$ and $S_T$ from the notation for these indices and moments. The simulated share-inversion estimator is
\[
\widehat\theta_T^{\mathrm{sim}}\in\arg\min_{\theta\in\Theta}\overline m_T^{\mathrm{sim}}(\theta)^{\prime}\widehat W_T\overline m_T^{\mathrm{sim}}(\theta).
\]
Equivalently, we may estimate $\theta$ and $\delta_1$ jointly subject to the simulated share equations:
\begingroup
\small
\begin{equation}
\begin{aligned}
&(\widehat\theta_T^{\mathrm{sim}},\widehat\delta_{1,T}^{\mathrm{sim}})\in\arg\min_{\theta\in\Theta,\;\delta_1\in\mathbb R^{JT}}\quad\overline m_T(\theta,\delta_1)^{\prime}\widehat W_T\overline m_T(\theta,\delta_1)
\\
&\text{subject to}\quad s_t=\sigma_{t,R_T,S_T}^{\mathrm{sim}}\bigl(\delta_{0,t}(\beta_0,\alpha),\delta_{1,t};\theta\bigr),\qquad t=1,\ldots,T.
\end{aligned}
\label{eq:simulated_gmm_mpec}
\end{equation}
\endgroup
Here, $\widehat\delta_{1,T}^{\mathrm{sim}}\in\mathbb R^{JT}$ stacks the estimated latent indices across all $T$ markets under the simulated share constraints. Replacing the exact shares with simulated shares generally changes the finite-sample estimator.

The next subsection states approximation conditions under which simulation preserves the limiting distribution of the exact estimator.

\subsection{Many-Market Asymptotics}
\label{sec:many_market_asymptotics}

Following the many-market framework of \citet{freyberger2015}, we let \mbox{$T\to\infty$} while holding fixed the number of inside products $J$, the number of instruments $q$, and the dimension $d_\theta$ of the structural parameter vector $\theta$.\footnote{Freyberger allows the number of products to vary across markets while remaining uniformly bounded. Here, every market has $J$ inside products.} We consider market shares at the population level, abstracting from consumer-sampling error. We establish exact-GMM asymptotics, then give conditions under which simulation preserves the limiting distribution. Norms are Euclidean for vectors and the associated operator norms for matrices.

\paragraph{Exact-GMM asymptotics.}
Throughout this subsection, we maintain the following model conditions: almost surely, Assumptions~\ref{ass:affine_independence}, \ref{ass:riblp_independence}, \ref{ass:known_prior}, and~\ref{ass:riblp_fullsupport} hold for every $\theta\in\Theta$, and $P^{0*}(u_{0,it};\theta)$ is continuous in $u_{0,it}$ at every such $\theta$. Theorem~\ref{thm:riblp_delta1_injective} therefore gives share injectivity and a positive-definite latent-index share Jacobian.

Write $\sigma_t(\delta;\theta):=\sigma_t(\delta_{0,t}(\beta_0,\alpha),\delta;\theta)$, where $\delta\in\mathbb R^J$ denotes a candidate value of the market-specific latent index $\delta_{1,t}$. The dependence on $\theta$ includes any specified parameter dependence of the subjective prior. For given observed shares $s_t$ and candidate $\theta$, the recovered index $\delta_{1,t}(\theta)$ solves $\sigma_t(\delta_{1,t}(\theta);\theta)=s_t$. The solution is unique whenever it exists. Existence at every candidate parameter remains a separate requirement, as in Assumption~A.1 of \citet[Appendix~B]{freyberger2015}. The next lemma establishes regularity of the recovered indices and moments from properties of the share mapping.

\begin{lemma}[Share inversion and moment regularity]
\label{lem:inverse_moment_regularity}
Fix a market $t$ satisfying the maintained model conditions above. Then the following statements hold.
\begin{enumerate}
\renewcommand{\labelenumi}{(\roman{enumi})}
\setlength{\itemsep}{0.35em}
\setlength{\parsep}{0pt}
\item \emph{(Latent-index smoothness).} For every fixed $\theta\in\Theta$, the share mapping $\delta\mapsto\sigma_t(\delta;\theta)$ is infinitely differentiable, with a symmetric positive-definite Jacobian $\partial\sigma_t/\partial\delta^{\prime}$. Derivatives with respect to $\delta$ of every positive order are uniformly bounded across indices, parameters, and markets, with bounds depending only on that order, $J$, and $\lambda$.
\item \emph{(Continuity of recovered indices and moments).} If $\sigma_t$ is jointly continuous on $\mathbb R^J\times\Theta$ and the share equations have a solution for every $\theta\in\Theta$, then $\delta_{1,t}(\theta)$ and $m_t(\theta)$ are continuous on $\Theta$.
\item \emph{(Differentiability of recovered indices and moments).} Let $\mathcal N\subset\operatorname{int}\Theta$ be a closed ball. If the premises of part~(ii) hold and $\sigma_t$ is jointly continuously differentiable near each pair $(\delta_{1,t}(\theta),\theta)$ with $\theta\in\mathcal N$, then $\delta_{1,t}(\theta)$ and $m_t(\theta)$ are continuously differentiable on an open neighborhood of $\mathcal N$, with
\begingroup
\small
\[
\frac{\partial\delta_{1,t}(\theta)}{\partial\theta^{\prime}}
=-\left.
\left[\frac{\partial\sigma_t(\delta;\theta)}{\partial\delta^{\prime}}\right]^{-1}
\frac{\partial\sigma_t(\delta;\theta)}{\partial\theta^{\prime}}
\right|_{\delta=\delta_{1,t}(\theta)},\qquad \theta\in\mathcal N.
\]
\endgroup
\end{enumerate}
\end{lemma}

Appendix~\ref{appendix:inverse_moment_regularity} proves the lemma. Part~(i) concerns latent-index smoothness at fixed $\theta$; parts~(ii)--(iii) derive continuity and differentiability of the recovered indices and moments as $\theta$ varies. We now impose the required share regularity, together with sampling and moment conditions.

\begin{assumption}[Exact-GMM regularity]
\label{ass:exact_gmm_asymptotics}
\leavevmode
\begin{enumerate}
\renewcommand{\labelenumi}{(\roman{enumi})}
\setlength{\itemsep}{0.45em}
\setlength{\parsep}{0pt}
\item \emph{(Sampling and weighting).} Markets are independent and identically distributed, $\Theta$ is compact with $\theta^*\in\Theta$, and $\widehat W_T\overset{p}{\to}W$ for a deterministic, symmetric positive-definite matrix $W\in\mathbb R^{q\times q}$.
\item \emph{(Share continuity and solvability).} Shares are generated by the model at $\theta^*$ under the specified subjective prior. Almost surely, $\sigma_t$ is jointly continuous on $\mathbb R^J\times\Theta$, and the share equations have a solution $\delta_{1,t}(\theta)$ for every $\theta\in\Theta$.
\item \emph{(Global moment bound).} The moment envelope $\sup_{\theta\in\Theta}\lVert m_t(\theta)\rVert$ has finite expectation.
\item \emph{(Local regularity and second moments).} For some fixed closed ball $\mathcal N\subset\operatorname{int}\Theta$ of positive radius centered at $\theta^*$, $\sigma_t$ is jointly continuously differentiable in a neighborhood of each pair $(\delta_{1,t}(\theta),\theta)$ with $\theta\in\mathcal N$, almost surely. Both $\lVert m_t(\theta^*)\rVert$ and $\sup_{\theta\in\mathcal N}\lVert\partial m_t(\theta)/\partial\theta^{\prime}\rVert$ have finite second moments, and $\Omega^*:=\mathbb E[m_t(\theta^*)m_t(\theta^*)^{\prime}]$ is positive definite.
\end{enumerate}
\end{assumption}

Using the share-inversion representation, write $\mathcal M(\theta):=\mathbb E[m_t(\theta)]=J^{-1}\mathbb E[Z_{1,t}^{\prime}\xi_{1,t}(\theta)]$ for the population moment. Under Assumption~\ref{ass:riblp_global_identification}, $\theta^*$ is the unique solution to $\mathcal M(\theta)=0$. Lemma~\ref{lem:inverse_moment_regularity}(ii) supplies moment continuity, so parts~(i)--(iii) of Assumption~\ref{ass:exact_gmm_asymptotics}, together with global identification, yield consistency. For asymptotic normality, Lemma~\ref{lem:inverse_moment_regularity}(iii) supplies moment differentiability under the local share smoothness in part~(iv). We also need the second-moment bounds and positive-definite covariance in that part and the local-rank condition in Assumption~\ref{ass:riblp_local_identification}.

To clarify when these regularity requirements hold, we provide sufficient conditions for selected parts of Assumption~\ref{ass:exact_gmm_asymptotics} in Appendix~\ref{appendix:exact_gmm_sufficient_conditions}. First, for part~(ii), Lemma~\ref{lem:inverse_existence_sufficient} establishes share-equation solvability under its stated curvature and residual bounds. Lemma~\ref{lem:share_continuity_primitives} establishes joint share continuity under continuity and integrability conditions on the coefficient and prior transformations.

Second, Lemma~\ref{lem:fixed_prior_smoothing} establishes the joint share differentiability required in part~(iv) under a parameter-independent prior and the stated shock-density and coefficient-regularity conditions. The tuna specification in Section~\ref{sec:tuna_application} satisfies these sufficient conditions. For the parameter-dependent priors allowed in Section~\ref{sec:model} and used in the Monte Carlo designs of Section~\ref{sec:monte_carlo}, joint share differentiability remains a maintained condition.

Finally, Lemma~\ref{lem:moment_envelopes} gives bounds that verify the moment-envelope requirements in parts~(iii)--(iv) when the stated expectations across markets are finite.

Lemma~\ref{lem:inverse_moment_regularity}(iii) and the moment-derivative bound in Assumption~\ref{ass:exact_gmm_asymptotics}(iv) permit differentiation under the expectation, giving
\[
D^*:=\left.\frac{\partial\mathcal M(\theta)}{\partial\theta^{\prime}}\right|_{\theta=\theta^*}=\mathbb E\!\left[\left.\frac{\partial m_t(\theta)}{\partial\theta^{\prime}}\right|_{\theta=\theta^*}\right]\in\mathbb R^{q\times d_\theta}.
\]
Assumption~\ref{ass:riblp_local_identification} requires $D^*$ to have full column rank; Appendix~\ref{appendix:joint_identification_rank} expresses it in terms of share derivatives.

\begin{theorem}[Exact GMM asymptotics]
\label{thm:riblp_gmm_asymptotics}
Under the maintained model conditions above, global identification (Assumption~\ref{ass:riblp_global_identification}), and Assumption~\ref{ass:exact_gmm_asymptotics}(i)--(iii), $\widehat\theta_T\overset{p}{\to}\theta^*$. If the local-rank condition (Assumption~\ref{ass:riblp_local_identification}) and Assumption~\ref{ass:exact_gmm_asymptotics}(iv) also hold, then
\[
\sqrt T\bigl(\widehat\theta_T-\theta^*\bigr)\overset{d}{\to}\mathcal N(0,V^*),
\]
where
\begingroup
\small
\begin{equation}
V^*=\bigl(\bigl(D^*\bigr)^{\prime}WD^*\bigr)^{-1}\bigl(D^*\bigr)^{\prime}W\Omega^*WD^*\bigl(\bigl(D^*\bigr)^{\prime}WD^*\bigr)^{-1}\in\mathbb R^{d_\theta\times d_\theta}.
\label{eq:gmm_asymptotic_variance}
\end{equation}
\endgroup
\end{theorem}

Appendix~\ref{appendix:exact_gmm_proof} proves consistency using the uniform law of large numbers in Lemma~2.4 and the extremum-estimator result in Theorem~2.1 of \citet{newey1994large}. With the additional local conditions, the proof applies their Theorem~3.4 to establish asymptotic normality.

\paragraph{Simulated-GMM asymptotics.}
Let $R_T,S_T\to\infty$. For each $T$, we reuse the same random numbers during optimization, recalculating coefficients and prior utilities at each candidate $\theta$. We use the simulated moments $m_t^{\mathrm{sim}}(\theta)$ and $\overline m_T^{\mathrm{sim}}(\theta)$ defined above. The next assumption imposes unique solvability of the simulated share equations and controls the combined subjective-prior and population-integration error.

\begin{assumption}[Simulation approximation]
\label{ass:simulation_asymptotics}
\leavevmode
\begin{enumerate}
\renewcommand{\labelenumi}{(\roman{enumi})}
\setlength{\itemsep}{0.45em}
\setlength{\parsep}{0pt}
\item \emph{(Simulated solvability and continuity).} With probability approaching one, the simulated share equations have a unique solution $\delta_{1,t}^{\mathrm{sim}}(\theta)$ for every sample market and every $\theta\in\Theta$, and $m_t^{\mathrm{sim}}(\theta)$ is continuous on $\Theta$.
\item \emph{(Uniform moment approximation).} For some positive deterministic sequence $a_{R_T,S_T}\to0$,
\[
\sup_{\theta\in\Theta}\lVert\overline m_T^{\mathrm{sim}}(\theta)-\overline m_T(\theta)\rVert=O_p(a_{R_T,S_T}).
\]
Here, $a_{R_T,S_T}$ controls the stochastic order of the uniform simulation error in the sample moments.
\item \emph{(First-order negligible simulation error).} $\sqrt T\,a_{R_T,S_T}\to0$.
\item \emph{(Second-moment approximation).} For variance estimation, the simulated second moments satisfy
\begingroup
\small
\[
\sup_{\theta\in\mathcal N}\left\lVert\frac{1}{T}\sum_{t=1}^T\left[m_t^{\mathrm{sim}}(\theta)m_t^{\mathrm{sim}}(\theta)^{\prime}-m_t(\theta)m_t(\theta)^{\prime}\right]\right\rVert=o_p(1).
\]
\endgroup
\end{enumerate}
\end{assumption}

Probability statements include data and simulation randomness. Part~(i) makes the simulated share-inversion representation well defined with probability approaching one. Under the exact-GMM consistency conditions, part~(ii) makes the simulated criterion uniformly close to the exact criterion, yielding consistency. Together with part~(ii), part~(iii) makes the moment approximation error negligible at the $T^{-1/2}$ scale. Under the exact-GMM asymptotic-normality conditions, this preserves the limiting distribution. Part~(iv) additionally supports consistent variance estimation.

Appendix~\ref{appendix:share_approximation_bounds} provides sufficient conditions for parts~(ii) and~(iv) by translating share-approximation bounds into moment-approximation bounds. These conditions include moment bounds and a uniform positive lower bound on the smallest eigenvalue of the latent-index share Jacobian $\partial\sigma_t(\delta;\theta)/\partial\delta^{\prime}$. This approach parallels Assumptions~A.7--A.8 and~B.8 of \citet[Appendix~B]{freyberger2015}.

We maintain part~(i) separately because exact share invertibility does not guarantee solvability, uniqueness, and continuity of simulated inversion. We also impose $\sqrt T\,a_{R_T,S_T}\to0$ in part~(iii). A specific growth rule for $R_T$ and $S_T$ would require an integration-error rate expressed in these numbers of draws, which we do not derive here.

\begin{theorem}[Simulated GMM asymptotics]
\label{thm:simulated_gmm_mpec_asymptotics}
Under the conditions for consistency in Theorem~\ref{thm:riblp_gmm_asymptotics} and Assumption~\ref{ass:simulation_asymptotics}(i)--(ii), $\widehat\theta_T^{\mathrm{sim}}\overset{p}{\to}\theta^*$. Under the conditions for asymptotic normality in that theorem and Assumption~\ref{ass:simulation_asymptotics}(i)--(iii),
\[
\sqrt T\bigl(\widehat\theta_T^{\mathrm{sim}}-\theta^*\bigr)\overset{d}{\to}\mathcal N(0,V^*),
\]
where $V^*$ is defined in Equation~\eqref{eq:gmm_asymptotic_variance}.
\end{theorem}

Appendix~\ref{appendix:simulated_gmm_proof} proves consistency by uniform criterion convergence and Theorem~2.1 of \citet{newey1994large}. For asymptotic normality, the proof uses exact-moment smoothness and the simulation-error bound to verify the conditions of their nonsmooth minimum-distance Theorem~7.2. The argument does not require the simulated moments to be differentiable everywhere, allowing for changes in consumers' consideration sets. Appendix~\ref{appendix:approximate_minimization} gives conditions on numerical optimization error that preserve these asymptotic results.

To estimate the asymptotic variance $V^*$, let $\widehat D_T\in\mathbb R^{q\times d_\theta}$ be a consistent estimator of the population moment Jacobian $D^*$. For the market-level moment covariance $\Omega^*$, define the estimator $\widehat\Omega_T\in\mathbb R^{q\times q}$ by
\begingroup
\small
\begin{equation}
\widehat\Omega_T:=\frac{1}{T}\sum_{t=1}^T\bigl[m_t^{\mathrm{sim}}(\widehat\theta_T^{\mathrm{sim}})-\overline m_T^{\mathrm{sim}}(\widehat\theta_T^{\mathrm{sim}})\bigr]\bigl[m_t^{\mathrm{sim}}(\widehat\theta_T^{\mathrm{sim}})-\overline m_T^{\mathrm{sim}}(\widehat\theta_T^{\mathrm{sim}})\bigr]^{\prime}.
\label{eq:gmm_sample_jacobian_covariance}
\end{equation}
\endgroup
Under the conditions for asymptotic normality in Theorem~\ref{thm:simulated_gmm_mpec_asymptotics} and Assumption~\ref{ass:simulation_asymptotics}(iv), $\widehat\Omega_T\overset{p}{\to}\Omega^*$. A consistent estimator of $V^*$ is therefore
\begingroup
\small
\begin{equation}
\widehat V_T=(\widehat D_T^{\prime}\widehat W_T\widehat D_T)^{-1}\widehat D_T^{\prime}\widehat W_T\widehat\Omega_T\widehat W_T\widehat D_T(\widehat D_T^{\prime}\widehat W_T\widehat D_T)^{-1}.
\label{eq:gmm_variance_estimator}
\end{equation}
\endgroup
The estimated covariance matrix of $\widehat\theta_T^{\mathrm{sim}}$ is $\widehat V_T/T$. Appendix~\ref{appendix:gmm_variance_consistency} proves variance consistency and gives two constructions for $\widehat D_T$: a finite-difference approximation and an analytical Jacobian obtained by implicit differentiation. The analytical Jacobian, used in our numerical applications, requires the additional smoothness and derivative-approximation conditions given in that appendix.

\section{Computation}
\label{sec:computation}

This section describes how we compute the simulated RI--BLP GMM estimator introduced in Section~\ref{sec:estimation}. We first explain how to solve the consumer's rational-inattention problem to evaluate simulated shares, then how to combine GMM searches with share inversion and MPEC solves to obtain GMM estimates. Appendix~\ref{appendix:computational_details} gives the algorithmic details. Throughout this section and its appendix, we omit $R_T$ and $S_T$ from function subscripts.

\subsection{Solving the Rational-Inattention Problem}

For each market, population draw, and candidate parameter vector, we solve the simulated reduced problem in Equation~\eqref{eq:simulated_ri_problem} using Newton--BA, the KKT-guided hybrid of Newton--Raphson and the Blahut--Arimoto (BA) algorithm proposed by \citet{zhu2026RIestimation}. The population draw includes both random coefficients and additive shocks, as defined in Section~\ref{sec:estimation}.

Newton--BA uses Newton--Raphson for interior solutions and switches to BA near the boundary of the probability simplex. For the current candidate integrated CCP vector $P^0\in\Delta(\mathcal J)$, the simulated KKT ratio $\widehat\kappa_{jrt}(P^0)$ is obtained by replacing the integral in Equation~\eqref{eq:inner_kkt_ratio} with the average over the $S_T$ subjective-prior draws. This hybrid method combines Newton--Raphson's fast local convergence at regular interior solutions with BA's ability to approach boundary solutions.

We initialize the integrated CCP vector uniformly and apply Newton--Raphson to the simulated interior first-order conditions. A Newton candidate that passes the final simplex-feasibility and KKT checks is accepted without switching to BA. Otherwise, Newton continues until an integrated CCP reaches $\tau_{\mathrm{switch}}$ or below, Newton cannot continue reliably, or its iteration limit is reached. Any of these events triggers the switch to BA.

When the probability threshold triggers the switch to BA, we check the simulated KKT ratios $\widehat\kappa_{jrt}(P^0)$ for probabilities $P_j^0\leq\tau_{\mathrm{switch}}$. BA increases a positive probability when $\widehat\kappa_{jrt}(P^0)>1$. For those whose ratios exceed one by more than the inner KKT tolerance, we therefore shorten the final Newton step if necessary so that they start BA at or above $\tau_{\mathrm{switch}}$. If no adjustment is needed, BA starts from the last feasible Newton candidate. BA then applies multiplicative updates until the KKT tolerance is satisfied or its iteration limit is reached, without returning to Newton--Raphson.

The final integrated CCP vector is accepted only after passing the simplex-feasibility and KKT checks. Appendix~\ref{appendix:newton_ba} gives the algorithmic details, including numerical stabilization of the objective and CCP calculations.

Once the simulated integrated CCP vector $\widehat P_{rt}^{0*}$ is obtained, we use it to calculate the simulated realized CCPs $\widehat P_{jrt}^{*}$ defined in Section~\ref{sec:estimation}. Averaging these probabilities over the joint population draws gives the simulated share vector $\sigma_t^{\mathrm{sim}}$ in Equation~\eqref{eq:simulated_share}. At fixed $\theta$ and simulation draws, the inner RI problem does not depend on $\delta_{1,t}$. We therefore keep $\widehat P_{rt}^{0*}$ fixed when updating $\delta_{1,t}$ and recompute only the realized CCPs and market shares.

\subsection{Computing the GMM Estimates}
\label{subsec:gmm_computation}

We first obtain candidate estimates using share-inversion GMM. When further optimization is required, we then solve the share-constrained GMM problem in MPEC form, using these candidates as starting values. Both formulations are introduced in Section~\ref{sec:estimation}.

To reduce computational cost, we first obtain starting values using fewer simulation draws, then refine and validate the estimates using the final numbers of draws. Within each share-inversion GMM search or MPEC solve, we reuse the same random numbers. The estimation procedure has three steps:
\begin{enumerate}[label=\emph{Step \arabic*:},leftmargin=*,itemsep=0.4em,parsep=0pt]
\item \emph{Lower-draw initialization.} Run share-inversion GMM with fewer draws from the specified starting values. If the specification is exactly identified and the candidate passes the independent feasibility and near-zero-moment checks at these draws, retain it for Step~2. Otherwise, continue optimization with an MPEC solve, using the candidate to initialize IPOPT at the same draws.
\item \emph{Final-draw refinement.} Using the Step~1 candidates as starting values, update $\delta_1$ by share inversion with the final draws. If share matching succeeds, continue the GMM search over $\theta$ with these draws.
\item \emph{Final optimization and acceptance.} Using the final draws, accept a candidate if the specification is exactly identified and the candidate passes the independent feasibility and near-zero-moment checks below. Otherwise, solve the MPEC problem with IPOPT, starting from a candidate saved in the preceding steps, and independently check the result.
\end{enumerate}

Both GMM implementations match shares in logarithms to improve scaling across products with different share magnitudes:
\begingroup
\small
\begin{equation}
\ln s_{jt}=\ln\sigma_{j,t}^{\mathrm{sim}}\bigl(\delta_{0,t}(\beta_0,\alpha),\delta_{1,t};\theta\bigr),\qquad j=1,\ldots,J,\quad t=1,\ldots,T.
\label{eq:computation_log_share_constraints}
\end{equation}
\endgroup
For positive observed and simulated shares, these constraints are equivalent to the share-level restrictions. Evaluations with nonpositive simulated shares are rejected, not replaced by a probability floor.

Share-inversion GMM involves an outer parameter search over $\theta$ and an inner share inversion for $\delta_1$ at each candidate $\theta$. Both problems are solved by bounded trust-region nonlinear least squares:
\begin{enumerate}[label=\arabic*.,leftmargin=*,itemsep=0.4em,parsep=0pt]
\item \emph{Inner share inversion.} For a fixed candidate $\theta$, solve
\begingroup
\small
\[
\min_{\delta_1}\frac{1}{2}\sum_{t=1}^{T}\sum_{j=1}^{J}\left[\ln\sigma_{j,t}^{\mathrm{sim}}\bigl(\delta_{0,t}(\beta_0,\alpha),\delta_{1,t};\theta\bigr)-\ln s_{jt}\right]^2
\]
\endgroup
over the specified bounds on $\delta_1$, using the analytical log-share Jacobian with respect to $\delta_1$. Accept the recovered indices $\delta_{1,t}^{\mathrm{sim}}(\theta)$ only when share matching satisfies the prescribed tolerance.
\item \emph{Outer GMM search.} Use the recovered indices $\delta_{1,t}^{\mathrm{sim}}(\theta)$ to evaluate the sample IV moments $\overline m_T^{\mathrm{sim}}(\theta)$. Update $\theta$ using bounded trust-region nonlinear least squares on the weighted moments to minimize the GMM criterion $\overline m_T^{\mathrm{sim}}(\theta)^{\prime}\widehat W_T\overline m_T^{\mathrm{sim}}(\theta)$. Each new candidate $\theta$ requires repeating the inner share inversion before reevaluating the moments. The analytical moment Jacobian accounts for the response of $\delta_{1,t}^{\mathrm{sim}}(\theta)$ to $\theta$ by implicitly differentiating the share equations.
\end{enumerate}

Each trust-region search uses its analytical Jacobian to form a local quadratic approximation to the least-squares criterion. Each update is limited by a trust region. The region is adjusted by comparing the reduction predicted by the approximation with the actual reduction in the criterion after recomputing the shares or moments. For share inversion, this replaces the log-share fixed-point iteration of \citet{berry1995}.

For share-constrained GMM, IPOPT \citep{wachter2006implementation} minimizes the GMM criterion jointly over $\theta$ and $\delta_1$, subject to the share equations. IPOPT need not match shares at every iteration. We supply IPOPT with the analytical objective gradient and a sparse Jacobian of the log-share constraints. IPOPT uses a limited-memory quasi-Newton approximation to the Lagrangian Hessian.

Both GMM implementations use the same analytical derivatives of the simulated shares $\sigma_t^{\mathrm{sim}}$. For derivatives with respect to $\theta$, we hold $\delta_1$ fixed and differentiate the inner RI optimality equations to obtain the response of $\widehat P_{rt}^{0*}$. We then differentiate the realized CCPs $\widehat P_{jrt}^*$, accounting for changes in both the integrated CCPs and baseline utilities $u_{0,jt}^{(r)}$. For derivatives with respect to $\delta_1$, we hold $\theta$ and $\widehat P_{rt}^{0*}$ fixed and differentiate the realized-CCP formula directly. Averaging the CCP derivatives over population draws gives the share derivatives. Appendix~\ref{appendix:derivative_evaluation} gives the formulas in Equations~\eqref{eq:computation_inner_derivative}--\eqref{eq:computation_delta_derivative}, states their local validity conditions, and explains the numerical approximation used in implementation.

Before accepting any estimate, we independently reevaluate the candidate using the final draws. Every inner RI solution must pass the simplex-feasibility and KKT checks in Appendix~\ref{appendix:newton_ba}. The parameter and latent-index bounds must hold within tolerance, and the maximum absolute log-share residual in Equation~\eqref{eq:computation_log_share_constraints} must not exceed the prescribed tolerance $\tau_{\mathrm{share}}$. These feasibility checks apply to candidates from both formulations.

For estimates obtained from a final MPEC solve, we additionally require successful or acceptable IPOPT termination and outer KKT residuals within their prescribed tolerances. We independently check stationarity, bound dual feasibility, and complementary slackness using the final objective gradient, constraint Jacobian, and returned multipliers. These checks establish approximate stationarity, not global optimality.

In an exactly identified specification, a share-inversion GMM candidate may be accepted without a final MPEC solve. It must pass the common feasibility checks, with both its moment norm and GMM objective below their prescribed near-zero tolerances. This numerical rule is motivated by the fact that a feasible candidate with exactly zero sample IV moments has a GMM objective of zero, which cannot be reduced further. The tuna application in Section~\ref{sec:tuna_application} uses this acceptance rule.

\section{Monte Carlo Simulations}
\label{sec:monte_carlo}

This section assesses the finite-sample performance of the simulated RI--BLP GMM estimator using two Monte Carlo designs: one with common coefficients and one with random coefficients. We examine parameter recovery, confidence-interval coverage, and computational reliability.

\subsection{Simulation Design}
\label{subsec:mc_design}

We conduct $B=500$ replications for each $T\in\{25,50,100\}$, with $J=2$ inside products and an outside option. Within each replication, we generate the market-level variables for 100 markets and use the first $T$ for each sample size. Both designs use scalar characteristics, with true parameter values $(\beta_0^*,\alpha^*,\beta_1^*)=(0.75,1.00,0.50)$ and a known information-cost parameter $\lambda=1$. The known additive-shock distributions are $F_{\varepsilon_0}=\mathcal N(0,0.36I_J)$ and $F_{\varepsilon_1}=\mathcal N(0,2.56I_J)$, where $I_J$ is the $J$-dimensional identity matrix. The two shock vectors are mutually independent and independent of the market variables and random coefficients.

We draw $x_{0,jt},z_{jt}\sim\mathcal U[-1,1]$, $x_{1,jt}\sim\mathcal N(0,1)$, $\xi_{1,jt}\sim\mathcal N(0,0.09)$, and the price shock $\eta_{jt}\sim\mathcal N(0,0.01)$ independently across variables, products, and markets. The excluded shifter $z_{jt}$ enters the price equation
\begin{equation}
\ln p_{jt}=-0.60+0.15x_{0,jt}+0.10x_{1,jt}+\gamma_z z_{jt}+0.35\xi_{1,jt}+\eta_{jt}.
\label{eq:mc_price}
\end{equation}
Here $\gamma_z=0.25$ in the common-coefficient design and $\gamma_z=0.60$ in the random-coefficient design. The larger value provides more exogenous price variation to help distinguish mean price sensitivity from its dispersion. The term $0.35\xi_{1,jt}$ makes prices correlated with the unobserved demand shifter.

The common-coefficient specification imposes $\beta_{0,i}=\beta_0$ and $\alpha_i=\alpha$, so $\theta=(\beta_0,\alpha,\beta_1)^{\prime}$. The random-coefficient specification instead sets
\begin{equation}
\beta_{0,i}=\beta_0+\sigma_x\nu_{x,i},\qquad \alpha_i=\alpha\exp(\sigma_p\nu_{p,i}-\sigma_p^2/2),
\label{eq:mc_random_coefficients}
\end{equation}
where $\nu_{x,i}$ and $\nu_{p,i}$ are independent standard normal variables, independent of the market variables and additive shocks. In this specification, the heterogeneity parameters are the dispersion parameters $\vartheta=(\sigma_x,\sigma_p)^{\prime}$, so $\theta=(\beta_0,\alpha,\beta_1,\sigma_x,\sigma_p)^{\prime}$, with $\sigma_x^*=\sigma_p^*=0.40$. The correction $-\sigma_p^2/2$ ensures $\mathbb E[\alpha_i]=\alpha$. The parameters $\sigma_x$ and $\sigma_p$ are the standard deviations of $\beta_{0,i}$ and $\ln\alpha_i$, respectively.

Both Monte Carlo designs use a correct-belief benchmark. Specifically, at the true parameter $\theta^*$, the subjective prior equals the data-generating distribution of latent utility conditional on public information:
\[
\pi_{u_1}=F_{u_1\mid x_0,p,z}\bigl(\,\cdot\mid x_{0,t},p_t,z_t;\theta^*\bigr),
\]
where $z_t=(z_{jt})_{j=1}^{J}$ collects the excluded price shifters.

To construct this prior, we use the decomposition $u_{1,it}=\beta_1x_{1,t}+\xi_{1,t}+\varepsilon_{1,it}$. Equation~\eqref{eq:mc_price}, together with the known price-equation coefficients and shock variances, determines the conditional Gaussian distribution of $(x_{1,t},\xi_{1,t})$ given $(x_{0,t},p_t,z_t)$. The remaining shock $\varepsilon_{1,it}$ is independent and has distribution $F_{\varepsilon_1}$. During estimation, we substitute each candidate $\beta_1$ into this utility expression to obtain the corresponding prior. The price-equation parameters and shock distributions remain fixed.

For GMM estimation, we form the moments in Equation~\eqref{eq:market_moment} using 8 instruments in the common-coefficient design: a constant, $x_{0,jt}$, $x_{1,jt}$, $z_{jt}$, the rival product's $x_0$ and $x_1$,\footnote{With $J=2$, the rival is the other inside product in the same market: product 1 uses $(x_{0,2t},x_{1,2t})$, and product 2 uses $(x_{0,1t},x_{1,1t})$.} and centered versions of $x_{0,jt}^2$ and $z_{jt}^2$. The random-coefficient design adds $x_{0,jt}x_{1,jt}$, $z_{jt}x_{0,jt}$, $z_{jt}x_{1,jt}$, and centered $x_{1,jt}^2$, giving 12 instruments.\footnote{Because $\gamma_z$ and the instrument sets differ between the designs, comparisons do not isolate the effect of random coefficients.} Independence of $x_0$, $x_1$, and $z$ from $\xi_1$ ensures the IV restrictions before sample centering and scaling.

Before forming the sample moments, we center the squared terms at their sample means and then divide each nonconstant instrument by its sample standard deviation.\footnote{For example, the squared-$x_0$ instrument becomes $(x_{0,jt}^2-\overline{x_0^2})/\operatorname{sd}(x_0^2)$, where the sample mean and standard deviation use all $TJ$ product-market observations, with divisor $TJ$ for the variance. Appendix~\ref{appendix:sample_standardized_instruments} explains how the asymptotic analysis accommodates this centering and scaling.} We then use the identity weighting matrix $\widehat W_T=I_q$ in Equation~\eqref{eq:simulated_gmm_mpec}, giving equal weight to the resulting moment components. Here $q=8$ in the common-coefficient design and $q=12$ in the random-coefficient design.

In the common-coefficient and random-coefficient designs, we estimate the structural parameters by simulated GMM using the procedure in Section~\ref{sec:computation}. We use separate scrambled Sobol grids for the $R_T$ joint population draws and $S_T$ subjective-prior draws defined in Section~\ref{sec:estimation}, with the numbers of draws reported in the tables. The grids are shared across markets, with new scrambles for each replication. For each sample size $T$, estimation reuses the random numbers used to generate market shares at $\theta^*$, following \citet[Appendix~B.1]{dube2012improving}.\footnote{These experiments assess recovery and inference under the finite-draw model. They do not separately measure approximation error relative to exact market shares.} We require a final MPEC solve with IPOPT using these draws before accepting an estimate. Appendix~\ref{appendix:mpec_initialization} gives the Monte Carlo starting-value rules and search bounds.

For inference, we use the same weighting matrix in the sandwich estimator $\widehat V_T$ in Equation~\eqref{eq:gmm_variance_estimator}. We obtain the analytical moment Jacobian $\widehat D_T=\widehat D_T^{\mathrm{imp}}$ in Equation~\eqref{eq:simulated_moment_jacobian} by implicit differentiation of the simulated share equations. The nominal 95\% Wald interval for parameter $k$ is $\widehat\theta_{T,k}^{\mathrm{sim}}\pm1.96\sqrt{(\widehat V_T)_{kk}/T}$. Appendix~\ref{appendix:gmm_variance_consistency} gives additional smoothness and derivative-approximation conditions sufficient for this Jacobian's consistency.

We report bias, root mean squared error (RMSE), and empirical coverage using only computationally validated estimates, which pass the acceptance checks in Section~\ref{subsec:gmm_computation} at the tolerances in Table~\ref{tab:computational_tolerances}.\footnote{For each statistic, $B_{\mathrm{valid}}$ counts the included replications: \[\textstyle \mathrm{RMSE}_k=\bigl[B_{\mathrm{valid}}^{-1}\sum_{b=1}^{B_{\mathrm{valid}}}(\widehat\theta_k^{(b)}-\theta_k^*)^2\bigr]^{1/2},\quad \mathrm{Coverage}_k=(100/B_{\mathrm{valid}})\sum_{b=1}^{B_{\mathrm{valid}}}\mathbf 1\{|\widehat\theta_k^{(b)}-\theta_k^*|\leq1.96\,\widehat{\mathrm{se}}_k^{(b)}\}.\] Here $\widehat\theta_k^{(b)}$ and $\widehat{\mathrm{se}}_k^{(b)}$ are the estimate of parameter $k$ and its standard error in replication $b$, and $\mathbf 1\{\cdot\}$ is the indicator function.} We present these statistics for the common-coefficient design, followed by the random-coefficient design.

\begingroup
\setlength{\intextsep}{4pt}
\Needspace{6\baselineskip}
\subsection{Common-Coefficient Results}
\label{subsec:mc_common_results}

The computational validation rates are 100\%, 99.4\%, and 98.4\% for $T=25,50,100$, respectively. For the common coefficients $\beta_0$, $\alpha$, and $\beta_1$ (Table~\ref{tab:mc_common_coefficients}), RMSE approximately halves between $T=25$ and $T=100$, as both $R_T$ and $S_T$ also increase. This approximate halving is broadly consistent with the asymptotic $T^{-1/2}$ rate, since increasing $T$ from $25$ to $100$ gives $(100/25)^{-1/2}=1/2$. At $T=100$, coverage ranges from 93.7\% to 94.3\%, reasonably close to the nominal 95\%.

\begin{table}[!htbp]
\centering
\begin{minipage}{0.96\textwidth}
\begin{threeparttable}
\caption{Monte Carlo results: common coefficients}
\label{tab:mc_common_coefficients}
\footnotesize
\setlength{\tabcolsep}{3pt}
\renewcommand{\arraystretch}{1.05}
\begin{tabular*}{\linewidth}{@{\extracolsep{\fill}}l *{3}{S[table-format=-1.4] S[table-format=1.4] S[table-format=2.1]}@{}}
\toprule
& \multicolumn{3}{c}{$T=25$} & \multicolumn{3}{c}{$T=50$} & \multicolumn{3}{c}{$T=100$} \\
& \multicolumn{3}{c}{\footnotesize $R_T=64,\ S_T=128$} & \multicolumn{3}{c}{\footnotesize $R_T=128,\ S_T=256$} & \multicolumn{3}{c}{\footnotesize $R_T=256,\ S_T=512$} \\
\cmidrule(lr){2-4}\cmidrule(lr){5-7}\cmidrule(l){8-10}
Parameter & {Bias} & {RMSE} & {Coverage} & {Bias} & {RMSE} & {Coverage} & {Bias} & {RMSE} & {Coverage} \\
\midrule
$\beta_0$ (0.75) & -0.0011 & 0.0214 & 88.0 & -0.0004 & 0.0136 & 92.6 & 0.0001 & 0.0093 & 94.3 \\
$\alpha$ (1.00) & 0.0002 & 0.0166 & 92.8 & 0.0002 & 0.0121 & 93.8 & 0.0006 & 0.0083 & 93.7 \\
$\beta_1$ (0.50) & 0.0056 & 0.0227 & 90.0 & 0.0027 & 0.0159 & 92.4 & 0.0011 & 0.0109 & 94.3 \\
\bottomrule
\end{tabular*}
\begin{tablenotes}[flushleft]
\fontsize{9}{11}\selectfont
\raggedright
\item \emph{Notes:} Statistics use only computationally validated estimates from 500 replications per $T$. True values are in parentheses. Coverage is the percentage of nominal 95\% Wald intervals containing the true value.
\end{tablenotes}
\end{threeparttable}
\end{minipage}
\end{table}

\FloatBarrier
\Needspace{6\baselineskip}
\subsection{Random-Coefficient Results}
\label{subsec:mc_random_results}

The computational validation rates are 98.0\%, 99.0\%, and 98.8\% for $T=25,50,100$, respectively. Table~\ref{tab:mc_random_coefficients} shows improved parameter recovery at larger sample sizes. At $T=100$, RMSE is below 3\% of the true value for each of the mean utility coefficients $\beta_0$, $\alpha$, and $\beta_1$.

The dispersion parameters $\sigma_x$ and $\sigma_p$ remain less precisely estimated, although their RMSE and the magnitude of their downward bias decline as $T$ increases. At $T=100$, the remaining downward bias is more pronounced for $\sigma_x$, at 5.1\% of its true value. The Wald intervals provide reasonably accurate coverage, with 94.9\% for $\sigma_p$ and some overcoverage for $\sigma_x$ at 97.8\%.

\begin{table}[!htbp]
\centering
\begin{minipage}{0.96\textwidth}
\begin{threeparttable}
\caption{Monte Carlo results: random coefficients}
\label{tab:mc_random_coefficients}
\footnotesize
\setlength{\tabcolsep}{3pt}
\renewcommand{\arraystretch}{1.05}
\begin{tabular*}{\linewidth}{@{\extracolsep{\fill}}l *{3}{S[table-format=-1.4] S[table-format=1.4] S[table-format=2.1]}@{}}
\toprule
& \multicolumn{3}{c}{$T=25$} & \multicolumn{3}{c}{$T=50$} & \multicolumn{3}{c}{$T=100$} \\
& \multicolumn{3}{c}{\footnotesize $R_T=64,\ S_T=128$} & \multicolumn{3}{c}{\footnotesize $R_T=128,\ S_T=256$} & \multicolumn{3}{c}{\footnotesize $R_T=256,\ S_T=512$} \\
\cmidrule(lr){2-4}\cmidrule(lr){5-7}\cmidrule(l){8-10}
Parameter & {Bias} & {RMSE} & {Coverage} & {Bias} & {RMSE} & {Coverage} & {Bias} & {RMSE} & {Coverage} \\
\midrule
$\beta_0$ (0.75) & -0.0053 & 0.0408 & 90.0 & -0.0032 & 0.0320 & 91.3 & -0.0030 & 0.0220 & 94.1 \\
$\alpha$ (1.00) & 0.0018 & 0.0301 & 94.1 & 0.0000 & 0.0215 & 92.5 & -0.0003 & 0.0143 & 95.1 \\
$\beta_1$ (0.50) & 0.0049 & 0.0284 & 89.8 & 0.0024 & 0.0201 & 92.5 & 0.0005 & 0.0144 & 95.3 \\
$\sigma_x$ (0.40) & -0.0393 & 0.1528 & 94.7 & -0.0289 & 0.1232 & 97.2 & -0.0206 & 0.0857 & 97.8 \\
$\sigma_p$ (0.40) & -0.0226 & 0.1436 & 94.1 & -0.0091 & 0.0930 & 97.0 & -0.0023 & 0.0616 & 94.9 \\
\bottomrule
\end{tabular*}
\begin{tablenotes}[flushleft]
\fontsize{9}{11}\selectfont
\raggedright
\item \emph{Notes:} Statistics use only computationally validated estimates from 500 replications per $T$. True values are in parentheses. Coverage is the percentage of nominal 95\% Wald intervals containing the true value. Coverage excludes one estimate at a parameter bound when $T=25$.
\end{tablenotes}
\end{threeparttable}
\end{minipage}
\end{table}

The Monte Carlo results show more precise recovery of mean utility coefficients than of dispersion parameters. We next apply the common-coefficient RI--BLP specification to Dominick's canned-tuna data, avoiding the additional uncertainty associated with estimating random-coefficient dispersion.
\FloatBarrier
\endgroup

\section{Empirical Application: Tuna Demand}
\label{sec:tuna_application}

We apply RI--BLP to weekly tuna sales to study the implications of endogenous information acquisition for estimated price responses. Using logit as a benchmark, we compare own-price elasticities and the implied shifts in demand across products and the outside option.

\subsection{Data}

We use the Dominick's Finer Foods scanner data studied by \citet{nevohatzitaskos2006}. We construct our estimation sample from the cleaned store--product--week panel provided by \citet{gandhi2023zeroes} in their replication package. A market $t$ is a calendar week aggregated across stores. A product $j$ is a packaged tuna item identified by a universal product code (UPC).\footnote{A brand can have several UPCs, representing different package sizes or varieties. We study 20 UPCs across six brands, treating each UPC as a separate product.}

The observed share $s_{jt}$ is product $j$'s units sold divided by weekly customer visits across stores. The outside share, $s_{0t}=1-\sum_{j=1}^{J}s_{jt}$, represents purchases of omitted tuna products or no tuna purchase. Because shoppers may buy multiple units in a single visit, units sold per visit only approximate purchase probabilities.

We select the $J=20$ best-selling products among the 49 UPCs, ranked by total units sold across all stores and weeks in the source data. To match the strictly positive model shares implied by Theorem~\ref{thm:riblp_delta1_injective}, we retain only weeks with $s_{jt}>0$ for all $j\in\mathcal J$ and observed prices for every selected product. This leaves $T=234$ weeks and 4,680 product--week observations. The selected products account for an average of 81.6\% of weekly tuna unit sales in this sample. Appendix~\ref{app:tuna_sample} details sample construction and tuna unit-sales coverage for alternative product sets.

\subsection{Estimation}

We use the same sample, price variable, and controls for logit and RI--BLP, with coefficients estimated separately. In the logit benchmark, outside mean utility is normalized to zero, but taste shocks $\varepsilon_{ijt}$ remain for all alternatives $j=0,\ldots,J$. Its share equation is
\begin{equation}
\ln(s_{jt}/s_{0t})
=\gamma_j+\tau_N\operatorname{NearLent}_t+\tau_L\operatorname{Lent}_t
+\tau_1\operatorname{Trend}_t-\alpha p_{jt}+\xi_{1,jt}.
\label{eq:tuna_logit_benchmark}
\end{equation}

In RI--BLP, consumer $i$ knows $u_{0,ijt}$ and acquires information about latent utility $u_{1,ijt}$. The data contain no suitable observed measure of $x_{1,jt}$, the product characteristic initially unknown to consumers. Omitting this term, we specify utilities for inside products $j=1,\ldots,J$ as
\begin{equation}
\begin{aligned}
u_{0,ijt}
&=\gamma_j+\tau_N\operatorname{NearLent}_t
+\tau_L\operatorname{Lent}_t+\tau_1\operatorname{Trend}_t
-\alpha p_{jt}+\varepsilon_{0,ijt},\\
u_{1,ijt}&=\xi_{1,jt}+\varepsilon_{1,ijt}.
\end{aligned}
\label{eq:tuna_riblp_utility}
\end{equation}
As in Section~\ref{sec:model}, we set the outside utility components to $u_{0,i0t}=u_{1,i0t}=0$.

In both models, product fixed effects $\gamma_j$ capture time-invariant UPC differences, including brand and package size. Lent is a high-demand period for tuna \citep{nevohatzitaskos2006}. We use the source data's four-week Lent indicator $\operatorname{Lent}_t$. The indicator $\operatorname{NearLent}_t$ equals one in the four weeks immediately before and the four weeks immediately after each designated Lent period, and zero otherwise. Their coefficients $\tau_L$ and $\tau_N$ measure utility shifts relative to ordinary weeks. $\operatorname{Trend}_t$ is the centered calendar-week index divided by its sample standard deviation.\footnote{For example, if the mean week index is $100$ and its standard deviation is $10$, week $90$ has trend $(90-100)/10=-1$ and week $110$ has trend $(110-100)/10=1$. A negative trend value means earlier than the sample's average week, and a positive value means later.} The trend coefficient $\tau_1$ captures gradual changes in utility over time beyond Lent-related seasonality.

Price $p_{jt}$ is a weighted average of store prices in dollars per ounce. Each store's weight equals its share of that UPC's total weekly revenue. The baseline price coefficient $\alpha>0$ is common across consumers and weeks. In both models, the baseline parameter vector is $\theta=(\gamma_1,\ldots,\gamma_J,\tau_N,\tau_L,\tau_1,\alpha)^{\prime}$. As a robustness check, Appendix~\ref{app:tuna_robustness} allows $\alpha$ to differ between Lent and non-Lent weeks, using an additional instrument.

The unobserved product--week demand shifter $\xi_{1,jt}$ may correlate with price. We address this endogeneity using IV estimation. Motivated by the cost-based IV approach of \citet[Appendix~C]{nevohatzitaskos2006}, we instrument price with acquisition cost per ounce, $c_{jt}$.\footnote{We infer store-level acquisition cost per ounce by multiplying regular package price by one minus the reported regular-profit fraction and dividing by package ounces. We obtain $c_{jt}$ by averaging these costs across stores for UPC $j$ in week $t$. Each store's weight is its share of that UPC's total weekly revenue. For example, costs of \$0.10 and \$0.20 per ounce with weights of 70\% and 30\% give $c_{jt}=0.7\times0.10+0.3\times0.20=0.13$ dollars per ounce.} In both models, the instrument vector $z_{1,jt}$ contains $c_{jt}$, UPC indicators, and the three time controls. Acquisition cost $c_{jt}$ explains 82.5\% of price variation not explained by these controls (partial $R^2=0.825$), supporting instrument relevance.\footnote{We regress price on the instrument vector: $p_{jt}=b^{\prime}z_{1,jt}+v_{jt}$, where $b$ collects regression coefficients and $v_{jt}$ is the residual. This is logit's first stage and an RI--BLP relevance diagnostic, not a separate GMM estimation step.} Instrument exclusion is a separate maintained assumption. Together with exogeneity of the controls, it is expressed as $\mathbb E[z_{1,jt}\xi_{1,jt}]=0$.

This instrument set provides 24 IV moments for the 24 baseline coefficients. We retain common coefficients because jointly estimating an additional random-coefficient dispersion parameter would require further identifying restrictions to satisfy the rank condition in Assumption~\ref{ass:riblp_local_identification}. For the logit benchmark, we use the closed-form log-share inversion in Equation~\eqref{eq:tuna_logit_benchmark} and estimate its coefficients by two-stage least squares.

In RI--BLP, we fix information cost $\lambda=1$ and assume centered Gaussian distributions for the subjective prior $\pi_{u_1}$ and population shocks $\varepsilon_0$ and $\varepsilon_1$. The baseline uses $\pi_{u_1}=\mathcal N(0,22I_J)$, $F_{\varepsilon_0}=\mathcal N(0,I_J)$, and $F_{\varepsilon_1}=\mathcal N(0,2.56I_J)$. The prior does not depend on $\theta$, satisfying the parameter-independence condition in Lemma~\ref{lem:fixed_prior_smoothing}. It also does not vary with observed prices. The population-shock vectors $\varepsilon_{0,it}$ and $\varepsilon_{1,it}$ are independent.

We estimate RI--BLP by simulated GMM with the fixed weighting matrix $\widehat W_T=\bigl[(TJ)^{-1}\sum_{t=1}^T Z_{1,t}^{\prime}Z_{1,t}\bigr]^{-1}$, where $Z_{1,t}$ has rows $z_{1,jt}^{\prime}$. For this fixed sample, we write $R$ and $S$ for $R_T$ and $S_T$. Numerical integration uses $R=512$ joint draws of $(\varepsilon_{0,it},\varepsilon_{1,it})$ from $F_{\varepsilon}$ and $S=1024$ draws from the subjective prior $\pi_{u_1}$. The two draw sets use separate fixed scrambled Sobol sequences, and the same prior draws are reused across population draws. Estimation follows the computational procedure in Section~\ref{subsec:gmm_computation}. Appendix~\ref{app:tuna_numerical} gives the empirical starting values, tolerances, and derivative checks. Appendix~\ref{app:tuna_integration} examines sensitivity to integration draws.

For inference in both models, we allow the market-level IV moments to be correlated across weeks. We use the GMM sandwich formula in Equation~\eqref{eq:gmm_variance_estimator}. We obtain $\widehat\Omega_T$ by estimating the moments' long-run covariance with the Newey--West heteroskedasticity and autocorrelation consistent (HAC) estimator, using eight calendar-week lags \citep{newey1987simple}. For RI--BLP, we use the analytical moment Jacobian in Equation~\eqref{eq:simulated_moment_jacobian}. Appendix~\ref{appendix:gmm_variance_consistency} gives the conditions for inference under serial dependence. Table~\ref{tab:tuna_main_results} reports coefficient estimates and own-price elasticity summaries.

\begin{table}[!htbp]
\centering
\caption{Demand estimates and own-price elasticities}
\label{tab:tuna_main_results}
\begin{minipage}{0.70\textwidth}
\footnotesize
\begin{tabular*}{\linewidth}{@{\extracolsep{\fill}}lcc@{}}
\toprule
 & Logit & RI--BLP \\
\midrule
Near Lent, $\tau_N$ & $0.107$ & $0.054$ \\
 & (0.072) & (0.033) \\
Lent, $\tau_L$ & $0.209^{***}$ & $0.104^{***}$ \\
 & (0.065) & (0.030) \\
Week trend, $\tau_1$ & $-0.084^{**}$ & $-0.037^{**}$ \\
 & (0.036) & (0.016) \\
Price coefficient, $\alpha$ & $2.675^{***}$ & $1.755^{***}$ \\
 & (0.334) & (0.202) \\
\midrule
Mean own-price elasticity & $-0.491$ & $-0.532$ \\ 
Share with own elasticity below $-1$ & 2.3\% & 2.3\% \\ 
\bottomrule
\end{tabular*}
\par\smallskip\fontsize{9}{11}\selectfont\raggedright
\emph{Notes:} 20 products, 234 weeks. Equal product--week weights for elasticities. Parentheses: HAC(8) standard errors. $^{***}p<0.01$, $^{**}p<0.05$, $^*p<0.10$.
\end{minipage}
\end{table}

\FloatBarrier
Table~\ref{tab:tuna_main_results} reports positive price coefficients, significant at the 1\% level, for both models. Higher prices therefore reduce utility through $-\alpha p_{jt}$. To compare the effects on demand, we next examine own-price elasticities.

\Needspace{6\baselineskip}
\subsection{Own-price elasticities}

In both logit and RI--BLP, own-price elasticity is defined using model-implied share $\sigma_{jt}$ as
\begin{equation}
\eta_{jjt}:=\frac{\partial\sigma_{jt}}{\partial p_{jt}}\frac{p_{jt}}{\sigma_{jt}}.
\label{eq:tuna_own_elasticity}
\end{equation}
It approximates the percentage demand change from a 1 percent own-price increase. If $\eta_{jjt}<-1$, demand is elastic and declines more than proportionately to price. If $-1<\eta_{jjt}<0$, demand is inelastic and declines less than proportionately. For RI--BLP, we compute the price responses using the simulated shares $\sigma_{j,t}^{\mathrm{sim}}$ and their analytical price derivatives at the estimated parameters, as described in Appendix~\ref{app:tuna_numerical}. When varying product $j$'s price $p_{jt}$, we hold the prices of the remaining products and all recovered demand shifters $\xi_{1,t}$ fixed.

Table~\ref{tab:tuna_main_results} reports mean own-price elasticities of $-0.491$ for logit and $-0.532$ for RI--BLP.\footnote{\citet[Table~7]{gandhi2023zeroes} report mean own-price elasticities of $-0.57$ for logit estimated using positive market shares and $-1.51$ for their bounds estimator, which includes observations with zero market shares. However, our estimates may still be affected by excluding observations with zero market shares.} The RI--BLP mean is 8.3\% larger in magnitude, but both models imply inelastic demand on average in the baseline sample.

\begin{figure}[!htb]
\centering
\includegraphics[width=0.86\textwidth]{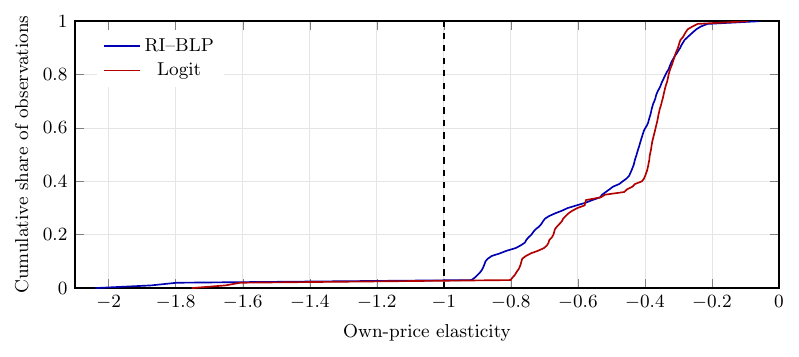}
\caption{Cumulative distributions of own-price elasticities}
\label{fig:tuna_elasticity_cdf}
\begin{minipage}{0.94\textwidth}
\footnotesize\raggedright
\emph{Notes:} We calculate one own-price elasticity for each product in each week: $20\times234=4{,}680$ elasticities per model. The dashed vertical line marks unit-elastic demand ($\eta_{jjt}=-1$).
\end{minipage}
\end{figure}

Beyond the averages, Figure~\ref{fig:tuna_elasticity_cdf} compares the distributions of own-price elasticities. Each curve gives the fraction of observations at or below each elasticity. Only 2.3 percent of observations have elastic demand ($\eta_{jjt}<-1$) in either model, so inelastic responses dominate the baseline sample. In addition, RI--BLP yields a wider range of own-price elasticities than logit.

Own-price elasticities depend substantially on the weeks included. Appendix~\ref{app:tuna_product_set} reports estimates for the same 20-product set using the 107 weeks with positive sales for all 30 products in the expanded-product comparison. Re-estimation changes the mean own-price elasticity from $-0.532$ to $-1.483$ for RI--BLP and from $-0.491$ to $-1.527$ for logit. In the restricted sample, both models therefore imply elastic rather than inelastic demand on average.

\FloatBarrier
\Needspace{6\baselineskip}
\subsection{Cross-price elasticities and diversion}

We next consider how a product's price affects demand for other alternatives. The following two measures apply to both models. For origin product $j$ and destination product $k\ne j$, cross-price elasticity is
\[
\eta_{kjt}:=\frac{\partial\sigma_{kt}}{\partial p_{jt}}\frac{p_{jt}}{\sigma_{kt}}.
\]
It approximates the percentage change in product $k$'s demand from a 1 percent increase in product $j$'s price. If $\eta_{kjt}>0$, raising product $j$'s price increases demand for product $k$. If $\eta_{kjt}<0$, it decreases demand for product $k$.

Diversion instead measures the destination's share change relative to the origin's own-share loss, not the absolute change in sales. To avoid dividing by very small weekly own-share losses, we aggregate price derivatives across weeks. The resulting derivative-weighted (DW) diversion is
\begin{equation}
D^{\mathrm{DW}}_{j\rightarrow k}
:=-\frac{\sum_{t=1}^{T}\partial\sigma_{kt}/\partial p_{jt}}
{\sum_{t=1}^{T}\partial\sigma_{jt}/\partial p_{jt}},\qquad k\ne j.
\label{eq:tuna_derivative_weighted_diversion}
\end{equation}
The destination can include the outside option, $k=0$, which includes omitted tuna products. The outside share is $\sigma_{0t}=1-\sum_{j=1}^{J}\sigma_{jt}$. A ratio of $0.05$ means five units of destination share gained per 100 origin-share units lost. A ratio of $-0.05$ means five additional units lost. Table~\ref{tab:tuna_substitution} reports an independence of irrelevant alternatives (IIA) check, cross-price elasticities, and outside diversion.

\begin{table}[!htbp]
\centering
\caption{Cross-price elasticities and outside diversion}
\label{tab:tuna_substitution}
\begin{minipage}{0.74\textwidth}
\footnotesize
\begin{tabular*}{\linewidth}{@{\extracolsep{\fill}}lcc@{}}
\toprule
 & Logit & RI--BLP \\ 
\midrule
IIA check & 0.0000 & 0.0145 \\
Mean cross-price elasticity & $0.0010$ & $-0.0070$ \\ 
Negative cross-price elasticities & 0.0\% & 97.7\% \\ 
Mean outside diversion & 0.941 & 1.388 \\ 
\bottomrule
\end{tabular*}
\par\smallskip\fontsize{9}{11}\selectfont\raggedright
\emph{Notes:} For the IIA check, we calculate the standard deviation of $\partial\ln\sigma_{kt}/\partial p_{jt}$ across the other inside products for each origin product and week, then average these standard deviations. Cross-price summaries weight all ordered inside-product pairs and weeks equally. Outside diversion averages Equation~\eqref{eq:tuna_derivative_weighted_diversion} across origin products.
\end{minipage}
\end{table}

Logit's IIA statistic is zero, as expected, because a given price change produces equal percentage responses across the other products. RI--BLP does not impose this equality, and its nonzero statistic shows that fitted percentage responses differ across destinations.

Raising one product's price by 1\% implies an average demand increase for another included product of approximately 0.0010\% under logit, but a decrease of 0.0070\% under RI--BLP. These small pairwise averages are not changes in total demand. Logit imposes positive cross-price responses when $\alpha>0$,\footnote{For logit, $\eta_{kjt}=\alpha p_{jt}\sigma_{jt}>0$ for $k\ne j$.} whereas 97.7\% of estimated RI--BLP responses are negative.

These RI--BLP responses reflect two channels. First, holding the integrated CCPs $P^{0*}$ fixed, a higher price makes the affected product less attractive and shifts demand toward other alternatives through the generalized-logit formula in Equation~\eqref{eq:logitccp}. This is the usual substitution effect. Second, consumers reoptimize information acquisition, changing $P^{0*}$ and generating an additional demand response. The negative cross-price responses reported here mean that the second channel reduces other products' demand by more than the first channel increases it.

One possible economic interpretation is that a higher price reduces the benefit of acquiring information before choosing among the included products. Consumers may then shift toward the outside option rather than another inside product.

Consistent with these losses, mean outside diversion is $1.388$ under RI--BLP, compared with $0.941$ under logit. Since shares sum to one, mean combined diversion to the other inside products is $1-1.388=-0.388$ for RI--BLP and $1-0.941=0.059$ for logit. Thus, on average across price origins, the other inside products collectively lose share under RI--BLP but gain share under logit.

\begin{figure}[!htb]
\centering
\includegraphics[width=\textwidth]{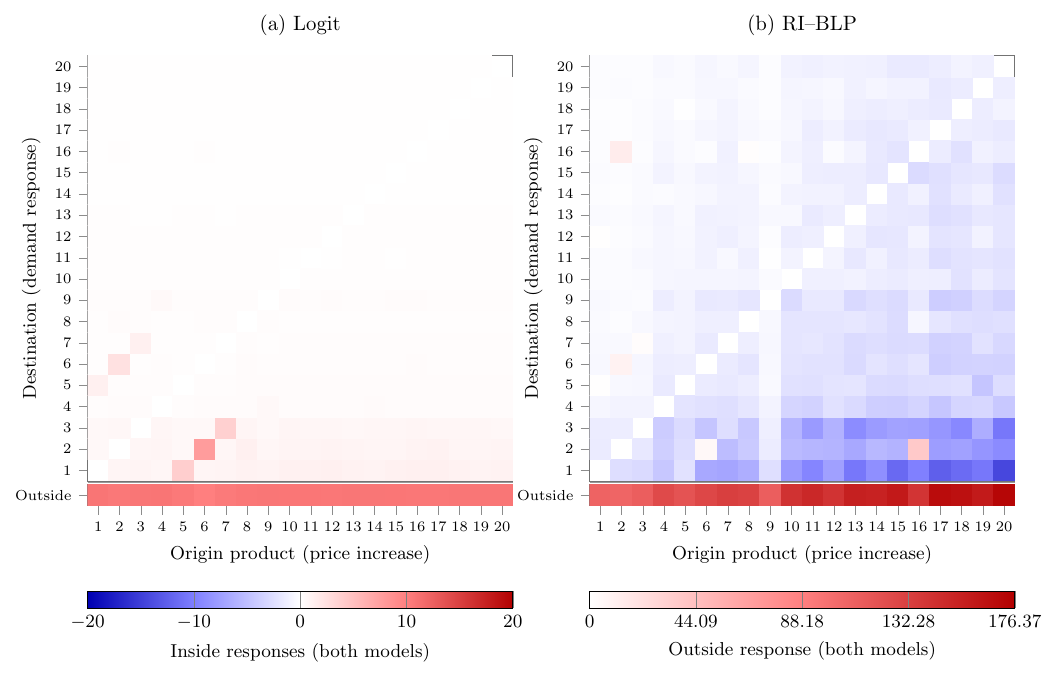}
\caption{Diversion across products and the outside option}
\label{fig:tuna_substitution_heatmaps}
\begin{minipage}{0.96\textwidth}
\footnotesize\raggedright
\emph{Notes:} Product numbers follow the sales ranking, with 1 denoting the best-selling UPC. Colours show $100D^{\mathrm{DW}}_{j\to k}$, the destination-share change per 100 own-share units lost. Inside and outside responses use separate colour scales, shared by both models. White denotes zero and diagonal cells are omitted. Only inside-product prices are varied.
\end{minipage}
\end{figure}

Figure~\ref{fig:tuna_substitution_heatmaps} illustrates these patterns for all 20 products and the outside option. Columns (horizontal axis) identify the product whose price rises, and rows (vertical axis) identify the destination whose share changes, with the outside option at the bottom. For example, column 1, row 2 shows how product 2 responds when product 1's price rises. Blue indicates losses and red gains. Darker colours mean larger absolute diversion ratios. Consistent with the averages above, logit shows gains for other included products (red cells), whereas RI--BLP shows mostly losses (blue cells).

These broad patterns persist in the integration-draw, Lent, and calibration checks in Appendices~\ref{app:tuna_integration}, \ref{app:tuna_robustness}, and~\ref{app:tuna_sensitivity}. In the sample and product-set comparisons in Appendix~\ref{app:tuna_product_set}, approximately 98\% of RI--BLP cross-price elasticities among the common 20 products remain negative, and mean diversion outside this common product set remains above one. Individual diversion ratios are less stable: their magnitudes and some signs vary across draw settings and samples, and destination rankings remain sensitive to integration draws.

\FloatBarrier
\subsection{Within-brand demand responses}
\label{sec:tuna_product_responses}

We next examine whether these demand patterns also arise among similar products. Table~\ref{tab:tuna_within_brand} lists the four same-brand, same-size pairs among the 20 products whose mean prices differ by less than 1 percent.

\begin{table}[!htbp]
\centering
\caption{Prices and own-price elasticities within brands}
\label{tab:tuna_within_brand}
\begin{minipage}{0.90\textwidth}
\footnotesize
\setlength{\tabcolsep}{4pt}
\begin{tabular*}{\linewidth}{@{\extracolsep{\fill}}lcccc@{}}
\toprule
 & & & \multicolumn{2}{c}{Mean own-price elasticity}\\
\cmidrule(l){4-5}
Products & Size & Mean prices & Logit & RI--BLP\\
 & (oz) & (cents/oz) & & \\
\midrule
1 and 5 & 6.12 & $13.10,\ 13.09$ & $-0.347,\ -0.349$ & $-0.322,\ -0.277$\\
2 and 6 & 6.00 & $13.16,\ 13.25$ & $-0.350,\ -0.354$ & $-0.336,\ -0.391$\\
3 and 7 & 6.12 & $12.77,\ 12.82$ & $-0.340,\ -0.342$ & $-0.353,\ -0.380$\\
4 and 9 & 6.50 & $11.47,\ 11.58$ & $-0.306,\ -0.309$ & $-0.303,\ -0.357$\\
\bottomrule
\end{tabular*}
\par\smallskip\fontsize{9}{11}\selectfont\raggedright
\emph{Notes:} Paired entries follow the product order in the first column. Means weight the 234 weeks equally.
\end{minipage}
\end{table}

\begin{samepage}
Mean own-price elasticities are nearly equal within each pair under logit but differ more under RI--BLP. Under RI--BLP, product 6's mean elasticity is 16 percent larger in magnitude than product 2's. Product 1's mean elasticity is also 16 percent larger in magnitude than product 5's. Thus, similar prices within a brand need not imply similar own-price elasticities.
\par\end{samepage}

For cross-product responses, logit implies positive diversion in both directions for every pair. In the baseline RI--BLP derivative-weighted results, only products 2 and 6 show positive diversion in both directions among the four pairs (Figure~\ref{fig:tuna_substitution_heatmaps}). Pairs 1/5 and 3/7 have mixed signs, whereas products 4 and 9 have negative diversion in both directions. These pair-specific magnitudes and some signs vary with integration draws (Appendix~\ref{app:tuna_integration}) and the products and weeks included (Appendix~\ref{app:tuna_product_set}). Under the fitted RI--BLP model, a price increase can reduce demand even for a similarly priced product of the same brand and size. A retailer cannot assume that this other product will offset part of the affected product's demand loss, as logit implies.

\FloatBarrier
\section{Conclusion}
\label{sec:conclusion}
RI--BLP incorporates endogenous information acquisition into aggregate demand estimation while allowing for unobserved preference heterogeneity and endogenous prices. We establish conditions for uniqueness of share inversion and develop a simulated RI--BLP GMM estimator. Under the stated identification, regularity, and simulation-accuracy conditions, the estimator is consistent and asymptotically normal as the number of markets grows with the number of products fixed.

The Monte Carlo experiments show accurate recovery of the mean utility coefficients $\beta_0$, $\alpha$, and $\beta_1$, with confidence-interval coverage generally close to 95\% in larger samples. The dispersion parameters $\sigma_x$ and $\sigma_p$ capture differences across consumers in preferences for product characteristics and price sensitivity. They remain less precisely estimated than the mean utility coefficients, with some downward bias and overcoverage for $\sigma_x$. These findings motivate further work on finite-sample estimation and inference for preference heterogeneity. Extending the simulation-error analysis of \citet{freyberger2015} to RI--BLP could provide a basis for developing bias corrections and variance adjustments tailored to integration under subjective beliefs and over population heterogeneity.

The tuna application shows why similar average own-price elasticities need not imply similar substitution patterns. Under the fitted RI--BLP model, a price increase can reduce demand even for similar products and shift more demand toward the outside option (choosing none of the included products). These findings illustrate how endogenous information acquisition can change cross-price responses even without random coefficients.

The broad empirical patterns remain relatively stable for the numbers of draws examined, while individual diversion ratios and destination rankings are more sensitive. Further work on numerical integration should therefore assess and improve accuracy specifically for these product-pair responses.

\clearpage
\bibliographystyle{apalike}
\bibliography{referenceBLP}
\clearpage

\begin{appendices}
\numberwithin{equation}{section}
\numberwithin{table}{section}
\numberwithin{figure}{section}

\section{Information Structure}
\label{appendix:information_structure}

\begin{table}[!htbp]
\centering
\caption{Information structure in the RI--BLP model}
\label{tab:information_structure}
\begin{minipage}{0.84\textwidth}
\footnotesize
\setlength{\tabcolsep}{4pt}
\begin{tabular*}{\linewidth}{@{\extracolsep{\fill}}lcccc@{}}
\toprule
Variable & Firm & Consumer & Consumer & Econometrician \\
 & $\tau=0$ & $\tau=1$ & $\tau=2$ & $\tau=3$ \\
\midrule
$x_{0,jt}$ & Observed & Observed & Observed & Observed \\
$p_{jt}$ & Chosen & Observed & Observed & Observed \\
$(\beta_{0,i}^{\prime},\alpha_i)^{\prime}$ & Unobserved & Observed & Observed & Unobserved \\
$\varepsilon_{0,ijt}$ & Unobserved & Observed & Observed & Unobserved \\
$u_{0,ijt}$ & Partial & Observed & Observed & Partial \\
\midrule
$x_{1,jt}$ & Observed & Uncertain & May learn & Observed \\
$\xi_{1,jt}$ & Observed & Uncertain & May learn & Unobserved \\
$\varepsilon_{1,ijt}$ & Unobserved & Uncertain & May learn & Unobserved \\
$u_{1,ijt}$ & Partial & Uncertain & May learn & Partial \\
\midrule
$s_{jt}$ & Not realized & Not realized & Generated & Observed \\
\bottomrule
\end{tabular*}
\par\smallskip\fontsize{9}{11}\selectfont\raggedright
\emph{Notes:} The table describes the specification with $\xi_{0,jt}=\varnothing$. The stages are firm pricing at $\tau=0$, consumers' selection of an information-acquisition strategy and product-choice rule at $\tau=1$, information receipt and product choice at $\tau=2$, and the Econometrician's observation of market outcomes at $\tau=3$.
\end{minipage}
\end{table}

At $\tau=1$, $\pi_{u_1}$ describes consumers' uncertainty about $u_{1,it}$. ``May learn'' allows partial learning about this utility without separately observing each component. ``Partial'' means that only some determinants of the utility component are observed.

The Econometrician knows the joint additive-shock distribution $F_{\varepsilon}$ (Assumption~\ref{ass:riblp_fullsupport}) but does not observe individual shock realizations. The Econometrician observes aggregate market shares, not individual CCPs. Rows indexed by $j$ concern inside products $j=1,\ldots,J$; the outside-option utility components are set to zero.

\FloatBarrier
\section{Supplement to Section~\ref{sec:share_inversion_identification}}
\label{appendix:share_inversion_identification}

\subsection{Proof of Theorem~\ref{thm:riblp_delta1_injective}}
\label{appendix:riblp_delta1_injective}

Fix a market $t$, a subjective prior $\pi_{u_1}$, a candidate parameter vector $\theta$, and the baseline index $\delta_{0,t}(\beta_0,\alpha)$. Following \citet{kristensenzhu2026RI}, write the regular set in our utility notation as
\[
\mathcal V_{0,\mathrm{reg}}:=\bigl\{u_0\in\mathbb R^J:P^{0*}(u_0;\theta)\in\operatorname{int}\Delta(\mathcal J)\bigr\}.
\]
On this set, every integrated CCP, including that of the outside option, is positive. With baseline utilities ranging over $\mathbb R^J$, \citet[Theorem~2.1]{kristensenzhu2026RI} ensures that $\mathcal V_{0,\mathrm{reg}}$ is nonempty under our common-prior specification and Assumption~\ref{ass:affine_independence}. Continuity of $P^{0*}$ makes this set open.

For each coefficient vector $(\beta_{0,i},\alpha_i)$, the event $u_{0,it}=\delta_{0,t}(\beta_0,\alpha)+\mu_{0,it}+\varepsilon_{0,it}\in\mathcal V_{0,\mathrm{reg}}$ corresponds to a nonempty open set of baseline-shock realizations. The full-support condition in Assumption~\ref{ass:riblp_fullsupport} implies that the marginal distribution of $\varepsilon_{0,it}$ has full support on $\mathbb R^J$, so this event has positive probability for every coefficient vector. By Assumption~\ref{ass:riblp_independence}, it also has positive probability under $F_{\beta_0,\alpha}(\cdot;\theta)\otimes F_{\varepsilon}$.

On this event, Equation~\eqref{eq:logitccp} gives $P_j^*>0$ for every $j\in\mathcal J$, every $\varepsilon_{1,it}\in\mathbb R^J$, and every $\delta_{1,t}\in\mathbb R^J$. The realized CCPs are nonnegative elsewhere, so their population averages satisfy $\sigma_j(\cdot)>0$ for $j=1,\ldots,J$ and $1-\sum_{j=1}^J\sigma_j(\cdot)>0$. Thus, the inside-share vector lies in $\operatorname{int}\Delta(\mathcal J)$.

The prior is fixed as $\delta_{1,t}$ varies, so $P^{0*}(u_{0,it};\theta)$ does not change. Equation~\eqref{eq:logitccp} is therefore continuously differentiable in $\delta_{1,t}$, even when some integrated CCPs are zero. Each derivative is bounded in absolute value by $1/(4\lambda)$. Let $P^*:=(P_1^*,\ldots,P_J^*)^{\prime}$ denote the realized CCP vector evaluated at $u_{0,it}=\delta_{0,t}(\beta_0,\alpha)+\mu_{0,it}+\varepsilon_{0,it}$ and $u_{1,it}=\delta_{1,t}+\varepsilon_{1,it}$. Dominated convergence permits differentiation under the integral in Equation~\eqref{eq:market_share} and gives the continuous share Jacobian
\begingroup
\small
\begin{equation}
\frac{\partial\sigma_t\bigl(\delta_{0,t}(\beta_0,\alpha),\delta_{1,t};\theta\bigr)}{\partial\delta_{1,t}^{\prime}}
=\frac{1}{\lambda}\int\left[\operatorname{diag}(P^*)-P^*{P^*}^{\prime}\right]\,d\!\left(F_{\beta_0,\alpha}(\cdot;\theta)\otimes F_{\varepsilon}\right).
\label{eq:share_jacobian_positive}
\end{equation}
\endgroup
This $J\times J$ Jacobian is symmetric. To show that it is positive definite, fix any nonzero $a=(a_1,\ldots,a_J)^{\prime}\in\mathbb R^J$, set $a_0:=0$, and let $P_0^*:=1-\sum_{j=1}^JP_j^*$ denote the residual outside-option probability. Pointwise in $(\beta_{0,i},\alpha_i,\varepsilon_{it})$,
\begingroup
\small
\begin{equation}
a^{\prime}
\frac{\partial P^*}{\partial u_1^{\prime}}
a
=
\frac{1}{\lambda}
\left[
\sum_{j\in\mathcal J}P_j^*a_j^2
-
\left(
\sum_{j\in\mathcal J}P_j^*a_j
\right)^2
\right]
\geq0.
\label{eq:ccp_variance_form}
\end{equation}
\endgroup
The bracketed expression is the variance of $\{a_j:j\in\mathcal J\}$ under the realized choice probabilities. When $u_{0,it}\in\mathcal V_{0,\mathrm{reg}}$, all these probabilities are positive, so the variance can equal zero only if all $a_j$, including $a_0=0$, are equal, which contradicts $a\neq0$. Thus, the quadratic-form integrand is strictly positive on an event with positive probability and nonnegative otherwise. Its integral is therefore strictly positive, so Equation~\eqref{eq:share_jacobian_positive} implies that the share Jacobian is positive definite everywhere.

Finally, take distinct latent indices $\delta,\widetilde\delta\in\mathbb R^J$ and set $h:=\delta-\widetilde\delta$. Integrating the positive quadratic form in Equation~\eqref{eq:share_jacobian_positive} along the segment joining them gives
\[
h^{\prime}\bigl[\sigma_t(\delta_{0,t},\delta;\theta)-\sigma_t(\delta_{0,t},\widetilde\delta;\theta)\bigr]>0.
\]
The share vectors are therefore distinct, proving injectivity.\qed

\subsection{Moment Jacobian and the Local Rank Condition}
\label{appendix:joint_identification_rank}

For almost every market, suppose that $(\delta_{1,t},\theta)\mapsto\sigma_t(\delta_{0,t}(\beta_0,\alpha),\delta_{1,t};\theta)$ is jointly continuously differentiable near $(\delta_{1,t}^*,\theta^*)$. This is an additional requirement beyond Theorem~\ref{thm:riblp_delta1_injective}. Holding $\lambda$ fixed, define the share derivatives at the true values by
\begin{equation}
\begin{aligned}
D_{\delta_1,t}^{\sigma}
&:=
\left.
\frac{\partial\sigma_t(\delta_{0,t},\delta_{1,t};\theta)}{\partial\delta_{1,t}^{\prime}}
\right|_{(\delta_{0,t},\delta_{1,t},\theta)=(\delta_{0,t}^*,\delta_{1,t}^*,\theta^*)}
\in\mathbb R^{J\times J},
\\
D_{\theta,t}^{\sigma}
&:=
\left.
\frac{\partial\sigma_t\bigl(\delta_{0,t}(\beta_0,\alpha),\delta_{1,t}^*;\theta\bigr)}{\partial\theta^{\prime}}
\right|_{\theta=\theta^*}
\in\mathbb R^{J\times d_\theta}.
\end{aligned}
\label{eq:identification_share_derivatives}
\end{equation}
Here, $D_{\theta,t}^{\sigma}$ holds $\delta_{1,t}^*$ fixed but includes the parameter dependence of $\delta_{0,t}$, $\mu_{0,it}$, $F_{\beta_0,\alpha}$, and $\pi_{u_1}$. The shock distribution $F_{\varepsilon}$ remains fixed under Assumption~\ref{ass:riblp_fullsupport}.

Theorem~\ref{thm:riblp_delta1_injective} establishes that $D_{\delta_1,t}^{\sigma}$ is positive definite and hence nonsingular. Under the joint differentiability condition above, the implicit-function theorem gives a differentiable inverse near $\theta^*$ for almost every market, with
\begin{equation}
\left.
\frac{\partial\delta_{1,t}(\theta)}{\partial\theta^{\prime}}
\right|_{\theta=\theta^*}
=
-\bigl(D_{\delta_1,t}^{\sigma}\bigr)^{-1}D_{\theta,t}^{\sigma}
\in\mathbb R^{J\times d_\theta}.
\label{eq:identification_implicit_delta_derivative}
\end{equation}
If Assumption~\ref{ass:exact_gmm_asymptotics}(iv) also holds, its derivative bound permits differentiation under the expectation. Using $\xi_{1,t}(\theta)=\delta_{1,t}(\theta)-X_{1,t}\beta_1$, the population moment Jacobian $D^*$ can therefore be written as
\begingroup
\small
\begin{equation}
D^*=\frac1J\mathbb E\!\left[Z_{1,t}^{\prime}\left(-\bigl(D_{\delta_1,t}^{\sigma}\bigr)^{-1}D_{\theta,t}^{\sigma}-\frac{\partial(X_{1,t}\beta_1)}{\partial\theta^{\prime}}\right)\right]\in\mathbb R^{q\times d_\theta}.
\label{eq:joint_identification_jacobian}
\end{equation}
\endgroup
The factor $1/J$ matches Equation~\eqref{eq:market_moment} and does not affect rank. Multiplication by $-\bigl(D_{\delta_1,t}^{\sigma}\bigr)^{-1}$ converts parameter-induced share changes into changes in the recovered latent index. Full column rank requires the mean utility coefficients and heterogeneity parameters $\vartheta$ to have locally distinguishable effects on the population moments.

Instrument orthogonality implies $\mathcal M(\theta^*)=0$. Under the smoothness and moment bounds above, the population moment function $\mathcal M$ is continuously differentiable near $\theta^*$. Full column rank of $D^*$ gives $d_\theta$ moment components with a nonsingular Jacobian. The inverse-function theorem applied to these components implies that $\theta^*$ is the unique solution to $\mathcal M(\theta)=0$ in a neighborhood of $\theta^*$. This rank requirement remains Assumption~\ref{ass:riblp_local_identification}, not a consequence of share injectivity.

\section{Proofs for Section~\ref{sec:estimation}}
\label{app:estimation_proofs}

We maintain the model conditions, many-market setting, and shorthand $\sigma_t(\delta;\theta)$ from Section~\ref{sec:many_market_asymptotics}.

\subsection{Proof of Lemma~\ref{lem:inverse_moment_regularity}}
\label{appendix:inverse_moment_regularity}

\emph{Part~(i): Latent-index smoothness.} Fix $\theta$ and individual heterogeneity. The subjective prior and integrated CCPs do not depend on $\delta$, so the realized CCPs in Equation~\eqref{eq:logitccp} have the logit derivative
\[
\frac{\partial P_j^*}{\partial\delta_k}=\lambda^{-1}P_j^*(\mathbf 1\{j=k\}-P_k^*),\qquad j,k=1,\ldots,J.
\]
This holds even when an integrated CCP is zero. Higher derivatives are polynomials in probabilities, multiplied by powers of $\lambda^{-1}$, and hence uniformly bounded. Dominated convergence permits differentiation under the share integral to every order. Symmetry and positive definiteness follow from Theorem~\ref{thm:riblp_delta1_injective}, proving part~(i).

\emph{Part~(ii): Continuity.} Fix $\theta_0\in\Theta$, write $d=\delta_{1,t}(\theta_0)$, and choose any radius $r>0$. For each unit vector $v\in\mathbb R^J$, positive definiteness along the segment from $d$ to $d+rv$ gives
\[
v^{\prime}\bigl[\sigma_t(d+rv;\theta_0)-s_t\bigr]>0.
\]
The left-hand side is continuous in $v$, so its minimum over the unit sphere is positive. Joint continuity preserves this strict inequality uniformly on the sphere for $\theta$ sufficiently close to $\theta_0$. The solution $\delta_{1,t}(\theta)$ must then lie inside the ball of radius $r$ about $d$. Otherwise, along the ray from $d$ through $\delta_{1,t}(\theta)$, strict monotonicity implied by the positive-definite Jacobian would contradict $\sigma_t(\delta_{1,t}(\theta);\theta)=s_t$. Thus $\delta_{1,t}(\theta)\to d$ as $\theta\to\theta_0$. Equation~\eqref{eq:market_moment} then gives continuity of $m_t(\theta)$.

\emph{Part~(iii): Differentiability.} Fix $\theta_0\in\mathcal N$. At $(\delta_{1,t}(\theta_0),\theta_0)$, the share equations hold and the index Jacobian is nonsingular by part~(i). The joint share differentiability assumed in part~(iii) permits application of the implicit-function theorem, yielding a continuously differentiable solution on an open neighborhood of $\theta_0$ contained in $\operatorname{int}\Theta$. Theorem~\ref{thm:riblp_delta1_injective} makes this the unique solution there. Differentiating $\sigma_t(\delta_{1,t}(\theta);\theta)=s_t$ gives the derivative formula in part~(iii), and the expression for $m_t$ gives its continuous differentiability. Uniqueness makes these local solutions agree on overlaps, so their union gives continuously differentiable recovered indices and moments on an open neighborhood of $\mathcal N$.\qed

\subsection{Sufficient Conditions}
\label{appendix:exact_gmm_sufficient_conditions}

The following lemmas address selected requirements of Assumption~\ref{ass:exact_gmm_asymptotics} in order: share-equation solvability and joint share continuity in part~(ii), joint share differentiability in part~(iv), and moment envelopes in parts~(iii)--(iv). Each lemma states the additional conditions under which its conclusion holds.

\begin{lemma}[Share-equation solvability]
\label{lem:inverse_existence_sufficient}
Maintain the conditions of Theorem~\ref{thm:riblp_delta1_injective} at every $\theta\in\Theta$. Fix a market, a reference index $d_t\in\mathbb R^J$, and a radius $r_t>0$. Suppose that, for every $\theta\in\Theta$, the smallest eigenvalue of $\partial\sigma_t(\delta;\theta)/\partial\delta^{\prime}$ is at least $\underline c_t>0$ throughout the closed ball $\lVert\delta-d_t\rVert\leq r_t$, and
\[
\sup_{\theta\in\Theta}\lVert\sigma_t(d_t;\theta)-s_t\rVert<\underline c_t r_t.
\]
Then the share equations have a unique solution $\delta_{1,t}(\theta)$, which lies in the open ball $\lVert\delta-d_t\rVert<r_t$, for every $\theta\in\Theta$.
\end{lemma}

\begin{proof}
Fix $\theta$ and write $g(\delta)=\sigma_t(\delta;\theta)-s_t$. If $\lVert h\rVert=r_t$, integration of the index Jacobian along the segment from $d_t$ to $d_t+h$ gives
\[
h^{\prime}g(d_t+h)\geq\underline c_t r_t^2-r_t\lVert g(d_t)\rVert>0.
\]
Let $B$ be the closed ball and $\operatorname{proj}_B$ its Euclidean projection. Brouwer's theorem gives a fixed point of $\delta\mapsto\operatorname{proj}_B[\delta-g(\delta)]$. The projection inequality implies $g(\delta)^{\prime}(d_t-\delta)\geq0$, excluding a boundary fixed point by the strict outward inequality. An interior fixed point satisfies $g(\delta)=0$. Uniqueness follows from Theorem~\ref{thm:riblp_delta1_injective}.
\end{proof}

When shares are generated by the model at $\theta^*$, one may take $d_t=\delta_{1,t}^*$. The lemma supplies Assumption~\ref{ass:exact_gmm_asymptotics}(ii)'s existence requirement if its conditions hold almost surely across markets.

Joint regularity in structural parameters requires more than latent-index smoothness, as also reflected in Assumption~B.3 of \citet[Appendix~B]{freyberger2015}. The next two lemmas fix a market and use the uniqueness of the reduced-problem solution implied by Assumption~\ref{ass:affine_independence}. Represent coefficient heterogeneity by $\nu\sim F_\nu$, independent of $\varepsilon_{it}\sim F_\varepsilon$, with both laws independent of $\theta$. Write $u_{0,it}=a_t(\nu;\theta)+\varepsilon_{0,it}$, where $a_t(\nu;\theta)\in\mathbb R^J$ is baseline utility excluding additive shocks. This representation includes common coefficients.

\begin{lemma}[Joint share continuity]
\label{lem:share_continuity_primitives}
Suppose $a_t(\nu;\theta)$ is continuous in $\theta$ for $F_\nu$-almost every $\nu$. Represent the subjective prior by $u_1=h_t(\omega;\theta)$, where $\omega$ has a parameter-independent law $F_\omega$. Suppose $h_t(\omega;\theta)$ is continuous in $\theta$ almost surely and, on each compact parameter neighborhood $K$, $\sup_{\theta\in K}\lVert h_t(\omega;\theta)\rVert$ has finite expectation under $F_\omega$. Then $\sigma_t(\delta;\theta)$ is jointly continuous in $(\delta,\theta)$.
\end{lemma}

\begin{proof}
The logarithmic aggregator in Equation~\eqref{eq:Q0_reduced}, multiplied by $\lambda$, lies between the smallest and largest utilities. On compact sets of $u_0$ and $\theta$, its absolute value is bounded by a constant plus $\sup_{\theta\in K}\lVert h_t(\omega;\theta)\rVert$. Dominated convergence gives joint continuity of $Q^0(P^0;u_0;\theta)$ in $(P^0,u_0,\theta)$. Compactness of the simplex and uniqueness of the maximizer then imply continuity of $P^{0*}(u_0;\theta)$ in $(u_0,\theta)$, and Equation~\eqref{eq:logitccp} gives continuous realized CCPs. Their boundedness permits dominated convergence over $\nu$ and $\varepsilon$, proving joint share continuity.
\end{proof}

This lemma verifies the joint share-continuity requirement in Assumption~\ref{ass:exact_gmm_asymptotics}(ii). Gaussian priors with continuously parameterized means and covariances satisfy its prior conditions. For example, in Section~\ref{sec:monte_carlo}, the conditional law of $(x_{1,t},\xi_{1,t},\varepsilon_{1,it})$ is fixed while $u_{1,it}=\beta_1x_{1,t}+\xi_{1,t}+\varepsilon_{1,it}$ varies continuously with $\beta_1$. On compact parameter sets, $\lVert u_{1,it}\rVert$ has an integrable envelope under this Gaussian law.

\begin{lemma}[Share smoothness under a parameter-independent prior]
\label{lem:fixed_prior_smoothing}
Suppose the subjective prior does not depend on $\theta$. Let $F_{\varepsilon}$ in Assumption~\ref{ass:riblp_fullsupport} have a continuously differentiable density $f_\varepsilon(\varepsilon_0,\varepsilon_1)$ on $\mathbb R^{2J}$ satisfying
\[
\int_{\mathbb R^{2J}}\lVert\nabla_{\varepsilon_0}f_\varepsilon(\varepsilon_0,\varepsilon_1)\rVert\,d\varepsilon_0\,d\varepsilon_1<\infty.
\]
On an open parameter domain, suppose $a_t(\nu;\theta)$ is continuously differentiable almost surely. Suppose also that, on each compact subset $K$ of that domain, $\mathbb E_\nu[\sup_{\theta\in K}\lVert\partial a_t(\nu;\theta)/\partial\theta^{\prime}\rVert]<\infty$. Then $\sigma_t(\delta;\theta)$ is jointly continuously differentiable on $\mathbb R^J$ times that domain. In particular, the density condition holds for any nonsingular joint Gaussian additive-shock distribution.
\end{lemma}

\begin{proof}
Since the prior does not depend on $\theta$, write $P_j^*(u_1\mid u_0)$ for the realized CCP, suppressing the prior and $\lambda$. After the change of variables $u_0=a_t(\nu;\theta)+\varepsilon_0$, the share is
\[
\sigma_{j,t}(\delta;\theta)=\mathbb E_\nu\!\left[\int_{\mathbb R^{2J}}P_j^*(\delta+\varepsilon_1\mid u_0)f_\varepsilon(u_0-a_t(\nu;\theta),\varepsilon_1)\,du_0\,d\varepsilon_1\right].
\]
All dependence on $\theta$ is now in the translated density. Its integrable gradient permits differentiation of translations in $L^1$. Since $0\leq P_j^*\leq1$, differentiation under the integral and the expectation gives
\begingroup
\small
\[
\frac{\partial\sigma_{j,t}(\delta;\theta)}{\partial\theta^{\prime}}
=-\mathbb E_\nu\!\left[\int P_j^*(\delta+\varepsilon_1\mid u_0)\nabla_{\varepsilon_0}f_\varepsilon(u_0-a_t(\nu;\theta),\varepsilon_1)^{\prime}\frac{\partial a_t(\nu;\theta)}{\partial\theta^{\prime}}\,du_0\,d\varepsilon_1\right].
\]
\endgroup
The norm of this derivative is bounded by the density-gradient integral in the lemma times $\mathbb E_\nu\lVert\partial a_t/\partial\theta^{\prime}\rVert$. Continuity of translations of an integrable function in $L^1$, continuity of $a_t$ and its derivative, and the assumed envelope give joint continuity in $(\delta,\theta)$. Here continuity in $\delta$ also uses the bounded logit derivatives from Lemma~\ref{lem:inverse_moment_regularity}(i). The same argument applied to those derivatives proves joint continuity of the index Jacobian. For a nonsingular Gaussian density, its gradient is the density multiplied by a linear function of $(\varepsilon_0,\varepsilon_1)$, whose absolute first moment is finite.
\end{proof}

This lemma verifies the joint share-differentiability requirement in Assumption~\ref{ass:exact_gmm_asymptotics}(iv), allowing changing consideration sets and dependence between $\varepsilon_0$ and $\varepsilon_1$. The tuna application in Section~\ref{sec:tuna_application} satisfies the lemma's prior-independence and shock-density conditions. In this specification, $a_t(\nu;\theta)$ is independent of $\nu$ and linear in $\theta$, so its derivative-envelope bound is finite at each fixed market. The next lemma provides sufficient conditions for the moment bounds in parts~(iii) and~(iv).

\begin{lemma}[Moment envelopes]
\label{lem:moment_envelopes}
Suppose the conditions for continuity of $\delta_{1,t}(\theta)$ in Lemma~\ref{lem:inverse_moment_regularity}(ii) hold and $\Theta$ is compact. Then $\sup_{\theta\in\Theta}\lVert m_t(\theta)\rVert\leq M_{0,t}$, where
\[
M_{0,t}:=\frac{\lVert Z_{1,t}\rVert}{J}\left[\sup_{\theta\in\Theta}\lVert\delta_{1,t}(\theta)\rVert+\lVert X_{1,t}\rVert\sup_{\theta\in\Theta}\lVert\beta_1\rVert\right].
\]
If $\mathcal N\subset\operatorname{int}\Theta$ is a closed ball and $\sigma_t$ is jointly continuously differentiable near each $(\delta_{1,t}(\theta),\theta)$ with $\theta\in\mathcal N$, define the scalar curvature lower bound $c_t$ and parameter-derivative upper bound $H_t$ by
\[
\begin{aligned}
c_t&:=\inf_{\theta\in\mathcal N}\lambda_{\min}\!\left(\left.\frac{\partial\sigma_t(\delta;\theta)}{\partial\delta^{\prime}}\right|_{\delta=\delta_{1,t}(\theta)}\right),\\
H_t&:=\sup_{\theta\in\mathcal N}\left\lVert\left.\frac{\partial\sigma_t(\delta;\theta)}{\partial\theta^{\prime}}\right|_{\delta=\delta_{1,t}(\theta)}\right\rVert,
\end{aligned}
\]
where $\lambda_{\min}$ denotes the smallest eigenvalue. Then $c_t>0$, $H_t<\infty$, and
\[
\sup_{\theta\in\mathcal N}\left\lVert\frac{\partial m_t(\theta)}{\partial\theta^{\prime}}\right\rVert\leq M_{1,t}:=\frac{\lVert Z_{1,t}\rVert}{J}\left(\frac{H_t}{c_t}+\lVert X_{1,t}\rVert\right).
\]
Consequently, $\mathbb E[M_{0,t}]<\infty$ supplies Assumption~\ref{ass:exact_gmm_asymptotics}(iii). If shares are generated by the model at $\theta^*$, then $\mathbb E[M_{1,t}^2]<\infty$ and $\mathbb E\lVert Z_{1,t}^{\prime}\xi_{1,t}\rVert^2<\infty$ supply the two moment bounds in part~(iv).
\end{lemma}

\begin{proof}
Compactness and continuity make the two suprema defining $M_{0,t}$ finite. Continuity and positive definiteness give $H_t<\infty$ and $c_t>0$. The expression for $m_t$ gives the first bound. Lemma~\ref{lem:inverse_moment_regularity}(iii) gives continuously differentiable recovered indices on a neighborhood of $\mathcal N$. Differentiating $m_t$, using that lemma's derivative formula and $\lVert\partial(X_{1,t}\beta_1)/\partial\theta^{\prime}\rVert\leq\lVert X_{1,t}\rVert$, gives the second bound. Model-generated shares at $\theta^*$ give $m_t(\theta^*)=J^{-1}Z_{1,t}^{\prime}\xi_{1,t}$, establishing the moment claims. Market-specific finiteness of these bounds does not imply integrability across markets.
\end{proof}

\subsection{Proof of Theorem~\ref{thm:riblp_gmm_asymptotics}}
\label{appendix:exact_gmm_proof}

\emph{Consistency.} Define the sample and population criteria by
\[
\mathcal L_T(\theta):=\overline m_T(\theta)^{\prime}\widehat W_T\overline m_T(\theta),\qquad \mathcal L(\theta):=\mathcal M(\theta)^{\prime}W\mathcal M(\theta).
\]
Lemma~\ref{lem:inverse_moment_regularity}(ii) gives continuous moments. Under the iid sampling, compactness, and moment-envelope conditions in Assumption~\ref{ass:exact_gmm_asymptotics}(i)--(iii), Lemma~2.4 of \citet{newey1994large} gives $\sup_{\theta\in\Theta}\lVert\overline m_T(\theta)-\mathcal M(\theta)\rVert=o_p(1)$ and continuity of $\mathcal M$. Weighting-matrix convergence implies uniform convergence of $\mathcal L_T$ to $\mathcal L$. Positive definiteness of $W$ and global identification make $\theta^*$ the unique minimizer of $\mathcal L$. Compactness and Theorem~2.1 of \citet{newey1994large} yield consistency.

\emph{Asymptotic normality.} For subsequent local arguments in this appendix, let $\mathcal N$ denote a smaller concentric closed ball of positive radius contained in the interior of the ball supplied by Assumption~\ref{ass:exact_gmm_asymptotics}(iv). Lemma~\ref{lem:inverse_moment_regularity}(iii) gives continuously differentiable moments near $\mathcal N$. The derivative-envelope bound in Assumption~\ref{ass:exact_gmm_asymptotics}(iv) permits differentiation under the expectation. Define $D_T(\theta):=\partial\overline m_T(\theta)/\partial\theta^{\prime}$. Lemma~2.4 of \citet{newey1994large}, applied to the derivatives, gives
\[
\sup_{\theta\in\mathcal N}\left\lVert D_T(\theta)-\frac{\partial\mathcal M(\theta)}{\partial\theta^{\prime}}\right\rVert=o_p(1),
\]
with a continuous population derivative. The moments $m_t(\theta^*)$ have mean zero and finite variance, so the central limit theorem gives $\sqrt T\,\overline m_T(\theta^*)\overset{d}{\to}\mathcal N(0,\Omega^*)$. Together with consistency, interiority, weighting-matrix convergence, and full column rank of $D^*$, these properties verify the hypotheses of Theorem~3.4 of \citet{newey1994large}, yielding Equation~\eqref{eq:gmm_asymptotic_variance}.\qed

\subsection{Sample-Standardized Instruments}
\label{appendix:sample_standardized_instruments}

The iid argument above uses the original instruments. We now allow sample centering and scaling, as implemented in Section~\ref{sec:monte_carlo}. With a constant instrument included, these transformations give moments $C_Tm_t(\theta)$, where $C_T\in\mathbb R^{q\times q}$ is data-dependent but constant in $\theta$. Using these transformed moments with weight $\widehat W_T$ is equivalent to using the original moments with weight $C_T^{\prime}\widehat W_TC_T$. The sandwich formulas agree for these two equivalent criteria. If $C_T\overset{p}{\to}C$ for a deterministic nonsingular $C$, the effective weight converges to $C^{\prime}WC$. The preceding consistency and asymptotic-normality results then apply with this limiting weight and the remaining conditions unchanged. For sample means and standard deviations, finite instrument second moments and positive variances of scaled components suffice. Multiplication by $C_T=O_p(1)$ also preserves the simulation bounds.

\subsection{Proof of Theorem~\ref{thm:simulated_gmm_mpec_asymptotics}}
\label{appendix:simulated_gmm_proof}

\emph{Consistency.} On the event in Assumption~\ref{ass:simulation_asymptotics}(i), the share-constrained problem is equivalent to minimizing $\mathcal L_T^{\mathrm{sim}}(\theta):=\overline m_T^{\mathrm{sim}}(\theta)^{\prime}\widehat W_T\overline m_T^{\mathrm{sim}}(\theta)$ over $\Theta$. Continuity and compactness give a minimizer. On the event's vanishing-probability complement, assign the estimator any fixed value in $\Theta$. Assumption~\ref{ass:simulation_asymptotics}(ii) and the exact-GMM proof in Appendix~\ref{appendix:exact_gmm_proof} give $\sup_{\theta\in\Theta}|\mathcal L_T^{\mathrm{sim}}(\theta)-\mathcal L(\theta)|=o_p(1)$. Global identification and Theorem~2.1 of \citet{newey1994large} yield consistency.

\emph{Asymptotic normality.} Parts~(ii)--(iii) of Assumption~\ref{ass:simulation_asymptotics} give
\[
\sqrt T\sup_{\theta\in\Theta}\left\lVert\overline m_T^{\mathrm{sim}}(\theta)-\overline m_T(\theta)\right\rVert=O_p\!\left(\sqrt T\,a_{R_T,S_T}\right)=o_p(1).
\]
The exact-moment central limit theorem thus gives $\sqrt T\,\overline m_T^{\mathrm{sim}}(\theta^*)\overset{d}{\to}\mathcal N(0,\Omega^*)$.

To verify the local increment condition of Theorem~7.2 of \citet{newey1994large}, use convexity of $\mathcal N$ and $\mathcal M(\theta^*)=0$. Integration of the exact-moment derivatives along the segment from $\theta^*$ to $\theta$, followed by the triangle inequality for the simulation errors at its endpoints, gives
\begingroup
\small
\[
\begin{aligned}
&\sup_{\theta\in\mathcal N}\frac{\left\lVert\overline m_T^{\mathrm{sim}}(\theta)-\overline m_T^{\mathrm{sim}}(\theta^*)-\mathcal M(\theta)\right\rVert}{T^{-1/2}+\lVert\theta-\theta^*\rVert}\\
&\quad\leq\sup_{\theta\in\mathcal N}\left\lVert D_T(\theta)-\frac{\partial\mathcal M(\theta)}{\partial\theta^{\prime}}\right\rVert+2\sqrt T\sup_{\theta\in\Theta}\left\lVert\overline m_T^{\mathrm{sim}}(\theta)-\overline m_T(\theta)\right\rVert=o_p(1).
\end{aligned}
\]
\endgroup
The terms vanish by the exact-Jacobian uniform law of large numbers and the simulation-error bound, verifying the local increment condition on every shrinking neighborhood of $\theta^*$. Consistency, exact minimization, the moment central limit theorem, and the stated identification and weighting conditions then give the result by Theorem~7.2 of \citet{newey1994large}.\qed

\subsection{Approximate Minimization}
\label{appendix:approximate_minimization}

The exact- and simulated-GMM theorems use global minimizers. For a numerical estimate $\widetilde\theta_T^{\mathrm{sim}}\in\Theta$, we allow optimization error while using the criterion $\mathcal L_T^{\mathrm{sim}}(\theta)$ constructed from the recovered indices $\delta_{1,t}^{\mathrm{sim}}(\theta)$.

Under the conditions for consistency in Theorem~\ref{thm:simulated_gmm_mpec_asymptotics}, uniform convergence gives consistency if the numerical estimate's criterion value exceeds the global minimum by $o_p(1)$. Under that theorem's conditions for asymptotic normality, Theorem~7.2 of \citet{newey1994large} gives the same limiting distribution if
\[
\mathcal L_T^{\mathrm{sim}}(\widetilde\theta_T^{\mathrm{sim}})-\inf_{\theta\in\Theta}\mathcal L_T^{\mathrm{sim}}(\theta)=o_p(T^{-1}).
\]
The same conclusions hold for exact GMM with $\mathcal L_T$ in place of $\mathcal L_T^{\mathrm{sim}}$.

\subsection{Simulation Error Bounds}
\label{appendix:share_approximation_bounds}

We translate share-approximation bounds into sufficient conditions for Assumption~\ref{ass:simulation_asymptotics}(ii) and~(iv). Write $\sigma_t^{\mathrm{sim}}(\delta;\theta)$ analogously to $\sigma_t(\delta;\theta)$, suppressing the numbers of draws. The three lemmas below bound index and moment errors, give a sufficient uniform-curvature condition, and establish moment approximation under an assumed share-error rate.

The simulated-inversion requirements in Assumption~\ref{ass:simulation_asymptotics}(i) and the rate condition $\sqrt T\,a_{R_T,S_T}\to0$ in part~(iii) are maintained separately.

\begin{lemma}[Share-approximation bound]
\label{lem:share_approximation_bound}
Suppose the conditions of Theorem~\ref{thm:riblp_delta1_injective} hold and the exact and simulated share equations have solutions $\delta_{1,t}(\theta)$ and $\delta_{1,t}^{\mathrm{sim}}(\theta)$ for every $\theta\in\Theta$ and sample market. Along the segment joining these two indices, suppose the smallest eigenvalue of $\partial\sigma_t(\delta;\theta)/\partial\delta^{\prime}$ is at least $c_\sigma>0$, uniformly in $t$, $\theta$, and $T$. Define
\[
e_T:=\sup_{\theta\in\Theta}\max_{1\leq t\leq T}\left\lVert\sigma_t^{\mathrm{sim}}\bigl(\delta_{1,t}^{\mathrm{sim}}(\theta);\theta\bigr)-\sigma_t\bigl(\delta_{1,t}^{\mathrm{sim}}(\theta);\theta\bigr)\right\rVert.
\]
Then, for every $t$ and $\theta$,
\begin{align}
\left\lVert\delta_{1,t}^{\mathrm{sim}}(\theta)-\delta_{1,t}(\theta)\right\rVert&\leq\frac{e_T}{c_\sigma},\label{eq:simulation_inverse_bound}\\
\left\lVert m_t^{\mathrm{sim}}(\theta)-m_t(\theta)\right\rVert&\leq\frac{\lVert Z_{1,t}\rVert e_T}{Jc_\sigma}.\label{eq:simulation_market_moment_bound}
\end{align}
Consequently,
\[
\sup_{\theta\in\Theta}\left\lVert\overline m_T^{\mathrm{sim}}(\theta)-\overline m_T(\theta)\right\rVert\leq\frac{e_T}{Jc_\sigma}\frac1T\sum_{t=1}^T\lVert Z_{1,t}\rVert.
\]
\end{lemma}

\begin{proof}
Fix $t$ and $\theta$ and set $h:=\delta_{1,t}^{\mathrm{sim}}(\theta)-\delta_{1,t}(\theta)$. Integration of the symmetric share Jacobian along the segment gives
\[
h^{\prime}\left[\sigma_t\bigl(\delta_{1,t}^{\mathrm{sim}}(\theta);\theta\bigr)-\sigma_t\bigl(\delta_{1,t}(\theta);\theta\bigr)\right]\geq c_\sigma\lVert h\rVert^2.
\]
The exact and simulated share equations imply that the bracket has norm at most $e_T$. The Cauchy--Schwarz inequality gives $c_\sigma\lVert h\rVert^2\leq\lVert h\rVert e_T$, proving Equation~\eqref{eq:simulation_inverse_bound}. Equation~\eqref{eq:market_moment} gives $m_t^{\mathrm{sim}}(\theta)-m_t(\theta)=J^{-1}Z_{1,t}^{\prime}h$. Taking norms and averaging establishes the remaining bounds.
\end{proof}

The preceding bound requires uniform curvature along the index segments, beyond the pointwise positive definiteness in Theorem~\ref{thm:riblp_delta1_injective}. A sufficient condition is:

\begin{lemma}[Uniform share curvature]
\label{lem:uniform_share_curvature}
Suppose there are constants $\rho>0$ and $\eta>0$ such that, at every market, parameter, and index on the segments in Lemma~\ref{lem:share_approximation_bound}, the data-generating probability that all exact realized CCPs, including the outside probability, are at least $\eta$ is at least $\rho$. Then the eigenvalue bound in that lemma holds with $c_\sigma=\rho\eta^2/\lambda$.
\end{lemma}

\begin{proof}
For any $b\in\mathbb R^J$, the Cauchy--Schwarz inequality gives
\[
\sum_{j=1}^J P_j^*b_j^2-\left(\sum_{j=1}^J P_j^*b_j\right)^2
\geq P_0^*\sum_{j=1}^J P_j^*b_j^2.
\]
On the stated event this is at least $\eta^2\lVert b\rVert^2$. Elsewhere it is nonnegative. Taking data-generating expectations and dividing by $\lambda$, Equation~\eqref{eq:share_jacobian_positive} gives the eigenvalue bound $c_\sigma=\rho\eta^2/\lambda$.
\end{proof}

\begin{lemma}[Simulation-moment approximation]
\label{lem:simulation_moment_conditions}
Suppose the conditions of Lemma~\ref{lem:share_approximation_bound} hold with probability approaching one, $e_T=O_p(a_{R_T,S_T})$ for a deterministic sequence $a_{R_T,S_T}\to0$, and $T^{-1}\sum_{t=1}^T\lVert Z_{1,t}\rVert^2=O_p(1)$. Then Assumption~\ref{ass:simulation_asymptotics}(ii) holds. If also $T^{-1}\sum_t\sup_{\theta\in\mathcal N}\lVert m_t(\theta)\rVert^2=O_p(1)$, then part~(iv) holds. The latter moment bound follows from Lemma~\ref{lem:inverse_moment_regularity} and Assumption~\ref{ass:exact_gmm_asymptotics}(i), (ii), and~(iv).
\end{lemma}

\begin{proof}
The average of $\lVert Z_{1,t}\rVert$ is $O_p(1)$ by Cauchy--Schwarz. The final bound in Lemma~\ref{lem:share_approximation_bound} therefore proves part~(ii). For second moments, the market-level bound gives
\[
\frac1T\sum_{t=1}^T\sup_{\theta\in\mathcal N}\lVert m_t^{\mathrm{sim}}(\theta)-m_t(\theta)\rVert^2=O_p(a_{R_T,S_T}^2).
\]
Expanding $m_t^{\mathrm{sim}}(m_t^{\mathrm{sim}})^{\prime}-m_tm_t^{\prime}$ gives two cross terms and the outer product of the moment error. The assumed local moment bound and Cauchy--Schwarz bound the norm of the average difference uniformly by $O_p(a_{R_T,S_T})+O_p(a_{R_T,S_T}^2)=o_p(1)$, proving part~(iv). Finally, if $r_{\mathcal N}$ is the radius of $\mathcal N$, continuously differentiable exact moments satisfy
\[
\sup_{\theta\in\mathcal N}\lVert m_t(\theta)\rVert
\leq\lVert m_t(\theta^*)\rVert+r_{\mathcal N}\sup_{\theta\in\mathcal N}\left\lVert\frac{\partial m_t(\theta)}{\partial\theta^{\prime}}\right\rVert.
\]
The second-moment bounds in Assumption~\ref{ass:exact_gmm_asymptotics}(iv) make the square of this envelope integrable. The law of large numbers across markets gives the claimed $O_p(1)$ bound.
\end{proof}

\subsection{Variance Estimation}
\label{appendix:gmm_variance_consistency}

Suppose the conditions for asymptotic normality in Theorem~\ref{thm:simulated_gmm_mpec_asymptotics} and Assumption~\ref{ass:simulation_asymptotics}(iv) hold. The exact-moment envelope derived at the end of the proof of Lemma~\ref{lem:simulation_moment_conditions} has a finite second moment. Continuity from Lemma~\ref{lem:inverse_moment_regularity} and this bound give a uniform law of large numbers for $m_t(\theta)m_t(\theta)^{\prime}$. The assumed second-moment approximation and consistency of $\widehat\theta_T^{\mathrm{sim}}$ therefore imply convergence of the uncentered simulated second moment to $\Omega^*$. Moreover, $\overline m_T^{\mathrm{sim}}(\widehat\theta_T^{\mathrm{sim}})\overset{p}{\to}\mathcal M(\theta^*)=0$, so centering does not change this limit and $\widehat\Omega_T\overset{p}{\to}\Omega^*$. For any $\widehat D_T\overset{p}{\to}D^*$, the continuous-mapping theorem then gives $\widehat V_T\overset{p}{\to}V^*$.

\paragraph{Finite-difference Jacobian.}
A consistent estimator of the population moment Jacobian $D^*$ can be constructed by central finite differences, without analytically differentiating the simulated moments. Let $h_T>0$ be a deterministic step size satisfying $h_T\to0$ and $\sqrt T\,h_T\to\infty$. Let $e_k\in\mathbb R^{d_\theta}$ have a one in position $k$ and zeros elsewhere, so that $\widehat D_Te_k$ is the $k$th column of $\widehat D_T$. Using the same random numbers in both evaluations, with coefficients and prior utilities recalculated at each parameter value, define these columns by
\begingroup
\small
\[
\widehat D_Te_k:=\frac{\overline m_T^{\mathrm{sim}}(\widehat\theta_T^{\mathrm{sim}}+h_Te_k)-\overline m_T^{\mathrm{sim}}(\widehat\theta_T^{\mathrm{sim}}-h_Te_k)}{2h_T},\qquad k=1,\ldots,d_\theta.
\]
\endgroup
Consistency and $h_T\to0$ ensure that both evaluation points lie in $\mathcal N$ with probability approaching one. The simulation contribution to each difference quotient is $O_p(a_{R_T,S_T}/h_T)=o_p(1)$, since $a_{R_T,S_T}/h_T=(\sqrt T\,a_{R_T,S_T})/(\sqrt T\,h_T)\to0$. The exact difference quotient averages the $k$th column of $D_T$ along the segment between the two evaluation points. It converges to the corresponding column of $D^*$ by the uniform law of large numbers for $D_T$ and continuity of $\partial\mathcal M(\theta)/\partial\theta^{\prime}$. Thus, $\widehat D_T\overset{p}{\to}D^*$, as in Theorem~7.4 of \citet{newey1994large}.

\paragraph{Analytical Jacobian by implicit differentiation.}
The implementations in Sections~\ref{sec:monte_carlo} and~\ref{sec:tuna_application} use analytical moment Jacobians obtained by implicitly differentiating the simulated share equations. For this approach, we supplement Assumption~\ref{ass:simulation_asymptotics} with smoothness and derivative-approximation conditions. Suppose that, with probability approaching one, for every sample market and $\theta\in\mathcal N$, the simulated share mapping $\sigma_t^{\mathrm{sim}}(\delta;\theta)$ is jointly continuously differentiable near $(\delta_{1,t}^{\mathrm{sim}}(\theta),\theta)$ and has a nonsingular Jacobian with respect to $\delta$ there. Applying the implicit-function argument of Equation~\eqref{eq:identification_implicit_delta_derivative} to the simulated shares gives the derivative of $\delta_{1,t}^{\mathrm{sim}}(\theta)$ with respect to $\theta$. The resulting analytical moment Jacobian, evaluated at $\widehat\theta_T^{\mathrm{sim}}$, is
\begingroup
\small
\begin{equation}
\begin{aligned}
\widehat D_T^{\mathrm{imp}}&:=\left.\frac{\partial\overline m_T^{\mathrm{sim}}(\theta)}{\partial\theta^{\prime}}\right|_{\theta=\widehat\theta_T^{\mathrm{sim}}}\\
&=\left.\frac1{TJ}\sum_{t=1}^T Z_{1,t}^{\prime}\left[\frac{\partial\delta_{1,t}^{\mathrm{sim}}(\theta)}{\partial\theta^{\prime}}-\frac{\partial(X_{1,t}\beta_1)}{\partial\theta^{\prime}}\right]\right|_{\theta=\widehat\theta_T^{\mathrm{sim}}}.
\end{aligned}
\label{eq:simulated_moment_jacobian}
\end{equation}
\endgroup
In addition to this local smoothness, suppose the simulated moment derivatives approximate the exact derivatives uniformly:
\begin{equation}
\sup_{\theta\in\mathcal N}\left\lVert\frac{\partial\overline m_T^{\mathrm{sim}}(\theta)}{\partial\theta^{\prime}}-D_T(\theta)\right\rVert=o_p(1).
\label{eq:implicit_jacobian_approximation}
\end{equation}
Consistency of $\widehat\theta_T^{\mathrm{sim}}$ places it in $\mathcal N$ with probability approaching one. Equation~\eqref{eq:implicit_jacobian_approximation} therefore gives $\widehat D_T^{\mathrm{imp}}-D_T(\widehat\theta_T^{\mathrm{sim}})=o_p(1)$. The exact-Jacobian uniform law of large numbers and continuity of the population derivative give $D_T(\widehat\theta_T^{\mathrm{sim}})\overset{p}{\to}D^*$. Hence $\widehat D_T^{\mathrm{imp}}\overset{p}{\to}D^*$, and using it in Equation~\eqref{eq:gmm_variance_estimator} preserves variance consistency. For the numerical approximation in Appendix~\ref{appendix:derivative_evaluation}, variance consistency additionally requires the computed moment Jacobian to differ from $\widehat D_T^{\mathrm{imp}}$ by $o_p(1)$.

\paragraph{Serially dependent markets.}
The GMM arguments above extend to serially dependent markets if the required uniform laws of large numbers and central limit theorem hold, with the remaining regularity and simulation-approximation conditions maintained \citep{hansen1982large}. The variance formulas then use the long-run covariance of the market-level moments in place of $\Omega^*$. Consistent inference requires a consistent moment Jacobian and HAC covariance estimate. For simulated GMM, the covariance approximation must cover the lagged products used in HAC as well as the unlagged products in Assumption~\ref{ass:simulation_asymptotics}(iv).

\section{Supplement to Section~\ref{sec:computation}}
\label{appendix:computational_details}

We use the simulated CCPs and utility draws from Section~\ref{sec:estimation}. Sections~\ref{sec:monte_carlo} and~\ref{sec:tuna_application} specify the draw construction and numbers of draws for the Monte Carlo experiments and the tuna application, respectively.

\subsection{KKT-Guided Newton--BA}
\label{appendix:newton_ba}

Fix $t$, $r$, and $\theta$. The model uses the $J$-dimensional inside-probability vector $P^0\in\Delta(\mathcal J)$, whereas the implementation stores all $J+1$ probabilities, including the residual outside probability. Indices $j\in\mathcal J$ include the outside option.

To evaluate the inner objective and KKT ratios stably, we subtract the largest scaled payoff for each prior draw before exponentiating:
\begingroup
\small
\[
\widetilde r_{jrt}^{(s)}:=\exp\!\left[\frac{u_{0,jt}^{(r)}+u_{1,jt}^{\pi,(s,r)}}{\lambda}-\max_{\ell\in\mathcal J}\frac{u_{0,\ell t}^{(r)}+u_{1,\ell t}^{\pi,(s,r)}}{\lambda}\right],\qquad j\in\mathcal J.
\]
\endgroup
The maximum includes the outside option. To evaluate the objective in Equation~\eqref{eq:simulated_ri_problem}, we add the subtracted maximum to $\ln(\sum_{\ell\in\mathcal J}P_\ell^0\widetilde r_{\ell rt}^{(s)})$ for each draw, then multiply by $\lambda$ and average over prior draws. The shift is independent of $P^0$, and the common exponential factor cancels from the KKT ratios.

Using these stabilized payoffs, the sample-average approximation to $\kappa_j$ in Equation~\eqref{eq:inner_kkt_ratio} is
\begin{equation}
\widehat\kappa_{jrt}(P^0):=\frac{1}{S_T}\sum_{s=1}^{S_T}\frac{\widetilde r_{jrt}^{(s)}}{\sum_{\ell\in\mathcal J}P_\ell^0\widetilde r_{\ell rt}^{(s)}},\qquad j\in\mathcal J.
\label{eq:simulated_kkt_ratio}
\end{equation}
By concavity of the simulated objective, an optimum $\widehat P_{rt}^{0*}$ satisfies the necessary and sufficient KKT conditions
\begingroup
\small
\begin{equation}
\widehat\kappa_{jrt}(\widehat P_{rt}^{0*})\leq1,\qquad \widehat P_{jrt}^{0*}\bigl[\widehat\kappa_{jrt}(\widehat P_{rt}^{0*})-1\bigr]=0,\qquad j\in\mathcal J.
\label{eq:simulated_inner_kkt}
\end{equation}
\endgroup

\paragraph{Step 1: Newton iteration.}
We start with $P_j^0=1/(J+1)$ for all $j\in\mathcal J$ and solve the $J$ interior first-order conditions $\widehat\kappa_{jrt}(P^0)=\widehat\kappa_{0rt}(P^0)$ for $j=1,\ldots,J$. Newton steps are taken only while all probabilities exceed $\tau_{\mathrm{switch}}$. We repeatedly halve a proposed step until all probabilities remain positive and the objective increases, allowing for numerical rounding. We switch to BA if the Newton system is singular or ill-conditioned, the direction is nonfinite or does not increase the objective locally, no acceptable step is found, or the iteration limit is reached. BA starts from the last feasible Newton iterate.

\paragraph{Step 2: Switching and BA initialization.}
A Newton candidate passing the final checks in Step~4 is accepted without BA. Otherwise, an accepted Newton step from $P^{0,\mathrm{old}}$ to $P^{0,\mathrm{new}}$ that lowers any probability to $\tau_{\mathrm{switch}}$ or below triggers the switch. The KKT ratios at $P^{0,\mathrm{new}}$ then guide the choice of BA's starting vector. Each ratio is the multiplier in the BA update in Equation~\eqref{eq:ba_update_riblp}, so a ratio above one increases the corresponding positive probability in the next update.

Let $\mathcal J_{\mathrm{switch}}$ be the set of alternatives satisfying both $P_j^{0,\mathrm{new}}\leq\tau_{\mathrm{switch}}$ and $\widehat\kappa_{jrt}(P^{0,\mathrm{new}})>1+\tau_{\mathrm{KKT}}$, where $\tau_{\mathrm{KKT}}$ is the inner KKT tolerance. If this set is empty, BA starts from $P^{0,\mathrm{new}}$. Otherwise, we shorten the final Newton step if necessary, using the fraction $\rho_{\mathrm{switch}}$:
\begingroup
\small
\[
\rho_{\mathrm{switch}}:=\min_{j\in\mathcal J_{\mathrm{switch}}}\frac{P_j^{0,\mathrm{old}}-\tau_{\mathrm{switch}}}{P_j^{0,\mathrm{old}}-P_j^{0,\mathrm{new}}}.
\]
\endgroup
BA then starts from $P^{0,\mathrm{old}}+\rho_{\mathrm{switch}}(P^{0,\mathrm{new}}-P^{0,\mathrm{old}})$. Since the old probabilities exceed $\tau_{\mathrm{switch}}$, we have $0<\rho_{\mathrm{switch}}\leq1$. This convex combination remains in the simplex and leaves every coordinate in $\mathcal J_{\mathrm{switch}}$ at or above the threshold, with at least one at equality.

\paragraph{Step 3: BA iteration.}
Starting from the probability vector selected in Step~1 or Step~2, BA updates all alternatives by
\begin{equation}
P_j^{0,(k+1)}=P_j^{0,(k)}\widehat\kappa_{jrt}\bigl(P^{0,(k)}\bigr),\qquad j\in\mathcal J.
\label{eq:ba_update_riblp}
\end{equation}
Here $k$ indexes iterations. Once BA begins, the algorithm does not return to Newton--Raphson. The identity $\sum_{j\in\mathcal J}P_j^0\widehat\kappa_{jrt}(P^0)=1$ ensures that the updated probabilities remain nonnegative and sum to one, apart from numerical rounding errors. We neither floor probabilities nor normalize during ordinary BA updates. BA continues until the KKT tolerance is satisfied or its iteration limit is reached.

After convergence, we may set the probabilities of alternatives whose KKT ratios are sufficiently below one to zero, normalize the remaining probabilities, and rerun BA. Any refined candidate must pass the checks in Step~4 over all alternatives. When $\theta$ changes, the new inner solve starts from the uniform vector in Step~1, so alternatives assigned zero probability in a previous solve are reconsidered.

\paragraph{Step 4: Final acceptance.}
Before acceptance, we normalize a copy of the final iterate to correct small rounding errors in the sum of probabilities and recompute all KKT ratios. Acceptance requires a finite candidate satisfying $P_j^0\geq0$ for all $j\in\mathcal J$ and $|\sum_{j\in\mathcal J}P_j^0-1|\leq\tau_{\mathrm{simplex}}$, where $\tau_{\mathrm{simplex}}$ is the simplex-feasibility tolerance. The candidate must also satisfy both KKT bounds
\begingroup
\small
\[
\max_{j\in\mathcal J}\widehat\kappa_{jrt}(P^0)\leq1+\tau_{\mathrm{KKT}},\qquad \max_{j\in\mathcal J}\left|P_j^0\bigl[\widehat\kappa_{jrt}(P^0)-1\bigr]\right|\leq\tau_{\mathrm{KKT}}.
\]
\endgroup

After acceptance, we calculate the realized CCPs $\widehat P_{jrt}^{*}$ using Equation~\eqref{eq:logitccp} with the simulated integrated CCPs $\widehat P_{jrt}^{0*}$. For numerical stability, we form $\ln\widehat P_{jrt}^{0*}+(u_{0,jt}^{(r)}+u_{1,jt}^{(r)})/\lambda$ for each alternative with positive integrated probability, including the outside option when applicable. We subtract the largest of these values, exponentiate, and divide by the sum. The resulting common exponential factor cancels between the numerator and denominator, leaving the CCPs unchanged. Alternatives with zero integrated probability receive zero realized probability.

\subsection{Analytical Derivative Evaluation}
\label{appendix:derivative_evaluation}

The formulas below differentiate the finite-draw share mapping analytically, holding the random numbers used to generate the draws and $\lambda$ fixed. Derivatives with respect to $\theta$ hold $\delta_1$ fixed but include the induced change in $\delta_{0,t}(\beta_0,\alpha)$. The outer GMM search also accounts for the response of the recovered indices to $\theta$.

\paragraph{Integrated CCPs.}
Fix a market $t$ and population draw $r$, and suppress these indices in the inner problem. Write $\mathcal J^+:=\{j\in\mathcal J:\widehat P_j^{0*}>0\}$ for the consideration set defined in Section~\ref{sec:model}. Choose a reference alternative $a\in\mathcal J^+$ and write $P_a^0=1-\sum_{\ell\in\mathcal J^+\setminus\{a\}}P_\ell^0$, with probabilities outside $\mathcal J^+$ fixed at zero. The remaining coordinates satisfy $\widehat\kappa_j-\widehat\kappa_a=0$ for $j\in\mathcal J^+\setminus\{a\}$.

Suppose the baseline utilities and subjective-prior utilities are continuously differentiable in $\theta$, the consideration set $\mathcal J^+$ is unchanged in a neighborhood of the candidate $\theta$, and the Jacobian of these reduced KKT equations is nonsingular. The positive integrated CCPs may vary within this neighborhood. For each parameter $\theta_k$, $k=1,\ldots,d_\theta$, differentiating the reduced KKT equations gives
\begingroup
\small
\begin{equation}
\sum_{\ell\in\mathcal J^+\setminus\{a\}}\frac{\partial(\widehat\kappa_j-\widehat\kappa_a)}{\partial P_\ell^0}\frac{\partial\widehat P_\ell^{0*}}{\partial\theta_k}=-\frac{\partial(\widehat\kappa_j-\widehat\kappa_a)}{\partial\theta_k},\qquad j\in\mathcal J^+\setminus\{a\}.
\label{eq:computation_inner_derivative}
\end{equation}
\endgroup
All KKT-ratio derivatives are evaluated at the inner solution after substituting for $P_a^0$. For the partial derivative on the right-hand side, probabilities are held fixed while baseline utilities and any parameter-dependent prior utilities vary. The reference derivative is $\partial\widehat P_a^{0*}/\partial\theta_k=-\sum_{\ell\in\mathcal J^+\setminus\{a\}}\partial\widehat P_\ell^{0*}/\partial\theta_k$. Integrated-CCP derivatives outside $\mathcal J^+$ are zero, as are all integrated-CCP derivatives when this set contains only one alternative.

In numerical evaluations, we solve the reduced KKT system on the subset of $\mathcal J^+$ whose KKT ratios are at least $1-10^{-4}$, choosing the reference alternative $a$ within this subset. Excluded probabilities are held at their computed values, with integrated-CCP derivatives set to zero. The reference probability is determined residually from all other probabilities, including those excluded from the derivative system. Excluding still-positive probabilities in this way is a numerical approximation. The realized-CCP derivatives below retain direct utility effects for every alternative in $\mathcal J^+$. A singular reduced KKT system is treated as a failed evaluation.

\paragraph{Simulated shares.}
Restoring the market and draw indices, write $\mathcal J_{rt}^+$ for the consideration set. For $j\in\mathcal J_{rt}^+$, let $w_{jrt}:=\ln\widehat P_{jrt}^{0*}+(u_{0,jt}^{(r)}+u_{1,jt}^{(r)})/\lambda$ denote the logarithm of the generalized-logit numerator. With $F_{\varepsilon}$ fixed by Assumption~\ref{ass:riblp_fullsupport} and the additive-shock draws held fixed, $u_{1,t}^{(r)}=\delta_{1,t}+\varepsilon_{1,t}^{(r)}$ does not vary with $\theta$ at fixed $\delta_{1,t}$. Hence $\partial w_{jrt}/\partial\theta_k=(\partial\widehat P_{jrt}^{0*}/\partial\theta_k)/\widehat P_{jrt}^{0*}+(\partial u_{0,jt}^{(r)}/\partial\theta_k)/\lambda$. Differentiating the generalized-logit formula gives
\begingroup
\small
\begin{equation}
\frac{\partial\widehat P_{jrt}^*}{\partial\theta_k}=\widehat P_{jrt}^*\left(\frac{\partial w_{jrt}}{\partial\theta_k}-\sum_{\ell\in\mathcal J_{rt}^+}\widehat P_{\ell rt}^*\frac{\partial w_{\ell rt}}{\partial\theta_k}\right),\qquad j\in\mathcal J_{rt}^+.
\label{eq:computation_realized_derivative}
\end{equation}
\endgroup
For the integrated-CCP derivatives entering $\partial w_{jrt}/\partial\theta_k$, we use Equation~\eqref{eq:computation_inner_derivative} with the numerical subset described above. The realized-CCP derivative is zero outside $\mathcal J_{rt}^+$. Averaging Equation~\eqref{eq:computation_realized_derivative} over population draws gives $\partial\sigma_{j,t}^{\mathrm{sim}}/\partial\theta_k=R_T^{-1}\sum_{r=1}^{R_T}\partial\widehat P_{jrt}^*/\partial\theta_k$.

At fixed $\theta$, the integrated CCPs do not depend on $\delta_{1,t}$, so the latent-index derivative is
\begingroup
\small
\begin{equation}
\frac{\partial\sigma_t^{\mathrm{sim}}}{\partial\delta_{1,t}^{\prime}}=\frac{1}{R_T}\sum_{r=1}^{R_T}\left[\operatorname{diag}(\widehat P_{rt}^*)-\widehat P_{rt}^*\widehat P_{rt}^{*\prime}\right]/\lambda,
\label{eq:computation_delta_derivative}
\end{equation}
\endgroup
where $\widehat P_{rt}^*:=(\widehat P_{jrt}^*)_{j=1}^J$ contains the inside-product realized CCPs.

\paragraph{Objective and constraint derivatives.}
The implementation minimizes one-half of the criterion in Equation~\eqref{eq:simulated_gmm_mpec}. The objective derivatives and outer MPEC KKT checks in Section~\ref{subsec:gmm_computation} use this normalization. The objective gradient follows directly from the affine moments in Equation~\eqref{eq:market_moment}, with $\widehat W_T$ held fixed. Each log-share derivative is the corresponding share derivative divided by $\sigma_{j,t}^{\mathrm{sim}}$. Market $t$'s constraints depend on the common parameters $\theta$ and its own latent indices $\delta_{1,t}$, but not on the latent indices of other markets. We supply IPOPT with these nonzero Jacobian blocks and their locations.

For the outer GMM search, implicit differentiation requires a locally continuously differentiable simulated share mapping and a nonsingular Jacobian with respect to $\delta_{1,t}$, as specified in Appendix~\ref{appendix:gmm_variance_consistency}. Under these conditions, differentiating the simulated share equations gives $\partial\delta_{1,t}^{\mathrm{sim}}(\theta)/\partial\theta^{\prime}$, as in Equation~\eqref{eq:identification_implicit_delta_derivative}. Substituting this derivative into Equation~\eqref{eq:simulated_moment_jacobian}, evaluated at the candidate $\theta$, gives the moment Jacobian used in the outer GMM search.

\subsection{Monte Carlo Implementation}
\label{appendix:mpec_initialization}

The Monte Carlo experiments in Section~\ref{sec:monte_carlo} follow Steps~1--3 in Section~\ref{subsec:gmm_computation}. Both designs are overidentified, so the exactly identified acceptance rule in that section does not apply. In each replication, the GMM parameter search starts from $\theta=(0.50,0.75,0.30)^{\prime}$ in the common-coefficient design and $\theta=(0.50,0.75,0.30,0.25,0.25)^{\prime}$ in the random-coefficient design.

For the initial share inversion, each $\delta_{1,jt}$ starts at the average of $\ln(s_{jt}/s_{0t})-\delta_{0,jt}$ and $\beta_1x_{1,jt}$, evaluated at the starting parameter vector. If necessary, we retry using $\ln(s_{jt}/s_{0t})-\delta_{0,jt}$ alone and then $\beta_1x_{1,jt}$ alone. Each of these index starts is clipped to $[-29,29]$, inside the search bounds given below.

The three steps use the following settings:
\begin{enumerate}[label=\emph{Step \arabic*:},leftmargin=*,itemsep=0.4em,parsep=0pt]
\item We use 32 population draws and 64 subjective-prior draws for the share-inversion GMM search. When this search returns a candidate, we attempt an MPEC solve with IPOPT at the same draws, using the candidate as the starting point. We retain the share-inversion candidate and any successful MPEC candidate as starting values for Step~2.
\item The final numbers of draws are $(R_T,S_T)=(64,128)$, $(128,256)$, and $(256,512)$ for $T=25$, $50$, and $100$, respectively. If no saved candidate matches shares with these draws, bounded trust-region nonlinear least squares jointly adjusts $\theta$ and $\delta_1$ to reduce squared log-share residuals, with a small quadratic penalty on departures of $\theta$ from its starting value. The penalty weight is reduced if necessary. This is a share-matching step, not GMM minimization. The outer GMM search resumes only if share inversion at fixed $\theta$ confirms share matching.
\item For the final MPEC solve, candidates matching shares to tolerance are ranked first by increasing GMM objective. Other finite candidates follow, ranked by share error and then objective. At most two distinct starting points are tried, with the second used only if the first solve fails the final checks.
\end{enumerate}
Failure of a preliminary search or lower-draw solve does not by itself reject a replication if a finite candidate remains for the final MPEC solve. The final-draw IPOPT solve and independent acceptance checks in Section~\ref{subsec:gmm_computation} are required even if the starting point is already accurate.

The numerical bounds in the Monte Carlo experiments are $\beta_0\in[-5,5]$, $\alpha\in[0.05,5]$, $\beta_1\in[-3,3]$, and $\delta_{1,jt}\in[-30,30]$. The random-coefficient design additionally imposes $\sigma_x\in[0,2]$ and $\sigma_p\in[0,1.5]$. These are computational search bounds, and the reported validation checks apply to this bounded problem.

Table~\ref{tab:computational_tolerances} gives the solver settings and tolerances used for the Monte Carlo acceptance checks.

\begin{table}[!htbp]
\centering
\caption{Monte Carlo computational settings and validation tolerances}
\label{tab:computational_tolerances}
\begin{minipage}{0.78\textwidth}
\footnotesize
\renewcommand{\arraystretch}{1.08}
\begin{tabularx}{\linewidth}{@{}>{\raggedright\arraybackslash}Xc@{}}
\toprule
Setting & Value \\
\midrule
Newton--BA switching threshold $\tau_{\mathrm{switch}}$ & $10^{-4}$ \\
Inner KKT tolerance $\tau_{\mathrm{KKT}}$ & $10^{-8}$ \\
Inner simplex tolerance $\tau_{\mathrm{simplex}}$ & $10^{-10}$ \\
Final log-share tolerance $\tau_{\mathrm{share}}$ & $2\times10^{-6}$ \\
Final bound-feasibility tolerance & $2\times10^{-6}$ \\
Final stationarity, bound-dual, and complementarity tolerances & $10^{-3}$ \\
\bottomrule
\end{tabularx}
\end{minipage}
\end{table}

For inference, we require the simulated share Jacobians with respect to $\delta_{1,t}$ to be nonsingular and $\widehat D_T^{\prime}\widehat W_T\widehat D_T$ to have condition number at most $10^{14}$. Estimated variances must be nonnegative.

\FloatBarrier
\Needspace{6\baselineskip}
\section{Supplementary Empirical Details}
\label{app:tuna}

\subsection{Sample construction}
\label{app:tuna_sample}

The input file, \texttt{data\_allstores.csv}, contains 959,331 store--product--week observations across 49 UPCs, 291 weeks, and 86 stores. We sum units and revenue by UPC and week, counting customer visits once per store--week. Products are ranked by total units sold across all stores and weeks, with ranks fixed across samples.

After combining sales across stores, we retain a week only if every selected UPC has a positive share and an observed price, with $0<s_{0t}<1$. A store-level zero alone does not exclude a week, and we never replace zero shares with positive values.

\begin{table}[!htbp]
\centering
\caption{Sample retention and tuna unit-sales coverage}
\label{tab:tuna_sample_audit}
\begin{minipage}{0.78\textwidth}
\footnotesize
\begin{tabular*}{\linewidth}{@{\extracolsep{\fill}}crrr@{}}
\toprule
Products $J$ & Retained weeks & Retention & Unit-sales coverage\\
 & & (\%) & (\%)\\
\midrule
5 & 290 & 99.7 & 46.7\\
10 & 234 & 80.4 & 62.2\\
20 & 234 & 80.4 & 81.6\\
30 & 107 & 36.8 & 89.8\\
\bottomrule
\end{tabular*}
\par\smallskip\fontsize{9}{11}\selectfont\raggedright
\emph{Notes:} Retention is the percentage of the 291 source weeks retained. Coverage is the selected products' percentage of weekly unit sales across all 49 UPCs, averaged over each row's retained weeks.
\end{minipage}
\end{table}

Table~\ref{tab:tuna_sample_audit} motivates the 20-product baseline: expanding from 10 to 20 products raises unit-sales coverage from 62.2\% to 81.6\% without losing weeks, whereas 30 products leave only 107 weeks. The baseline balances coverage against sample length. Appendix~\ref{app:tuna_product_set} compares 20 and 30 products on the same weeks to separate these effects.

\FloatBarrier
\subsection{Computation and numerical validation}
\label{app:tuna_numerical}

\paragraph{Estimation.}
We follow the computational procedure in Section~\ref{sec:computation}. The inner solver uses switching threshold $\tau_{\mathrm{switch}}=10^{-4}$, KKT tolerance $\tau_{\mathrm{KKT}}=10^{-10}$, and simplex tolerance $\tau_{\mathrm{simplex}}=10^{-12}$. These inner tolerances apply to estimation, inference, and demand responses. If the stopping criterion is met but the normalized output fails the final checks, BA continues from the iterate before normalization, using only the remaining iteration budget.

Starting coefficients come from logit IV estimates recomputed from the sample. We first estimate RI--BLP at $32/64$ draws, then use each accepted fit to initialize the next grid: $64/128$, $128/256$, $256/512$, and the $512/1{,}024$ baseline. The baseline initializes the $1{,}024/2{,}048$ fit and the full-sample sensitivity specifications. All starting estimates are generated within the same run. Each fit is refined and validated using its specified numbers of draws. The empirical bounds are $\alpha\in[0.01,20]$, $[-25,25]$ for each remaining coefficient, and $\delta_{1,jt}\in[-40,40]$.

\paragraph{Acceptance.}
The empirical specifications are exactly identified. Both full- and matched-sample fits use the acceptance rules in Section~\ref{subsec:gmm_computation}. The independent inner-KKT check allows a residual of at most $1.01\times10^{-10}$, while the solver targets $10^{-10}$. The bound-feasibility tolerance is $2\times10^{-6}$. For acceptance without a final MPEC solve, absolute log-share residuals must be at most $2\times10^{-7}$. This near-zero-moment rule also requires $\|\overline m_T(\theta,\delta_1)\|_2\leq10^{-7}$ and a GMM objective in Equation~\eqref{eq:simulated_gmm_mpec} of at most $2\times10^{-12}$.

For estimates obtained from the final MPEC solve, the log-share tolerance is $5\times10^{-7}$, and the outer MPEC KKT checks in Section~\ref{subsec:gmm_computation} use tolerance $10^{-3}$. The same inner and bound checks apply. All reported RI--BLP estimates pass the applicable acceptance checks. We compute inference only after acceptance, requiring $\widehat D_T^{\prime}\widehat W_T\widehat D_T$ to be numerically nonsingular and the estimated covariance matrix to be positive semidefinite up to numerical rounding.

\paragraph{Demand derivatives.}
For the common-coefficient model in Equation~\eqref{eq:tuna_riblp_utility}, the analytical price derivative is $\partial\sigma_{j,t}^{\mathrm{sim}}/\partial p_{kt}=-\alpha\,\partial\sigma_{j,t}^{\mathrm{sim}}/\partial\delta_{0,kt}$. We obtain the baseline-index derivatives by applying Equations~\eqref{eq:computation_inner_derivative} and~\eqref{eq:computation_realized_derivative} with respect to $\delta_{0,kt}$ in place of $\theta_k$, holding the estimated coefficients, $\delta_{1,t}$, the prior, and the simulation draws fixed. These derivatives include the response of optimal information acquisition. In the Lent-specific specification, $\alpha$ is replaced by the relevant week's price coefficient. All reported demand responses use these analytical derivatives. After verifying share matching, we use observed shares in the elasticity denominators.

\paragraph{Finite-difference checks.}
We validate selected columns of $\partial\sigma_t^{\mathrm{sim}}/\partial\delta_{0,t}^{\prime}$ using central finite differences in the first, middle, and last sample markets. Each check perturbs one component of $\delta_{0,t}$ in both directions and reoptimizes information acquisition at both points. The estimated coefficients, $\delta_{1,t}$, the prior, and the simulation draws remain fixed. Because finite differences can amplify inner-solver error, the perturbed-share evaluations use tighter tolerances: $10^{-12}$ for the KKT conditions and $10^{-14}$ for simplex feasibility. The analytical derivatives being checked retain the estimation tolerances given above.

Perturbation sizes follow the prespecified sequence $10^{-4},10^{-5},5\times10^{-6},2.5\times10^{-6},1.25\times10^{-6},6.25\times10^{-7}$. A check passes when two consecutive approximations agree with the analytical derivative and with each other within absolute tolerance $2\times10^{-8}$ plus relative tolerance $3\times10^{-4}$.\footnote{For each comparison, the relative tolerance is scaled by the larger maximum absolute entry of the two derivative columns. We check columns 1, 11, and 20 in the 20-product full-sample specifications. Matched-sample checks instead perturb selected prices using the same relative step sequence, with the analytical column setting the error scale. All evaluated steps are recorded. The check fails if an inner solve fails or no consecutive pair meets the agreement criteria.} These checks assess differentiation of the finite-draw share function. Appendix~\ref{app:tuna_integration} separately examines sensitivity to integration draws.

\paragraph{Validation results.}
For the baseline estimate, the maximum inner KKT residual is below $10^{-10}$, the maximum absolute log-share residual is $4.88\times10^{-15}$, and the moment norm is $2.44\times10^{-11}$. Across the baseline's accepted finite-difference comparisons, the maximum absolute discrepancy is $4.07\times10^{-11}$ and the maximum relative discrepancy is $4.11\times10^{-8}$. Derivative-weighted diversion sums to one across destinations to within $4.5\times10^{-16}$.

\FloatBarrier
\subsection{Sensitivity to integration draws}
\label{app:tuna_integration}

To assess simulation accuracy, Table~\ref{tab:tuna_integration} compares estimates as the numbers of population and prior draws, $R/S$, increase from $32/64$ to $1{,}024/2{,}048$, with $512/1{,}024$ as the baseline. Each larger draw set includes the smaller set. Population and subjective-prior draws use separate fixed scrambled Sobol sequences. The model and calibration remain fixed, and the coefficients and latent indices are re-estimated for each $R/S$ pair.

\begin{table}[!htbp]
\centering
\caption{Sensitivity to integration draws}
\label{tab:tuna_integration}
\begin{minipage}{0.80\textwidth}
\footnotesize
\setlength{\tabcolsep}{4pt}
\begin{tabular*}{\linewidth}{@{\extracolsep{\fill}}crrrrr@{}}
\toprule
Draws & $\widehat\alpha$ & Mean own & Negative & Outside & Top \\
$R/S$ & & elasticity & cross (\%) & diversion & match (\%) \\
\midrule
32/64 & 1.495 & $-0.506$ & 79.3 & 1.426 & -- \\
64/128 & 1.652 & $-0.527$ & 85.7 & 1.419 & 35 \\
128/256 & 1.694 & $-0.525$ & 89.4 & 1.354 & 30 \\
256/512 & 1.789 & $-0.542$ & 95.6 & 1.376 & 45 \\
512/1,024 & 1.755 & $-0.532$ & 97.7 & 1.388 & 45 \\
1,024/2,048 & 1.714 & $-0.511$ & 97.9 & 1.370 & 50 \\
\bottomrule
\end{tabular*}
\par\smallskip\fontsize{9}{11}\selectfont\raggedright
\emph{Notes:} $R/S$ denotes population/prior draws. Top match is the percentage of products with the same highest-diversion inside destination as in the preceding row. The highest ratio is used even when all are negative.
\end{minipage}
\end{table}
Table~\ref{tab:tuna_integration} shows that doubling the draws from $512/1{,}024$ to $1{,}024/2{,}048$ changes the price coefficient and aggregate demand summaries only modestly, within the prespecified tolerances.\footnote{The tolerances are a 5\% change in $\widehat\alpha$, 0.025 in mean own-price elasticity, 5 percentage points in the share of negative cross-price elasticities, and 0.10 in mean outside diversion.} The broad demand patterns persist, while top-destination agreement is 50\%, below the 80\% threshold.

\begin{table}[!htbp]
\centering
\caption{Selected diversion ratios with different numbers of draws}
\label{tab:tuna_integration_pairs}
\begin{minipage}{0.62\textwidth}
\footnotesize
\begin{tabular*}{\linewidth}{@{\extracolsep{\fill}}ccrr@{}}
\toprule
Origin & Destination & $512/1{,}024$ & $1{,}024/2{,}048$\\
\midrule
10 & 1 & $-7.923$ & $-8.604$\\
16 & 2 & $4.267$ & $-0.386$\\
1 & 5 & $0.056$ & $-0.021$\\
\bottomrule
\end{tabular*}
\par\smallskip\fontsize{9}{11}\selectfont\raggedright
\emph{Notes:} Values are $100D^{\mathrm{DW}}$ for RI--BLP. Origin identifies the product whose price changes and destination the product whose share responds. Column headings give $R/S$.
\end{minipage}
\end{table}

Table~\ref{tab:tuna_integration_pairs} shows greater sensitivity at the product-pair level: diversion from product 16 to product 2 and from product 1 to product 5 changes sign. The aggregate summaries are relatively stable, but some individual diversion ratios and destination rankings remain sensitive to integration draws.

\FloatBarrier

\subsection{Lent-specific price coefficients}
\label{app:tuna_robustness}

Motivated by the Lent comparisons in \citet{nevohatzitaskos2006} and \citet{gandhi2023zeroes}, we allow the price coefficient to differ during Lent. We replace $-\alpha p_{jt}$ with \mbox{$-(\alpha+\alpha_L\operatorname{Lent}_t)p_{jt}$}, so the price coefficient is $\alpha$ outside Lent and $\alpha+\alpha_L$ during Lent. The increment $\alpha_L$ is distinct from the Lent utility shift $\tau_L$.

Both models retain the baseline controls and instruments. We instrument the added price interaction $p_{jt}\operatorname{Lent}_t$ with $c_{jt}\operatorname{Lent}_t$, constructed by multiplying acquisition cost by the Lent indicator. The additional exclusion assumption is $\mathbb E[c_{jt}\operatorname{Lent}_t\xi_{1,jt}]=0$.

\begin{table}[!htbp]
\centering
\caption{Lent price-interaction estimates}
\label{tab:tuna_lent_coefficients}
\begin{minipage}{0.70\textwidth}
\footnotesize
\begin{tabular*}{\linewidth}{@{\extracolsep{\fill}}lcc@{}}
\toprule
 & Logit & RI--BLP \\
\midrule
Near Lent, $\tau_N$ & $0.107$ & $0.054$ \\
 & (0.072) & (0.033) \\
Lent, $\tau_L$ & $0.356^{**}$ & $0.229^{**}$ \\
 & (0.163) & (0.099) \\
Week trend, $\tau_1$ & $-0.084^{**}$ & $-0.037^{**}$ \\
 & (0.036) & (0.016) \\
Price coefficient outside Lent, $\alpha$ & $2.616^{***}$ & $1.704^{***}$ \\
 & (0.327) & (0.194) \\
Lent increment, $\alpha_L$ & $0.810$ & $0.691$ \\
 & (0.660) & (0.450) \\
\midrule
Price coefficient during Lent, $\alpha+\alpha_L$ & $3.426^{***}$ & $2.396^{***}$ \\
 & (0.762) & (0.519) \\
\bottomrule
\end{tabular*}
\par\smallskip\fontsize{9}{11}\selectfont\raggedright
\emph{Notes:} 20 products, 234 weeks, including 15 Lent weeks. Both models include UPC fixed effects. Parentheses contain HAC(8) standard errors. $^{***}p<0.01$, $^{**}p<0.05$, $^{*}p<0.10$.
\end{minipage}
\end{table}

Table~\ref{tab:tuna_lent_coefficients} reports positive estimates of $\alpha_L$ in both models. Their direction agrees with the greater price sensitivity during Lent reported by \citet[Table~C1]{nevohatzitaskos2006} and \citet[Table~8]{gandhi2023zeroes}, but neither increment is statistically significant at the 10\% level in our sample.

Allowing a Lent-specific price coefficient leaves the aggregate RI--BLP cross-price and outside-diversion results almost unchanged (Table~\ref{tab:tuna_saved_specification_responses} in Appendix~\ref{app:tuna_sensitivity}).

\FloatBarrier
\subsection{Calibration sensitivity}
\label{app:tuna_sensitivity}
\begingroup
\setlength{\intextsep}{4pt plus 1pt minus 1pt}
\setlength{\abovecaptionskip}{4pt}
\setlength{\abovedisplayskip}{4pt plus 1pt minus 1pt}
\setlength{\belowdisplayskip}{4pt plus 1pt minus 1pt}
\setlength{\abovedisplayshortskip}{4pt plus 1pt minus 1pt}
\setlength{\belowdisplayshortskip}{4pt plus 1pt minus 1pt}

We first vary the prior and both population-shock variances jointly, holding $\lambda=1$. In the order $\pi_{u_1}$, $F_{\varepsilon_0}$, and $F_{\varepsilon_1}$, the per-product variance triples are $(20,0.75,1.44)$ for the low calibration, $(22,1,2.56)$ for the baseline, and $(24,1.5,4)$ for the high calibration. Figure~\ref{fig:tuna_calibration_sensitivity} shows that higher variances increase both the price coefficient and mean own-price elasticity magnitude. Demand remains inelastic on average and more elastic than logit (Table~\ref{tab:tuna_main_results}).

\begin{figure}[!htbp]
\centering
\includegraphics[width=0.80\textwidth]{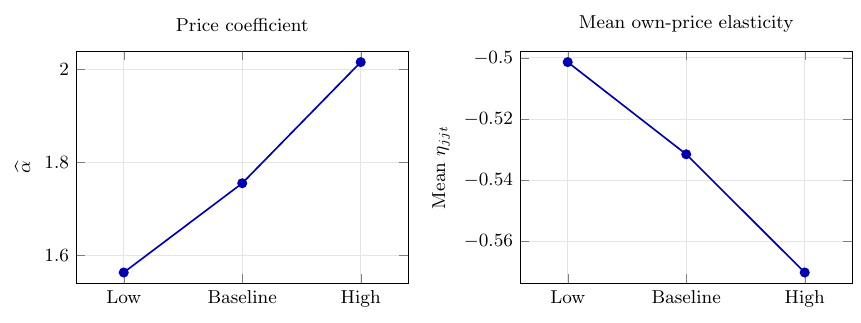}
\caption{Price coefficients and elasticities across calibrations}
\label{fig:tuna_calibration_sensitivity}
\begin{minipage}{0.94\textwidth}
\footnotesize\raggedright
\emph{Notes:} RI--BLP is re-estimated at each variance calibration using $R=512$ population draws and $S=1024$ prior draws.
\end{minipage}
\end{figure}

\begin{table}[!htbp]
\centering
\caption{Cross-price robustness: variance and Lent}
\label{tab:tuna_saved_specification_responses}
\begin{minipage}{0.78\textwidth}
\footnotesize
\begin{tabular*}{\linewidth}{@{\extracolsep{\fill}}lrrr@{}}
\toprule
Specification & Mean cross-price & Negative cross & Outside \\
 & elasticity & (\%) & diversion \\
\midrule
Baseline & $-0.00696$ & 97.7 & 1.388 \\
Low variance & $-0.00846$ & 98.6 & 1.508 \\
High variance & $-0.00507$ & 95.9 & 1.258 \\
Lent price interaction & $-0.00694$ & 97.7 & 1.392 \\
\bottomrule
\end{tabular*}
\par\smallskip\fontsize{9}{11}\selectfont\raggedright
\emph{Notes:} 20 products, 234 weeks, $R=512$, $S=1024$, $\lambda=1$. Negative cross (\%) is the percentage of inside cross-price elasticities below zero. Outside diversion averages Equation~\eqref{eq:tuna_derivative_weighted_diversion} across price origins.
\end{minipage}
\end{table}

Across the low- and high-variance calibrations in Table~\ref{tab:tuna_saved_specification_responses}, 95.9\%--98.6\% of cross-price elasticities are negative and mean outside diversion exceeds one. Magnitudes change, but the broad demand pattern persists.

We next vary information cost $\lambda$ while keeping the baseline variances fixed (Table~\ref{tab:tuna_information_cost}).

\begin{table}[!htbp]
\centering
\caption{Information-cost sensitivity}
\label{tab:tuna_information_cost}
\begin{minipage}{0.78\textwidth}
\footnotesize
\begin{tabular*}{\linewidth}{@{\extracolsep{\fill}}crrrr@{}}
\toprule
$\lambda$ & $\widehat\alpha/\lambda$ & Mean own & Negative & Outside \\
 & & elasticity & cross (\%) & diversion \\
\midrule
0.8 & 2.282 & $-0.586$ & 97.5 & 1.455 \\
0.9 & 1.987 & $-0.554$ & 97.6 & 1.425 \\
1.0 & 1.755 & $-0.532$ & 97.7 & 1.388 \\
1.1 & 1.564 & $-0.515$ & 97.5 & 1.347 \\
1.2 & 1.404 & $-0.504$ & 97.2 & 1.307 \\
\bottomrule
\end{tabular*}
\par\smallskip\fontsize{9}{11}\selectfont\raggedright
\emph{Notes:} Re-estimated with baseline variances and $512/1{,}024$ draws. $\widehat\alpha/\lambda$ is the price coefficient divided by information cost. Outside diversion: see Table~\ref{tab:tuna_saved_specification_responses}.
\end{minipage}
\end{table}

Table~\ref{tab:tuna_information_cost} shows that raising $\lambda$ from 0.8 to 1.2 reduces mean own-price elasticity in magnitude by about 14\% and mean outside diversion by about 10\%. Elasticity and diversion magnitudes are therefore sensitive to the calibration. Nevertheless, 97.2\%--97.7\% of cross-price elasticities remain negative. As in the variance checks, demand remains inelastic on average and mean outside diversion exceeds one.

\endgroup

\FloatBarrier
\subsection{Sensitivity to products and weeks}
\label{app:tuna_product_set}

The sample change is substantial: requiring positive sales for all 30 products leaves 107 of the baseline's 234 weeks. We re-estimate both models with 20 products on these weeks, then with 30, separating week and product-set changes. The RI calibration and $512/1{,}024$ draws remain fixed. This comparison does not correct for selection from excluding zero-share weeks.

\begin{table}[!htbp]
\centering
\caption{Product-set and sample comparison}
\label{tab:tuna_product_set}
\begin{minipage}{0.86\textwidth}
\footnotesize
\setlength{\tabcolsep}{4pt}
\begin{tabular*}{\linewidth}{@{\extracolsep{\fill}}lrrrr@{}}
\toprule
 & & Mean own & Negative & Comparable outside\\
Specification & Weeks & elasticity & cross (\%) & diversion $\times100$\\
\midrule
\multicolumn{5}{@{}l}{\emph{RI--BLP}}\\
$J=20$: baseline & 234 & $-0.532$ & 97.66 & 138.77\\
$J=20$: re-estimated & 107 & $-1.483$ & 97.80 & 137.29\\
$J=30$: re-estimated & 107 & $-1.292$ & 98.01 & 140.40\\
\addlinespace[4pt]
\multicolumn{5}{@{}l}{\emph{Logit}}\\
$J=20$: baseline & 234 & $-0.491$ & 0.00 & 94.08\\
$J=20$: re-estimated & 107 & $-1.527$ & 0.00 & 95.12\\
$J=30$: re-estimated & 107 & $-1.356$ & 0.00 & 95.12\\
\bottomrule
\end{tabular*}
\par\smallskip\fontsize{9}{11}\selectfont\raggedright
\emph{Notes:} Summaries cover products 1--20. Comparable outside diversion is averaged across these price origins and multiplied by 100. For $J=30$, the comparable outside category also includes products 21--30.
\end{minipage}
\end{table}

Table~\ref{tab:tuna_product_set} shows that restricting the 20-product sample to 107 weeks and re-estimating changes both models from inelastic to elastic demand on average. The mean own-price elasticity magnitudes are similar: $1.483$ (RI--BLP) and $1.527$ (logit). Expanding to 30 products on the same weeks reduces both magnitudes, but both remain above one.

Under RI--BLP, roughly 98\% of cross-price elasticities remain negative and mean comparable outside diversion stays above one (above 100 in the table) across all samples.

\begin{table}[!htbp]
\centering
\caption{RI--BLP diversion sign changes across samples}
\label{tab:tuna_sample_pair_signs}
\begin{minipage}{0.68\textwidth}
\footnotesize
\begin{tabular*}{\linewidth}{@{\extracolsep{\fill}}lr@{}}
\toprule
Comparison & Pairs changing sign\\
\midrule
234 to 107 weeks ($J=20$) & 3 of 380\\
20 to 30 products (107 weeks) & 1 of 380\\
\bottomrule
\end{tabular*}
\par\smallskip\fontsize{9}{11}\selectfont\raggedright
\emph{Notes:} Counts compare derivative-weighted diversion signs for products 1--20. Directions are counted separately, giving $20\times19=380$ pairs.
\end{minipage}
\end{table}

Table~\ref{tab:tuna_sample_pair_signs} records few diversion sign changes. The broad RI--BLP cross-price and comparable outside-diversion patterns remain stable across these comparisons. Own-price elasticities, however, are sensitive to the weeks included, and the baseline finding of inelastic demand on average does not hold in the restricted sample.
\FloatBarrier

\end{appendices}

\end{document}